\documentclass[fp,twocolumn]{jpsj3}

\usepackage{graphicx}
\usepackage{txfonts}
\usepackage{bm}
\usepackage[version=4]{mhchem}
\usepackage{color}

\renewcommand{\theenumii}{\textit{\theenumi.\arabic{enumii}}}

\renewcommand{\theenumiii}{\textit{\theenumii.\arabic{enumiii}}}

\title{Exotic Superconducting States in FeSe-based Materials}

\author{
Takasada Shibauchi$^1$\thanks{shibauchi@k.u-tokyo.ac.jp},
Tetsuo Hanaguri$^2$\thanks{hanaguri@riken.jp},
and Yuji Matsuda$^3$\thanks{matsuda@scphys.kyoto-u.ac.jp},
}

\inst{
$^1$Department of Advanced Materials Science, University of Tokyo, Kashiwa 277-8561, Japan \\
$^2$RIKEN Center for Emergent Matter Science, Wako 351-0198, Japan \\
$^3$Department of Physics, Kyoto University, Kyoto 606-8502, Japan 
}

\abst{
High-temperature superconductivity and a wide variety of exotic superconducting states discovered in FeSe-based materials have been at the frontier of research on condensed matter physics over the past decade.
Unique properties originating from the multiband electronic structure, strongly orbital-dependent phenomena, extremely small Fermi energy, electronic nematicity, and topological aspects give rise to many distinct and fascinating superconducting states.
Here, we provide an overview of our current understanding of the superconductivity of {\it bulk} FeSe-based materials, focusing on FeSe and the isovalent substituted FeSe$_{1-x}$S$_{x}$ and FeSe$_{1-x}$Te$_{x}$.
We discuss the highly nontrivial superconducting properties in FeSe, including extremely anisotropic pairing states, crossover phenomena from Bardeen--Cooper--Schrieffer (BCS) to Bose--Einstein condensation (BEC) states, a novel field-induced superconducting phase, and broken time-reversal symmetry.
We also discuss the evolution of the superconducting gap function with sulfur and tellurium doping, paying particular attention to the impact of quantum critical nematic fluctuations and the topological superconductivity.
FeSe-based materials provide an excellent playground to study various exotic superconducting states.
}

\kword{iron-based superconductors, electronic nematicity, BCS--BEC crossover, quantum criticality, FFLO state, time-reversal symmetry breaking, topological superconductivity}
\begin{document}
\maketitle

%%%%%%%%%%%%%%%%%%%%%%%%%%%%%%%%%%%%%%%%%%%%%%%%%%%%%%%%%%%%%%%%%%%%%%%%%%%%%%%
%%%%%%%%%%%%%%%%%%%%%%%%%%%%%%%%%%%%%%%%%%%%%%%%%%%%%%%%%%%%%%%%%%%%%%%%%%%%%%%

%\tableofcontents

\textbf{CONTENTS}

\begin{enumerate}

\item \textbf{Introduction}

\item \textbf{Electronic Structure and Phase Diagrams}
	\begin{enumerate}
	\item \textit{Band structure of FeSe}
	\item \textit{Electronic nematic state}
	\item \textit{Nematic quantum critical point (QCP)}
	\end{enumerate}

\item \textbf{Superconducting Gap Structure} 
	\begin{enumerate}
	\item \textit{Bulk measurements}
		\begin{enumerate}	
	    	\item \textit{Temperature dependence}
	    	\item \textit{Field dependence}
		\end{enumerate}	
	\item \textit{Angle-resolved photoemission spectroscopy (ARPES)}
	\item \textit{Scanning tunneling microscopy / spectroscopy (STM / STS)}
	\end{enumerate}

\item \textbf{BCS-BEC Crossover}
	\begin{enumerate}
	\item \textit{BCS-BEC crossover}
	\item \textit{Evidence for the crossover in FeSe}
    		\begin{enumerate}	
	    	\item \textit{Penetration depth}
	    	\item \textit{ARPES}
		\item \textit{Quasiparticle interference (QPI)}
		\item \textit{Quantum-limit vortex core}
		\end{enumerate}	
	\item 	\textit{Superconducting fluctuations, preformed pairs, and pseudogap}
		\begin{enumerate}
		\item \textit{Giant superconducting fluctuations}
		\item \textit{Pseudogap}
		\item \textit{Evolution of BCS-BEC crossover in FeSe$_{1-x}$S$_{x}$}
		\end{enumerate}
	\end{enumerate}

\item \textbf{Exotic Superconducting State Induced by Magnetic Field} 
	\begin{enumerate}
	\item \textit{Field-induced superconducting phase}
	\item \textit{FeSe in strong magnetic field}
	\item \textit{Fulde--Ferrell--Larkin--Ovchinnikov (FFLO) state}
	\item \textit{Highly spin-polarized field-induced state in the BCS-BEC crossover regime}
	\end{enumerate}

\item  \textbf{Superconductivity near the Nematic Critical Point}
	\begin{enumerate}
	\item \textit{Abrupt change in superconducting gap}
	\item \textit{Possible ultranodal pair state with Bogoliubov Fermi surface}
	\end{enumerate}

\item \textbf{Time-Reversal Symmetry (TRS) Breaking} 
	\begin{enumerate}
	\item \textit{Effect of nematic twin boundary}
	\item \textit{Evidence from gap structure}
	\item \textit{Evidence from muon spin rotation ($\mu$SR)}
	\item \textit{TRS breaking in the bulk}
	\end{enumerate}

\item \textbf{Topological Superconducting States} 
	\begin{enumerate}
	\item \textit{Topological quantum phenomena}
	\item \textit{Topological superconductivity and Majorana quasiparticles}
	\item \textit{Potential platforms for topological superconductivity}
	\item \textit{Basic properties of FeSe$_{1-x}$Te$_{x}$}
	\item \textit{Topological phenomena in FeSe$_{1-x}$Te$_{x}$}
		\begin{enumerate}
		\item \textit{Band structure and ARPES experiments}
		\item \textit{Majorana bound state (MBS) in the vortex core}
		\item \textit{Search for other Majorana features}
		\end{enumerate}
	\end{enumerate}

\item \textbf{Conclusion} 

\end{enumerate}

%%%%%%%%%%%%%%%%%%%%%%%%%%%%%%%%%%%%%%%%%%%%%%%%%%%%%%%%%%%%%%%%%%%%%%%%%%%%%%%
%%%%%%%%%%%%%%%%%%%%%%%%%%%%%%%%%%%%%%%%%%%%%%%%%%%%%%%%%%%%%%%%%%%%%%%%%%%%%%%

\section{Introduction}

The discovery of high-temperature superconductivity in iron-pnictide compounds has been a significant breakthrough in the condensed matter community.
In 2006, Hosono's research group discovered superconductivity at the superconducting transition temperature $T_{\rm c} \approx 6$~K in the iron-pnictide LaFePO~\cite{Kamihara06}.
By replacing phosphorus with arsenic and by partially substituting oxygen with fluorine, $T_{\rm c}$ increases to 26~K~\cite{Kamihara08}.
By replacing La with other rare-earth elements, $T_{\rm c}$ is raised to 56~K~\cite{Wang08}.
The iron-chalcogenide superconductors have also been extensively studied.
In particular, iron-selenide FeSe with $T_{\rm c}\approx 9$~K, discovered by Wu's group in 2008~\cite{Hsu08}, has drawn considerable  attention because $T_{\rm c}$ increases  to 37~K under pressure~\cite{Medvedev09}.
Moreover, $T_{\rm c}$ increases to more than 50~K in monolayer FeSe thin films grown on SrTiO$_3$~\cite{Qing12}.  
Thus, iron-pnictides/chalcogenides became a new class of high-$T_{\rm c}$ superconductors, knocking the cuprates off their pedestal as a unique class of high-$T_{\rm c}$ superconductors.
There is almost a complete consensus that high-$T_{\rm c}$ superconductivity in these iron-based superconductors cannot be explained theoretically by the conventional electron-phonon pairing mechanism.  Thus, the origin of superconductivity is unconventional~\cite{Mazin10,Hirschfeld11,Chubukov12}.

Iron-based superconductors are two-dimensional (2D) layered materials with metallic Fe-pnictogen/chalcogen tetrahedral FeAs$_4$ (or P/Se/Te) layers~\cite{Paglione10,Stewart11,Wang12,Wen11}.
The pnictogen/chalcogen atoms alternatively reside above and below the Fe layers and are located at the center of the Fe atom squares.
There are several different types of iron-pnictide superconductor, which are often abbreviated by the ratio of the elements in their parent compositions and are known, for example,  111, 122, and 1111 types, as well as iron-chalcogenide superconductors, such as 11 and 122 types.
The parent compounds are metals, in contrast to Mott insulators, the parent compounds of the cuprates. Moreover, whereas in cuprates the physics is captured by a single band originating from one $3d_{x^2-y^2}$-orbital per Cu site, iron-based superconductors have about six electrons occupying the nearly degenerate $3d$ Fe orbitals.
Therefore,  the systems are intrinsically multiband/multi orbital systems,  and the interorbital Coulomb interaction plays an essential role.

The Fermi surface of iron-pnictides mainly consists of $d_{xy}$, $d_{yz}$,  and $d_{xz}$ orbitals, forming well-separated hole pockets near the center of the Brillouin zone and electron pockets near the zone corners.
The parent compounds of iron-pnictides are spin-density-wave (SDW) metals, which exhibit a transition at $T_{\rm N}$~\cite{Dai15}.
Below $T_{\rm N}$, a stripe-type long-range antiferromagnetic (AFM) order sets in, which breaks the lattice fourfold ($C_4$) rotational symmetry. 
The high-$T_{\rm c}$ superconductivity appears when the SDW is suppressed by either chemical substitution or  pressure.
The highest $T_{\rm c}$ is often achieved in the vicinity of an AFM quantum critical point (QCP), where the SDW transition vanishes~\cite{Shibauchi14}.
Therefore, a pairing mechanism mediated by the exchange of AFM fluctuations, which stem from the intra-atomic Coulomb repulsion associated with the quasi-nesting between electron and hole pockets, has been widely discussed.
This scenario predicts the so-called $s_{\pm}$ order parameter, in which the sign of the superconducting gap changes between the electron and hole pockets of the Fermi surface~\cite{Kuroki09,Mazin10,Hirschfeld11,Chubukov12}.

On the other hand, the orbital degrees of freedom in iron-pnictides give rise to various phenomena.
Almost all families of iron-pnictides exhibit a tetragonal-to-orthorhombic structural transition at $T_{\rm s}$, which is accompanied by an orbital ordering that splits the degenerate $d_{xz}$ and $d_{yz}$ orbitals~\cite{Yi11}.
This transition that breaks the $C_4$ symmetry of the crystal lattice is referred to as an electronic nematic transition~\cite{Fernandes14}.
This nematic transition is believed to be a result of intrinsic electronic instability because its effect on the electronic properties is much larger than that expected from the observed structural distortion.
This nematic transition either precedes or is coincident with the SDW transition,  and the endpoint of the nematic transition is located very close to the AFM QCP.
Moreover, it has been reported that the electronic nematic order persists even above the superconducting dome in the tetragonal lattice phase in some of the iron-based superconductors~\cite{Kasahara12,Singh15,Thewalt18}.
It has been shown that the nematic susceptibility, which is measured as an induced resistivity anisotropy in response to an external strain~\cite{Chu12}, exhibits divergent behavior at $T \rightarrow 0$ while approaching the endpoint of the structural transition, indicating the presence of a nematic QCP~\cite{Kuo16}.
Consequently,  an alternative scenario of the pairing mechanism, which is mediated by orbital fluctuations, has been proposed~\cite{Kontani10}.
This scenario can hardly support the $s_{\pm}$ gap function.

\begin{figure}[t]
%    \vspace{1cm}
    \centering
    \includegraphics[width=\linewidth]{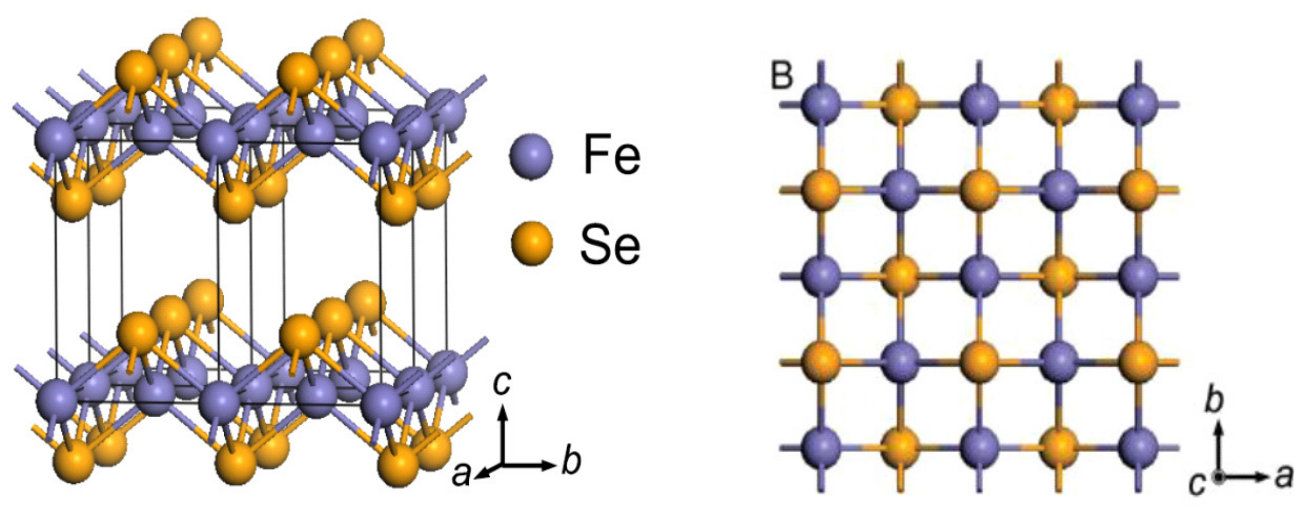}
    \caption{(Color online)
    Crystal structure of FeSe.
    Adopted from Ref.~\citen{Hsu08}.
    }
    \label{Crystal}
\end{figure}

Thus the question as to whether the nematic order is driven by spin and/or orbital degrees of freedom has been a topic of intense research, which is intimately related to the driving mechanism of iron-based superconductivity.
The nematic correlations intertwined with AFM order in iron-pnictides prevent us from identifying the essential role of nematic fluctuations, raising a fundamental question as to which fluctuations are the main driving force of the Cooper pairing. 
Identifying the relationship between nematic and SDW orders presents a ``chicken-or-egg" problem: Does the SDW order induce  the nematic order, or does the nematic order facilitate the SDW order?
Although an intricate coupling between magnetic and orbital degrees of freedoms is crucial to the understanding of the underlying physics responsible for their wide variety of exotic properties of iron-pnictides, these questions have been the topic of many debates~\cite{Fernandes14,Chubukov16a,Yamakawa16,Lee09}.
The nematic order is directly linked to the superconductivity because nematic instability is a characteristic feature of the normal state, upon which superconductivity emerges at lower temperatures.
Moreover, the nematicity not only in iron-based superconductors but also in cuprates  has now been extensively discussed ~\cite{Daou10,Hinkov08,Parker10,Kohsaka07,Sato17a,Murayama19,Ishida20}.

\begin{figure}[t]
    \centering
	\includegraphics[width=0.8\linewidth]{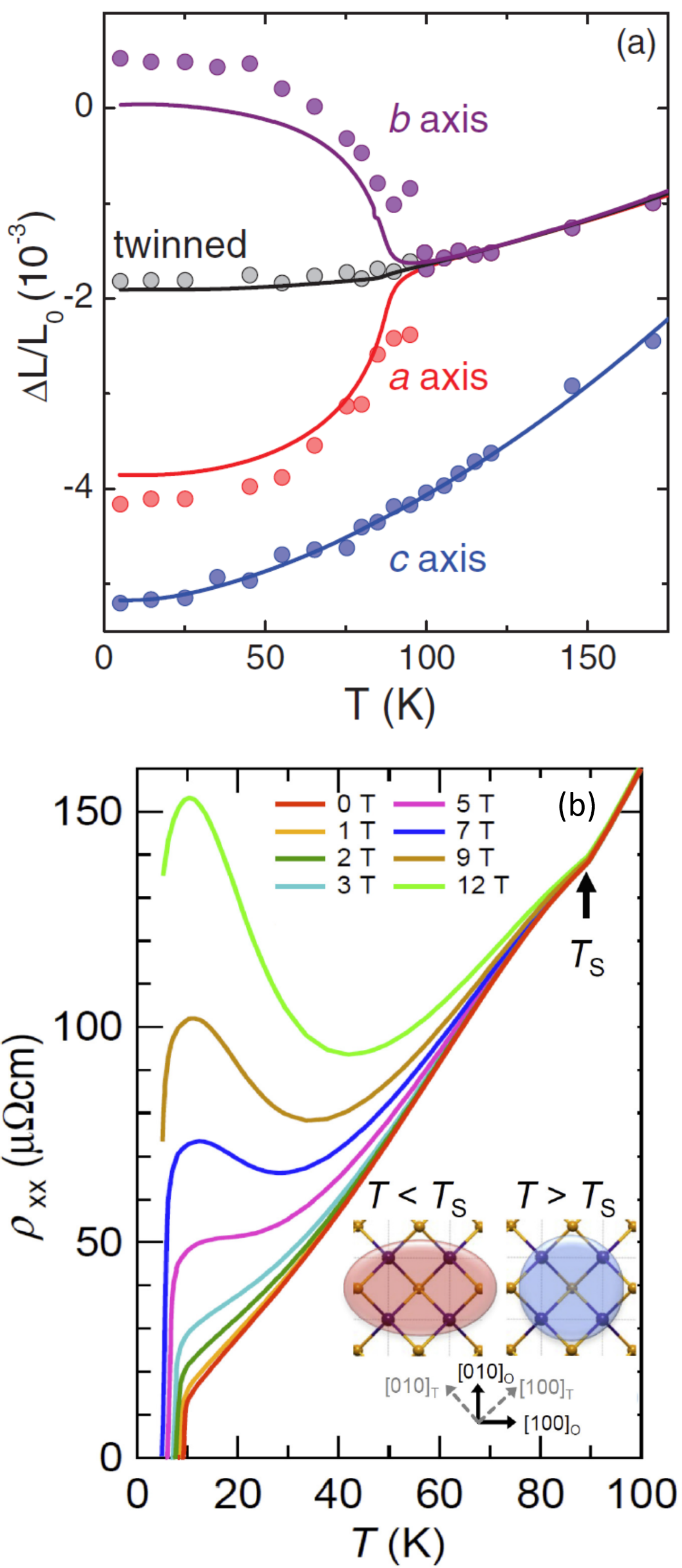}
	\caption{(Color online) (a) Temperature dependence of relative length changes along the three axes in FeSe.
	A tiny orthorhombic lattice distortion of $\sim 0.2$\% develops below the nematic phase transition at $T_{\rm s}\approx90$~K.
	%This is too small to account for the very large electronic anisotropy discussed in this review, and thus the nematicity is not due to the lattice instability but electronic in origin.
	Adopted from Ref.\citen{Bohmer13}.
    (b) Temperature dependence of the resistivity of FeSe for $\bm{H} \parallel c$~\cite{Kasahara14}.
    %Below $T_{\rm s}$, very large magnetoresistance is observed, which is a signature of long mean free path in compensated semimetals.
	The inset illustrates the appearance of the electronic nematicity.}
    \label{rho_T}
\end{figure}

Recently the iron-chalcogenide FeSe ($T_{\rm c}\approx 9$~K) and the isovalently substituted FeSe$_{1-x}$S$_{x}$ and FeSe$_{1-x}$Te$_{x}$ have become a central system in the research of exotic superconducting states.
FeSe is structurally the simplest among the iron-based superconductors, because it consists only of a stack of 2D FeSe layers weakly coupled by the van der Waals interaction (see Fig.~\ref{Crystal}).
The experimental progress has been markedly accelerated owing to the significant advances in the material quality of FeSe.
In particular, the chemical vapor transport technique has enabled us to grow high-quality millimeter-size single crystals of FeSe free of impurity phases~\cite{Bohmer13,Chareev13}, making detailed studies of the intrinsic bulk properties of FeSe possible.
FeSe is a strongly correlated semimetal, as revealed by the quasiparticle effective masses determined by   quantum oscillations, angle-resolved photoemission spectroscopy (ARPES), and heat capacity measurements, which are strongly enhanced from those calculated by density functional theory (DFT).

FeSe-based materials provide a unique opportunity to explore the effect of nematicity.  
FeSe also exhibits a nematic transition at $T_{\rm s}\approx 90$~K, as shown in Figs.~\ref{rho_T}(a) and \ref{rho_T}(b).
In contrast to iron-pnictides, however, FeSe exhibits no magnetic order down to $T\rightarrow 0$ despite its high $T_{\rm s}$, and its ground state is still an unconventional superconducting state.
The nematicity is markedly tunable by hydrostatic pressure $P$ and chemical substitution.
In FeSe, the structural (nematic) transition is rapidly suppressed by pressure, and $T_{\rm s}$ goes to zero at $P\approx2$~GPa.
The AFM static order is induced before the complete suppression of $T_{\rm s}$ and nematicity appears to coexist with the AFM order.
Nuclear magnetic resonance (NMR) and inelastic neutron scattering experiments revealed that the nematicity and magnetism are still highly entangled in FeSe.  

The nematicity in FeSe$_{1-x}$S$_{x}$ is also strongly suppressed with sulfur doping, and $T_{\rm s}$ goes to zero at $x\approx 0.17$.
The nematic fluctuations are strongly enhanced with sulfur doping, and the nematic susceptibility diverges towards $T=0$, revealing the presence of a nematic QCP at $x\approx 0.17~$~\cite{Hosoi16}.
Near the nematic QCP, no sizable AFM fluctuations are observed, indicating that the nematicity is disentangled from magnetic order.  

What distinuishes FeSe-based materials from other superconductors is the unique electronic structure, particularly the extremely shallow Fermi surface associated with the very small  number of carriers, multiband nature, and orbital-dependent electron correlations.
Remarkably, among the superconductivity, magnetism, and nematicity, all three of these most fundamental properties can be largely tuned.
Because of these properties, FeSe-based materials serve as not only a model system for  understanding   the effect of the nematicity on the normal and superconducting states, but also a new playground for exotic pairing states, which have been a long-standing issue of superconductivity.
  
In this review article, we shall attempt to bring   readers up to date with the rapidly expanding field of research on the superconductivity of FeSe-based materials, focusing on the experimental aspects of {\it bulk} properties.
Excellent reviews of research on atomic-layer thin films of FeSe have recently been published~\cite{Huang17,Wang17}.
For theoretical aspects,  readers can refer to another very recent review article~\cite{Kreisel20}. 
In Sect.~2, we discuss the electronic structure and phase diagram of FeSe briefly.
We then discuss several topics of active research among exotic superconducting states in FeSe-based materials, such as the superconducting gap structure (Sect.~3), the crossover phenomena from weak coupling Bardeen--Cooper--Schrieffer (BCS) to strong coupling Bose--Einstein condensation (BEC) states (Sect.~4), a magnetic-field-induced superconducting phase (Sect.~5), the superconducting state near the nematic QCP (Sect.~6), broken time-reversal symmetry (Sect.~7), and the topological superconducting state (Sect.~8).

%%%%%%%%%%%%%%%%%%%%%%%%%%%%%%%%%%%%%%%%%%%%%%%%%%%%%%%%%%%%%%%%%%%%%%%%%%%%%%%
%%%%%%%%%%%%%%%%%%%%%%%%%%%%%%%%%%%%%%%%%%%%%%%%%%%%%%%%%%%%%%%%%%%%%%%%%%%%%%%

\section{Electronic Structure and Phase Diagrams}

\subsection{Band structure of FeSe}

\begin{figure}[b]
	\centering
%	\vspace{1cm}
	\includegraphics[width=0.75\linewidth]{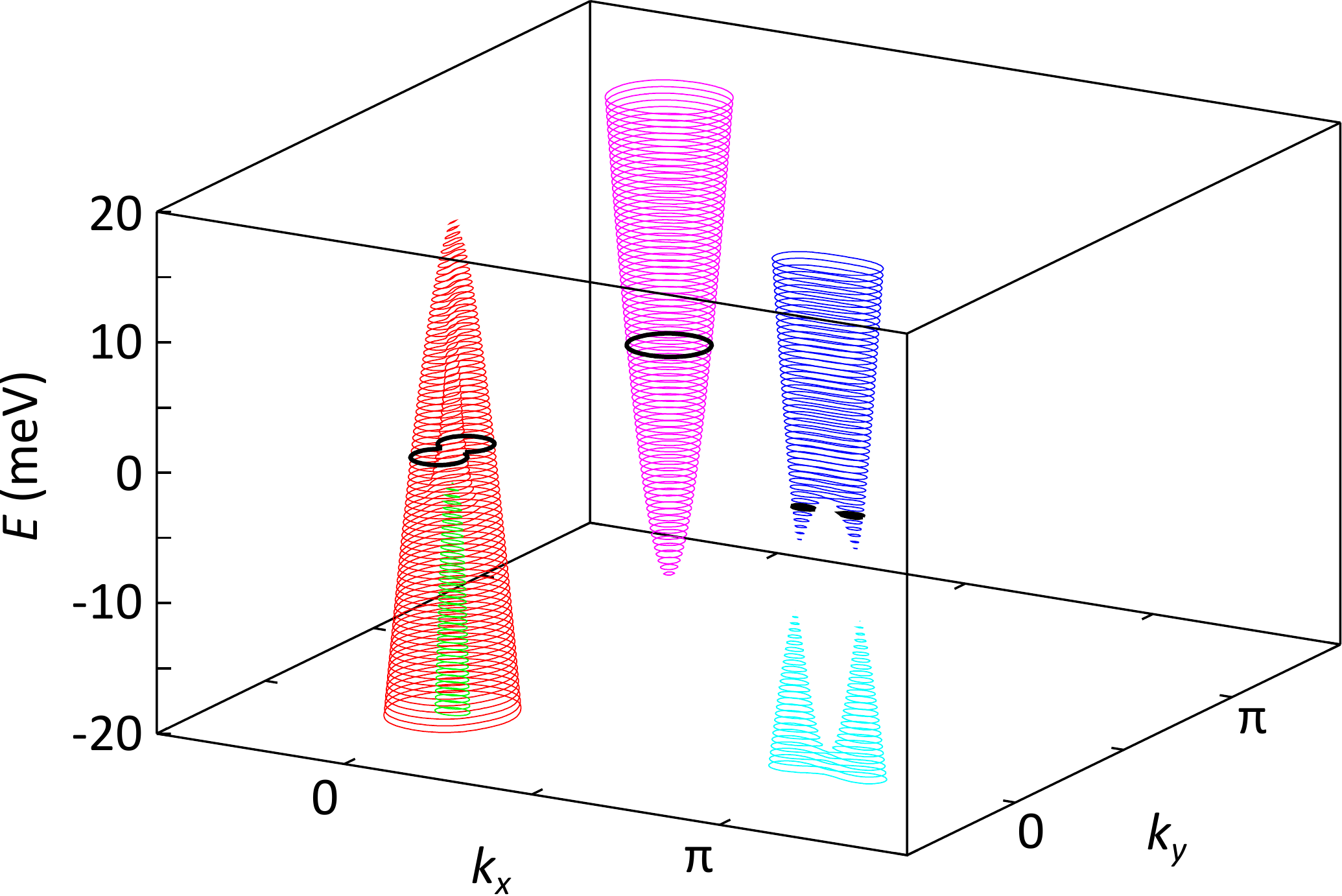}
	\caption{(Color online) 
	Schematic energy dispersion of the hole and electron pockets in the nematic phase of FeSe including spin-orbit interaction.
	The orbital-dependent energy shift is taken into account so that each band can be fitted to the ARPES data. By courtesy of Y.~Yamakawa and H.~Kontani. 
	}
	\label{FS_3D}
\end{figure}
	
First, we will not attempt to give an exhaustive survey of current research on the band structure of FeSe in the limited space here.
We will direct  readers to more extensive reviews~\cite{Pustovit16,Coldea17,Bohmer18}.
As FeSe is a compensated semimetal with equal numbers of electron and hole carriers, it is essentially a multiband superconductor.
As shown in Fig.\,\ref{rho_T}(b), a very large magnetoresistance is observed at low temperatures, which is a signature of a long mean free path in compensated semimetals~\cite{Kasahara14}. 
Similar to iron-pnictides, the Fermi surface of FeSe consists of $d_{xy}$, $d_{yz}$, and $d_{xz}$ orbitals, forming well-separated hole and electron pockets.
%The Fermi surface consists of well separated hole pockets near the center of the Brillouin zone and electron pockets near the zone corners.
Compared with the Fermi surface obtained by   DFT calculations, the actual Fermi surface in the tetragonal phase above $T_{\rm s}$ is much smaller and dispersions are significantly renormalized~\cite{Maletz14,Watson15}.

In the tetragonal phase,   DFT calculations show three hole pockets~\cite{Terashima14}, but ARPES measurements report two pockets, showing that one hole band shifts below the Fermi level.
Below the nematic transition at $T_{\rm s}\approx 90$~K, the orbital ordering lifts the degeneracy of $d_{xz}$ and $d_{yz}$ orbitals.
The splitting of the $d_{xz}$ and $d_{yz}$ energy bands is $\sim 50$~meV at the $M_{x}$ point in the unfolded Brillouin zone~\cite{Nakayama14,Shimojima14,Suzuki15,Zhang15,Watson16,Watson17}.
Then, one hole pocket in the tetragonal phase is shifted below the Fermi level below $T_{\rm s}$.
As a result, there is only one quasi-2D hole pocket around the $\Gamma$ point in the nematic phase.
The hole pocket is strongly distorted to an elliptical shape, which consists of a $d_{yz}$ orbital along the longer axis and a $d_{xz}$ orbital along the shorter axis of the ellipse.
Similar to iron-pnictides, the tiny orthorhombic distortion of $\sim 0.2$\% below $T_{\rm s}$ is too small to account for the very large electronic anisotropy, and thus the nematicity is not caused by the lattice instability but is electronic in origin. 

Although the shape and orbital character of the hole pocket appear to be well understood, a consensus about the shape of the electron pockets has not yet been reached.
This is mainly because both hole and electron pockets are extremely shallow and therefore high-resolution ARPES measurements are prerequisite to determining the detailed structure of the Fermi surfaces.
Unfortunately, such high-resolution ARPES measurements are available only in a limited momentum range around the $\Gamma$ point.
The schematic energy dispersion of the hole and electron pockets in the nematic phase including the spin-orbit interaction is shown in Fig.~\ref{FS_3D}~\cite{Yamakawa_PC}.
As a result of the finite spin-orbit interaction, Dirac cones around the $M_{x}$ point have a small gap, forming massive Dirac cones.
Moreover, the degeneracy at the $\Gamma$ point of the hole bands is lifted. In this energy shift, there is only one hole pocket around the $\Gamma$ point and one electron pocket around the $M_{y}$ point.

\begin{figure}[t]
	\centering
	%\vspace{-2cm}
	\includegraphics[width=1\linewidth]{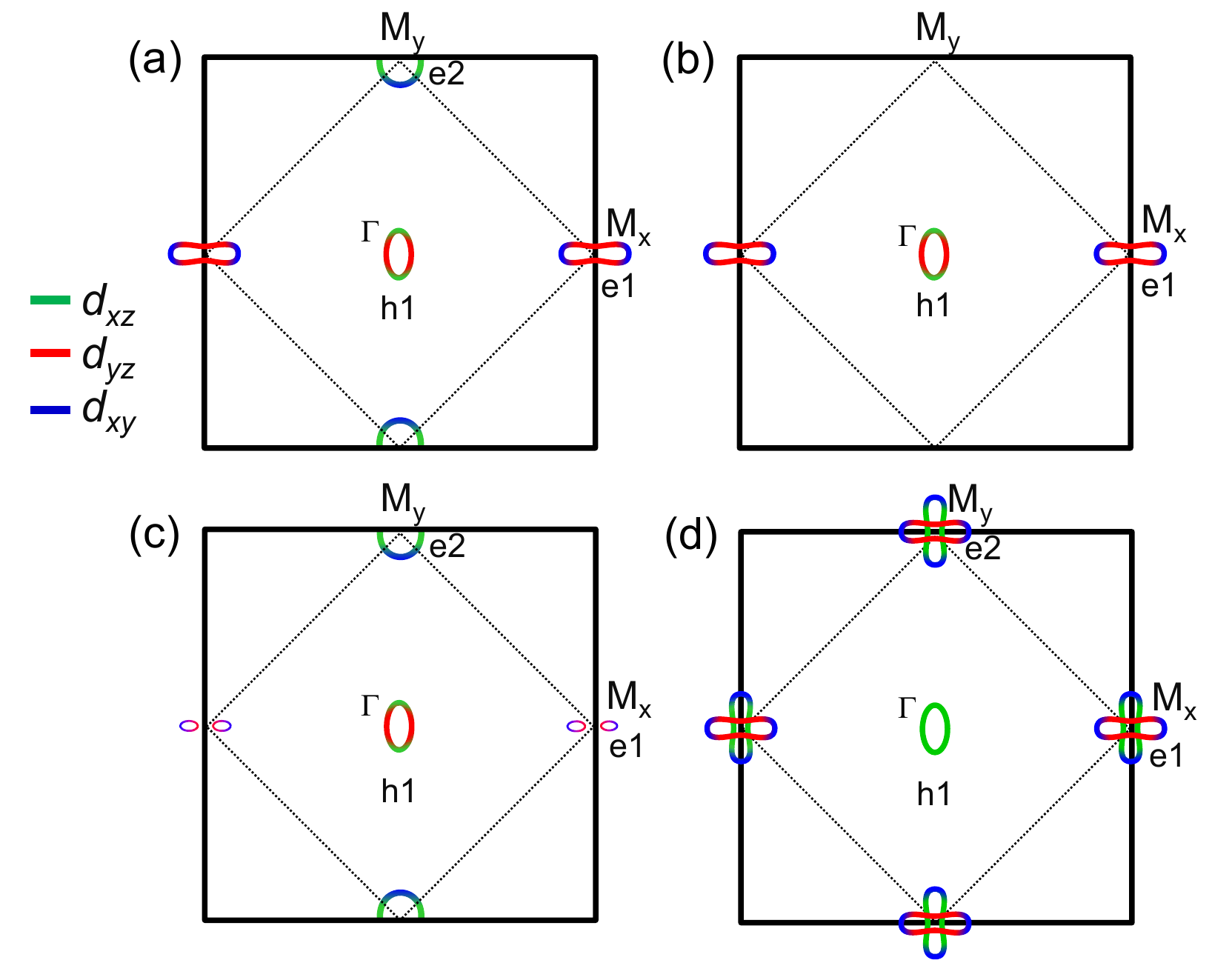}
	%\vspace{-3cm}
	\caption{(Color online) 
	Four possible Fermi surface structures of FeSe proposed on the basis of ARPES and quantum oscillations. Solid and dotted black lines represent the unfolded and folded Brillouin zone boundaries, respectively.
	In all the possibilities, there is one hole pocket around the $\Gamma$ pocket.
	For (a), (b), and (c), the hole pocket consists of $d_{yz}$ and $d_{xz}$ orbitals, while for (d), it consists of only a $d_{xz}$ orbital.
	The electron pocket around the $M_{x}$ point has a bow-tie shape for (a) and (b), while the electron pocket around the $M_{y}$ point is present for (a) and absent for (b).
	For (c), the electron pocket around the $M_{x}$ point is disconnected, forming a double Dirac-core-like structure. For (d), petal-like electron pockets appear both at the $M_{x}$ and $M_{y}$ points. Note that here the axes are defined as $b<a<c$, which is different from some experimental definitions. 
	}
	\label{FS}
\end{figure}
	
Figures~\ref{FS}(a)-\ref{FS}(d) illustrate four possible Fermi surface structures proposed on the basis of  ARPES and quantum oscillations.
Figures~\ref{FS}(a) and \ref{FS}(b) show the Fermi surfaces indicated  by some ARPES measurements, where the electron pocket has a bow-tie shape around the $M_{x}$ point~\cite{Suzuki15,Fanfarillo16,Watson17,Yi19}.
On the other hand, quantum oscillation measurements indicate the presence of a tiny pocket with a nontrivial Berry phase \cite{KasaharaSu}.
This suggests that the electron pocket around the $M_{x}$ point is disconnected, forming the Dirac-cone-like band structure, as shown in Fig.~\ref{FS}(c).
As shown in Fig.~\ref{FS_3D}, the electron pockets around the $M_{x}$ point shown in red have a fork-tailed shape near the Fermi level.
Therefore, the shape of the electron pocket strongly depends on the position of the Fermi level, which is expected to be sensitive to the number of carriers. Thus, a slight deviation from the compensation condition may change the shape of the electron pocket markedly.

In addition to these Fermi surfaces, petal-like electron pockets at both  $M_{x}$ and $M_{y}$ points have also been proposed on the basis of some ARPES experiments~\cite{Watson16,Fedorov16,Kushnirenko17}.
However, as these ARPES measurements were performed by using heavily twinned crystals, careful interpretation may be necessary.  Recent experiments   using nano-ARPES on a twinned crystal indicate a single electron pocket  at the $M_x$ point as shown in  Figs.\,\ref{FS} (a) and \ref{FS} (b) \cite{Rhodes20}.
The electron pocket at the $M_{y}$ point is also controversial for ARPES measurements using detwinned crystals.
Although a very tiny electron pocket at the $M_{y}$ point has been reported~\cite{Yi19} as illustrated in Fig.~\ref{FS}(a), the absence of such a pocket has also been reported~\cite{Huh20} as shown in Fig.~\ref{FS}(b).
It has been pointed out that the presence or absence of an electron pocket at the $M_{y}$ point strongly depends on the level of the $d_{xy}$ orbital.
As well as  the above four types of   Fermi surface, an additional inner hole pocket with $d_{xz}$ orbital character has been suggested very recently~\cite{Li19}.
However, we assume that there is only one hole pocket in the following arguments. 

Quantum oscillations of the resistivity, known as the Shubnikov-de Haas (SdH) effect, have been reported by several reserach groups~\cite{Watson15,Terashima14}.
Four branches are observed in the SdH oscillations.
It has been suggested that two of the branches correspond to the extremal orbits of the hole pocket and the other two  correspond to those of the electron pocket.
The cross section of each Fermi surface is extremely small, occupying at most 2--3\% of the Brillouin zone. 
The electronic specific heat coefficient $\gamma$ estimated from the effective masses and Fermi surface volume assuming a 2D cylindrical Fermi surface is close to the observed $\gamma$-value, suggesting that most parts of the Fermi surface are mapped out by the SdH measurements~\cite{Terashima14}.

\subsection{Electronic nematic state}

\begin{figure}[t]
	\centering
	\includegraphics[width=0.8\linewidth]{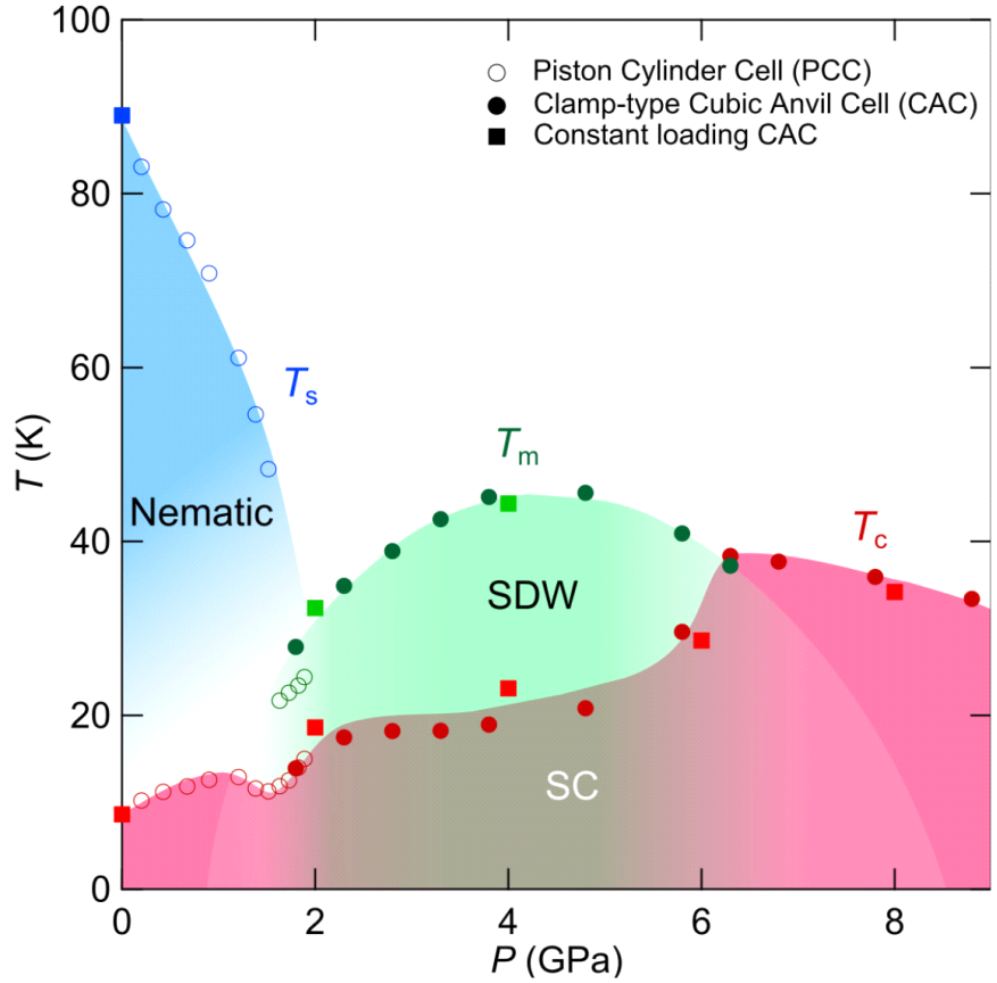}
	\caption{(Color online) 
	Pressure--temperature ($P$--$T$) phase diagram of FeSe.
	The structural or nematic ($T_{\rm s}$, blue), SDW ($T_{\rm m}$, green), and superconducting transition temperatures ($T_{\rm c}$, red) are shown as functions of hydrostatic pressure determined by the resistive anomalies measured in the piston-cylinder cell (PCC, open circles), clamp-type cubic anvil cell (CAC, closed circles), and constant-loading-type CAC (closed squares).
	Color shades for the nematic, SDW, and superconducting (SC) states are guides to the eyes.
	Adopted from Ref.~\citen{Sun16}.
	}
	\label{FeSe_p-T}
\end{figure}

Pressure and chemical substitution are nonthermal control parameters, which provide a continuous means to modify the electronic structure.
The ground state of iron-based materials can be significantly tuned by these parameters, which have been widely employed to access the QCP~\cite{Shibauchi14}.
Hydrostatic pressure experiments on FeSe have been performed by many groups~\cite{Medvedev09,Bendele12,Terashima15,Kothapalli16,Sun16,Matsuura17,Xiang17,Yip17,Gati19,Holenstein19, Reiss20}.
As shown in Fig.~\ref{FeSe_p-T}, the structural (nematic) transition at $T_{\rm s}$ is rapidly suppressed by pressure ($P<2$~GPa).
Before the complete suppression of $T_{\rm s}$, the static magnetic order is induced at $T_{\rm m}$, suggesting the presence of an overlap region of nematic and pressure-induced magnetic phases.
The presence of a static long-range magnetic order has been confirmed by muon spin rotation ($\mu$SR), M\"{o}ssbauer spectroscopy and NMR measurements, and the magnetic phase is most likely to be an SDW phase of the stripe type similar to that observed in iron-pnictides.
With increasing pressure, $T_{\rm m}$ increases and intersects with $T_{\rm s}$ at $\sim 2$~GPa.
At $\sim 4.2$~GPa, $T_{\rm m}$ peaks then vanishes abruptly at $\sim 6.3$~GPa.
Thus,  the pressure-induced magnetic state has a dome-like shape in the $P$--$T$ phase diagram. 
It has been reported that superconductivity is filamentary rather than a bulk phenomenon inside the magnetic dome, implying that the ground state in this dome pressure range is likely to be an AFM metal.
Near the end point of the magnetic order, $T_{\rm c}$ markedly increases up to $\sim 38$~K.

\begin{figure*}[t]
	\centering
	\includegraphics[width=0.8\linewidth]{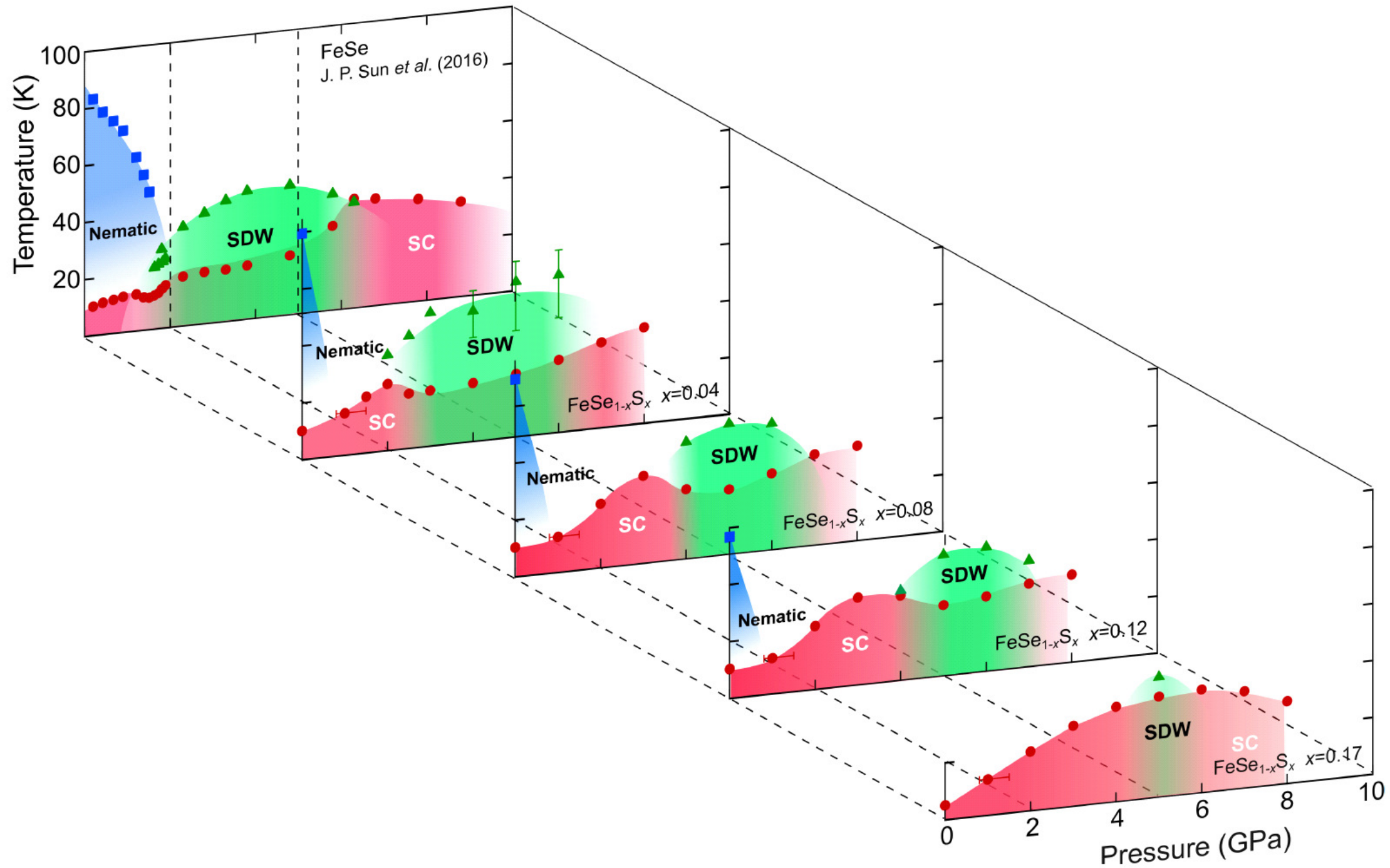}
	\caption{(Color online) 
	Temperature--pressure--concentration ($T$--$P$--$x$) phase diagram of FeSe$_{1-x}$S$_{x}$.
	The structural or nematic ($T_{\rm s}$, blue squares), SDW ($T_{\rm m}$, green triangles),  and superconducting transition temperatures ($T_{\rm c}$, red circles) determined by the resistivity anomalies are plotted against the hydrostatic pressure $P$ and sulfur content $x$.
	$T_{\rm c}$ is also determined by the magnetic susceptibility.
	Adopted from Ref.~\citen{Matsuura17}.
	}
	\label{FeSeS_pTx}
\end{figure*}

Electronic nematicity is a ubiquitous property of the iron-based superconductors.
There are two scenarios for the driving mechanism of the nematicity.
One route to the nematicity is via critical magnetic (spin) fluctuations and the other is via the critical orbital fluctuations.
In iron-pnictides, where the nematicity is always accompanied by the AFM order, strong AFM fluctuations are observed above $T_{\rm s}$~\cite{Dai15}.
Thus, the critical spin fluctuations have been suggested to be responsible for the nematicity~\cite{Fernandes14}.
It has been argued that this spin-nematic scenario envisaged in iron-pnictides may still be applicable to FeSe~\cite{Chubukov15,Glasbrenner15,Wang15a,Yu15}.
In fact, the magnitudes of the lattice distortion, elastic softening, and elasto-resistivity associated with the structural transition in FeSe are comparable to those of Fe-pnictides~\cite{Bohmer15,Watson15}.
It has been shown that the nematic order could be driven by the AFM spin fluctuations without the requirement of magnetic order~\cite{Wang15a,Chubukov16a,Yamakawa16}.
Spin excitations measured by inelastic neutron scattering experiments show that the dynamic susceptibility $\chi(\bm{q},\omega)$ peaks at $\bm{q}=(\pi,0)$, where $\bm{q}$ is the scattering vector and $\hbar \omega$ is the energy change between incident and outgoing neutrons. 
The results suggest the presence of both stripe and N\'{e}el spin fluctuations over a wide energy range even above $T_{\rm s}$~\cite{Wang16a,Wang16b}.
In the nematic state well below $T_{\rm s}$, the NMR spin-lattice relaxation rate $1/T_1$ shows the presence of strong AFM fluctuations down to $T_{\rm c}$~\cite{Imai09}.

There is an alternative scenario where charge or orbital degrees of freedom play a more predominant role than spins~\cite{Baek15,Watson15,Su15,Jiang16}.
In FeSe, where the nematic transition occurs without magnetic order, no sizable low-energy spin fluctuations are observed above $T_{\rm s}$ by NMR~\cite{Imai09,Baek15,Wiecki18}, in contrast to the iron-pnictides.
It has been shown that the orbital ordering is unequivocally the origin of the nematic order in FeSe~\cite{Nakayama14,Shimojima14,Zhang15}.
In the nematic phase where the splitting of the $d_{xz}$ and $d_{yz}$ energy bands occurs, a momentum-dependent orbital polarization has been found in ARPES measurements, indicating that the nematicity is most likely to be orbital in origin~\cite{Suzuki15}.
Moreover, NMR measurements report that the difference in static internal field in the $ab$ plane in the orthorhombic phase at the Se nucleus is predominantly from the Fe ion $3d$ electron orbitals, not from the electron spins~\cite{Cao19}.
Recent symmetry-resolved electronic Raman scattering measurements provide a direct experimental observation of critical fluctuations associated with electronic charge or orbital nematicity near $T_{\rm s}$~\cite{Massat16}.
These results put into question the spin-nematic scenario for the nematicity in FeSe.

The appearance of the AFM order induced at low pressures is consistent with the fact that the nematic order is close to the magnetic instability as suggested by {\it ab initio} calculations~\cite{Scherer17}.
It has been suggested that the pressure changes the Fermi surface topology of FeSe.  
A hole band with $d_{xy}$ orbital character is nearly 10~meV below the Fermi level around the ($\pi, \pi$) point in the unfolded Brillouin zone at ambient pressure.
The top of the hole band crosses the Fermi energy under pressure, giving rise to a hole pocket.
This hole pocket largely enhances the AFM nesting properties, leading to the appearance of the static AFM order~\cite{Yamakawa17}.
Therefore,  the spin and orbital degrees of freedom are highly entangled even in nonmagnetic FeSe, which makes it difficult to pin down the driving mechanism of the nematicity and superconductivity.

Until now, most studies have been conducted in thermal equilibrium, where the dynamical property and excitation can be masked by the coupling with the lattice. 
Recently, by using a femtosecond optical pulse, the ultrafast dynamics of electronic nematicity has been detected.
A short-life nematicity oscillation, which is related to the imbalance of Fe $d_{xz}$ and $d_{yz}$ orbitals, has been reported. 
Such real-time observations of the electronic nematic excitation that is instantly decoupled from the underlying lattice would be important for the future investigation of the nematicity~\cite{Shimojima19}.

\subsection{Nematic quantum critical point (QCP)}

Next we discuss why FeSe$_{1-x}$S$_{x}$ is a system  suitable for exploring the effect of nematicity disentangled from that of the static AFM order.
The nematicity can be tuned continuously by isoelectronic sulfur substitution and $T_{\rm s}$ vanishes at $x\approx 0.17$.
The compensated metal character should be unaffected by the isovalent substitution of Se.
Moreover, in contrast to the application of pressure, the sulfur substitution appears not to change the Fermi surface topology \cite{Coldea19,Hanaguri18}.
 NMR $1/T_1$ measurements reveal that the AFM fluctuations are slightly enhanced at a small $x$ but strongly suppressed by further sulfur substitution, leading to the absence of  AFM fluctuations near the nematic QCP~\cite{Baek15,Wiecki18}.
Figure~\ref{FeSeS_pTx} displays the temperature--pressure--concentration ($T$--$P$--$x$) phase diagram of FeSe$_{1-x}$S$_{x}$ in wide ranges of pressure up to 8--10~GPa and sulfur content ($0\leq x \leq 0.17$), where $T_{\rm s}$ and $T_{\rm m}$ are determined by the resistivity anomalies~\cite{Matsuura17}.
The $T_{\rm c}$ values are determined from   zero-resistivity as well as   magnetic susceptibility measurements.
As shown in Fig.~\ref{FeSeS_pTx}, the magnetic dome shrinks with increasing $x$ and disappears at $x\approx 0.17$~\cite{Matsuura17}.
This indicates that,  in contrast to the pressure, sulfur substitution moves the magnetic instability away from the nematic order.

\begin{figure}[b]
	\centering
	\includegraphics[width=0.75\linewidth]{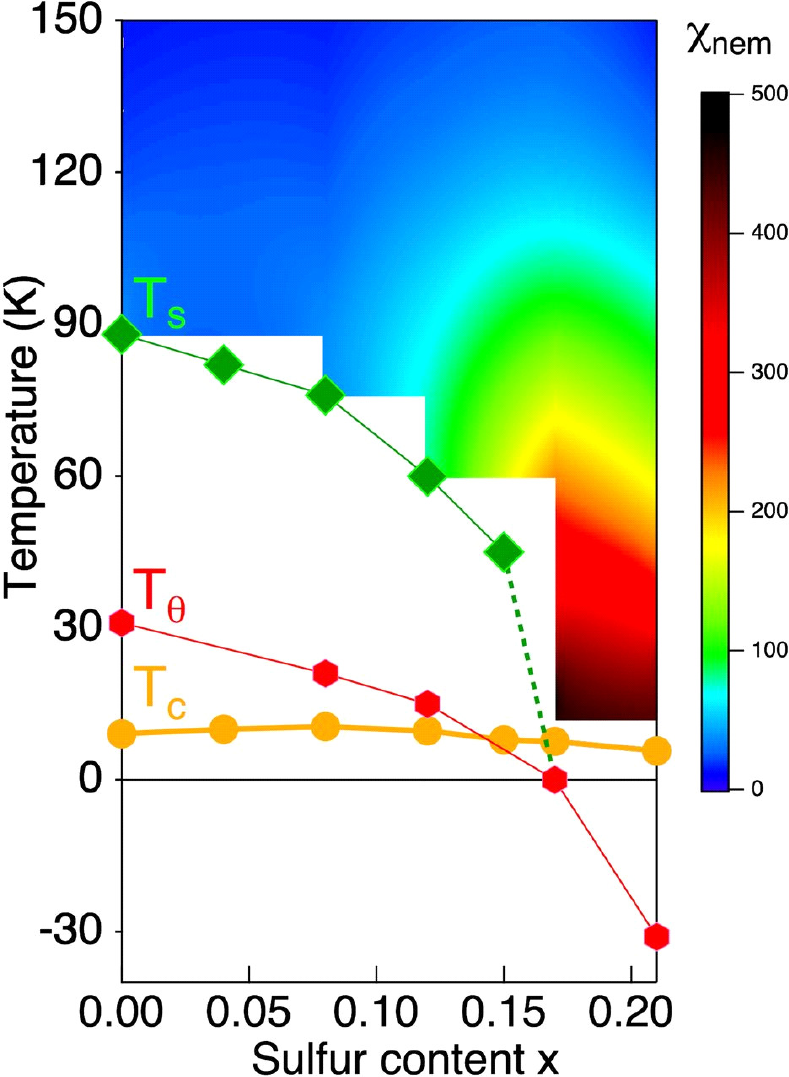}
	\caption{(Color online) 
	Phase diagram and quantum criticality of FeSe$_{1-x}$S$_{x}$.
	Temperature dependence of the nematic transition ($T_{\rm s}$, green diamonds) and the superconducting transition temperature ($T_{\rm c}$, orange circles) determined by the zero-resistivity criterion.
	The Curie-Weiss temperature is also plotted ($T_{\theta}$, red hexagons).
	The magnitude of $\chi_{\rm nem}$ in the tetragonal phase is superimposed on the phase diagram with a color contour (see the color bar for the scale).
	The lines are the guides for the eyes.
	Adopted from Ref.~\citen{Hosoi16}.
	}
	\label{Nematic_Sus}
\end{figure}

Thus, an important issue is how the nematic fluctuations evolve with the sulfur substitution.
An elegant way to  experimentally evaluate the nematic fluctuations has been developed by the Stanford group, which is based on   elasto-resistivity measurements  using a piezoelectric device~\cite{Chu12}.
Nematic susceptibility is defined as $\chi_{\rm nem}\equiv  {\rm d} \eta/{\rm d} \epsilon$, where $\eta=(\rho_{xx}-\rho_{yy})/\rho \sim \Delta\rho/\rho$ is the change in resistivity induced by the lattice strain $\epsilon$.
The nematic susceptibility can probe fluctuations associated with the phase transition, which bears some resemblance to the magnetic susceptibility $\chi_{\rm mag}={\rm d}M/{\rm d}H$ in the magnetic system.
It has been reported that the $\chi_{\rm nem}$ of FeSe$_{1-x}$S$_{x}$ exhibits   Curie-Weiss-like temperature dependence,
\begin{equation}
\chi_{\rm nem}(T) = \frac{a}{T - T_{\theta}} + \chi_0,
\end{equation}
where $a$ and $\chi_0$ are constants and $T_{\theta}$ corresponds to the Curie-Weiss temperature of the electronic system~\cite{Hosoi16}. As shown in Fig.~\ref{Nematic_Sus}, the nematic fluctuations are strongly enhanced by increasing the sulfur content $x$. Near $x\approx 0.17$,  $T_{\theta}$ goes to zero, indicating that the nematic susceptibility diverges towards $T=0$.
Moreover, quantum oscillations and quasiparticle interference (QPI) measurements revealed that the Fermi surface changes smoothly when crossing the nematic QCP.
These results reveal the presence of a nematic QCP at $x\approx 0.17$~\cite{Coldea19,Hanaguri18}.   

In the nematic phase, $T_{\rm c}$ increases gradually, peaks at $x\sim 0.08$ then decreases gradually with $x$.
Around the nematic QCP, $T_{\rm c}$ decreases and reaches $\sim 4$~K in the tetragonal phase~\cite{Hanaguri18,Sato18}.
Very recently, linear-in-temperature resistivity, which is a hallmark of the non-Fermi liquid property, has been reported at $x\approx 0.17$, indicating that the nematic critical fluctuations emanating from the QCP have a significant impact on the normal-state electronic properties~\cite{Licciardello19}.
We will show that the nematicity also strongly affects the superconductivity in Sect.~6.

%%%%%%%%%%%%%%%%%%%%%%%%%%%%%%%%%%%%%%%%%%%%%%%%%%%%%%%%%%%%%%%%%%%%%%%%%%%%%%%
%%%%%%%%%%%%%%%%%%%%%%%%%%%%%%%%%%%%%%%%%%%%%%%%%%%%%%%%%%%%%%%%%%%%%%%%%%%%%%%

\section{Superconducting Gap Structure}

\subsection{Bulk measurements}

\subsubsection{Temperature dependence}

One of the most important properties of unconventional superconductors is the characteristic structure of the superconducting gap $\Delta(\bm{k})$, which is intimately related to the pairing mechanism.
In phonon-mediated conventional superconductivity, the momentum-independent pairing interaction leads to BCS $s$-wave superconductivity with a constant $\Delta=1.76k_{\rm B}T_{\rm c}$, and thus the bulk physical quantities that are related to quasiparticle excitations show exponential temperature dependence at low temperatures.
For unconventional superconductors, however, the pairing interaction may have strong dependence on momentum $\bm{k}$, leading to anisotropic $\Delta(\bm{k})$, which sometimes has zeros (nodes) in certain $\bm{k}$ directions.
In such cases, the existence of low-lying excitations in the quasiparticle energy spectrum changes the exponential temperature dependence to power-law behavior.
Therefore, the low-temperature measurements of bulk quantities sensitive to low-energy quasiparticle excitations, such as magnetic penetration depth, specific heat, and thermal conductivity, are   important for studying the pairing mechanism of superconductors.

\begin{figure*}[t]
\centering
\includegraphics[width=0.95\linewidth]{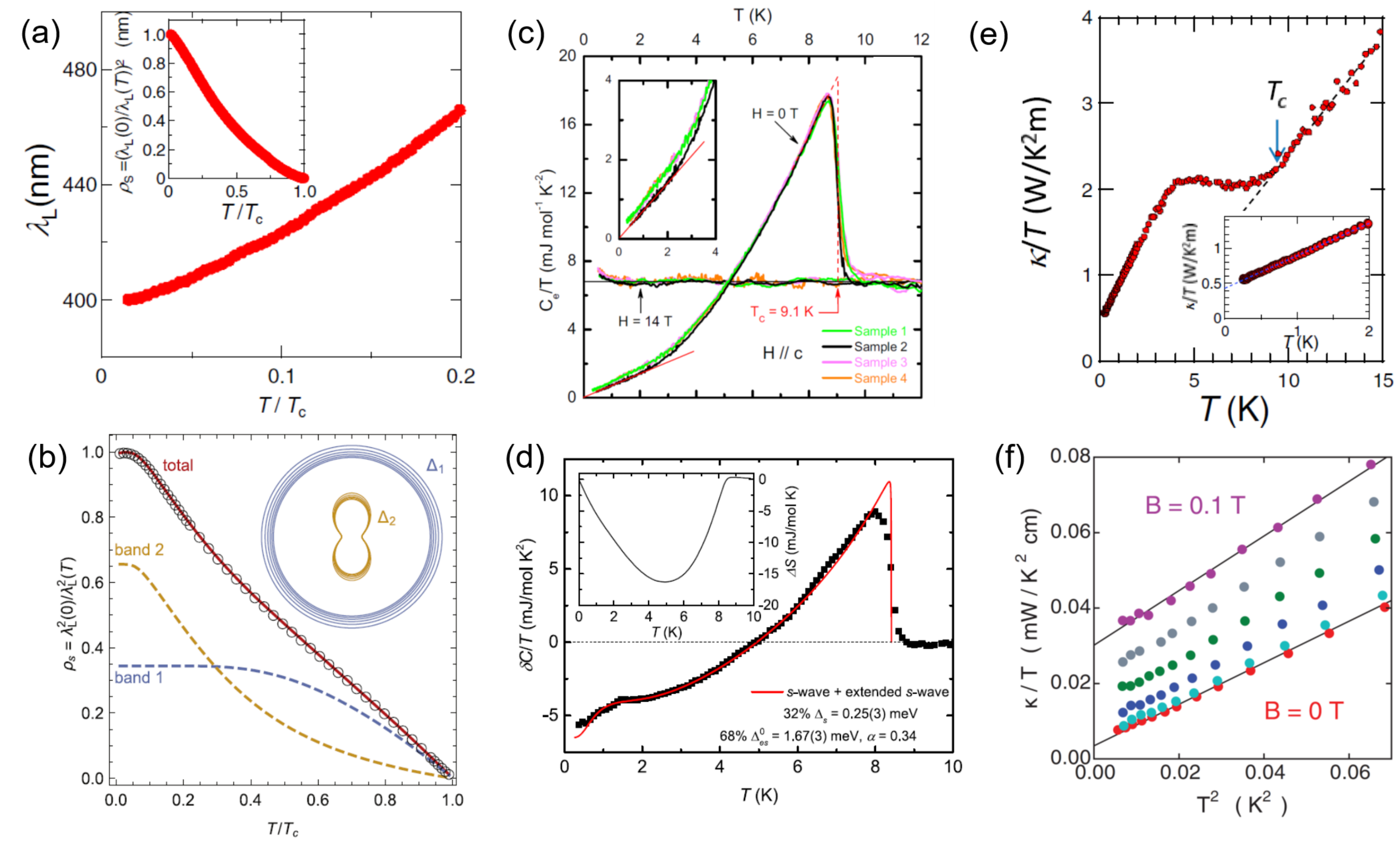}
\caption{(Color online) 
Temperature dependence of bulk quantities in FeSe single crystals.
(a) Temperature dependence of magnetic penetration depth at low temperatures, measured in the Meissner state by a tunnel diode oscillator technique.
Adopted from Ref.~\citen{Kasahara14}.
The inset shows the temperature dependence of superfluid density.
(b) Superfluid density measured by a cavity perturbation technique.
The data are fitted by a two-gap model (lines).
Adopted from Ref.~\citen{Li16}.
(c) Electronic specific heat divided by temperature $C_e/T$ as a function of temperature in the superconducting state at zero field and in the normal state at 14~T.
The inset is an expanded view at low temperatures.
Adopted from Ref.~\citen{Hardy19}.
(d) Difference in $C_e/T$ between the superconducting and normal states, fitted by a two-gap model (red line).
The inset shows the temperature dependence of the entropy change.
Adopted from Ref.~\citen{Jiao17}.
(e) Thermal conductivity divided by temperature $\kappa/T$ as a function of temperature.
The inset is an expanded view at low temperatures.
Adopted from Ref.~\citen{Kasahara14}.
(f) $\kappa/T$ as a function of $T^2$ at zero and low fields.
Here, the residual $\kappa_0/T$ at zero field is an order of magnitude smaller than that in (e). 
Adopted from Ref.~\citen{Bourgeois-Hope16}.
}
\label{BulkT}
\end{figure*}

The temperature dependence of the London penetration depth $\lambda(T)$, which is directly related to the number of superconducting electrons, is one of the sensitive probes of thermally excited quasiparticles.
When the gap $\Delta(\bm{k})$ has line (point) nodes, the low-energy quasiparticle excitation spectrum depends on the energy $E$ as $\propto E$ ($\propto E^2$), and thus $\Delta\lambda(T)=\lambda(T)-\lambda(0)$ is proportional to $T$ ($T^2$) at low temperatures unless the supercurrent direction is always perpendicular to the nodal directions.
The precision measurements of penetration depth in the Meissner state using the tunnel diode oscillator technique at 13~MHz have shown that $\Delta\lambda(T)$ in high-quality single crystals of FeSe has nonexponential, quasi-linear temperature dependence ($\sim T^{1.4}$) at low temperatures below $\sim0.2T_{\rm c}$, as shown in Fig.~\ref{BulkT}(a)~\cite{Kasahara14}.
This result suggests the presence of line nodes in the superconducting gap.
The deviation from the $T$-linear dependence may be attributed to the impurity scattering or multiband effect (combining nodal and full-gapped bands), which can increase the exponent $\alpha$ from unity in the power-law temperature dependence $T^\alpha$.

A surface impedance study at higher frequencies of 202 and 658~MHz using a cavity perturbation technique also revealed the strong temperature dependence of the superfluid density $\rho_s(T)=\lambda^2(0)/\lambda^2(T)$, but as shown in Fig.~\ref{BulkT}(b), it exhibits  flattening at the lowest temperatures, in contrast to the nodal gap behavior~\cite{Li16}.
The data can be fitted to two gaps with a small minimum gap $\Delta_{\rm min}\approx 0.25 k_{\rm B}T_{\rm c}$ in one band. 
The implication of the presence of two different results, nodal and gap minima, is that the pairing symmetry is anisotropic $s$ with accidental nodes (if present),  which will be discussed at the end of Sect.~3.1.2.

The heat capacity is the most fundamental thermodynamic quantity that can also probe the quasiparticle excitations in the superconducting state.
In the analysis of specific heat $C(T)$, the contribution from the phonons, which usually depends on temperature as $C\sim T^3$ at low temperatures, must be subtracted to extract the electronic contribution.
This can be done by comparing the data at zero field and above the upper critical field $H_{\rm c2}$ in the normal state.
The temperature dependence of electronic specific heat divided by temperature, $C_e/T(T)$, also shows exponential behavior in fully gapped superconductors and $T$-linear behavior for the gap structure with line nodes, as in the case of $\Delta\lambda(T)$.
Although the low-temperature data of $C_e/T(T)$ sometimes show multigap behaviors suggesting the presence of a small full gap~\cite{Jiao17,Chen17,Muratov18}, the most recent measurements using high-quality, vapor-grown single crystals clearly indicate the $T$-linear behavior at low temperatures down to $\sim 0.3$~K~\cite{Hardy19}, as shown in Figs.~\ref{BulkT}(c) and \ref{BulkT}(d).
This also suggests the presence of line nodes or tiny gap minima in the superconducting gap in FeSe, which is a similar situation to that in penetration depth studies. 

Another sensitive probe of quasiparticle excitations is the thermal conductivity $\kappa$, which can be measured in the superconducting state.
The temperature dependence of thermal conductivity also has phononic and electronic components, but in the zero-temperature limit, $\kappa/T$ gives very important information on the superconducting gap structure.
As $\kappa/T$ is proportional to $C/T$ as well as to the mean free path $\ell$, $\kappa/T(T\to0)$ always vanishes for fully gapped superconductors.
This is understood by the fact that in the zero-temperature limit, $C/T$ vanishes while $\ell$ is limited by the impurity scattering.
In contrast, the presence of nodes in the gap leads to small but finite low-energy states due to impurity scattering, thus giving rise to the finite residual $\kappa_0/T\equiv \kappa/T(T\to0)$.
When the impurity scattering is increased, the residual density of states (DOS) increases, while the mean free path decreases, resulting in essentially no change in residual $\kappa_0/T$, which is called universal residual thermal conductivity in nodal superconductors.
In FeSe bulk crystals, there are also two types of report with different conclusions as shown in Figs.~\ref{BulkT}(e) and \ref{BulkT}(f); one shows a sizable $\kappa_0/T$ suggesting a nodal state~\cite{Kasahara14,Sato18} and the other shows much smaller residual values of $\kappa_0/T$ from which the authors concluded a fully gapped state with small gap minima~\cite{Bourgeois-Hope16}. 

\subsubsection{Field dependence}

Not only the temperature dependence but also the field dependence gives important clues on the presence or absence of the superconducting gap nodes.
When the magnetic field $H$ is applied to induce vortices inside nodal superconductors, the supercurrent flowing around a vortex affects the quasiparticle energy spectrum through the Doppler shift mechanism  $E(\bm{k}) \rightarrow E(\bm{k}) - \hbar\bm{k} \cdot \bm{v}_{\rm s}$ (where $\bm{v}_{\rm s}$ is the supercurrent velocity around the vortex) and enhance the low-energy DOS.
This Doppler shift effect (or Volovik effect) can be seen by strong increases in specific heat and thermal conductivity in the zero-temperature limit with increasing magnetic field, and in the case of line nodes, the $\sqrt{H}$ dependence of $C_e/T$ and $\kappa_0/T$ is expected.
In the thermal conductivity study reporting the absence of $\kappa_0/T$, one of the measured crystals clearly showed a strong increase in $\kappa/T$ with the field at low temperatures~\cite{Bourgeois-Hope16}.
This indicates that the superconducting gap $\Delta(\bm{k})$ has very strong momentum dependence even though the gap does not have any zeros; in other words, the minimum gap is very small.
In the other study showing the presence of $\kappa_0/T$, however, $\kappa_0/T$ actually decreased with increasing $H$ at low fields~\cite{Kasahara14}.
This unusual behavior can be explained by the reduction in mean free path dominating the increase in DOS, suggesting that the quasiparticles are scattered by vortices.
Such scattering induced by vortices may be seen in very clean single crystals with very large $\ell$ values, and in fact, a similar reduction in $\kappa_0/T$ at low fields has been observed in very clean crystals of CeCoIn$_5$~\cite{KasaharaY05}.
In S-substituted crystals of FeSe$_{1-x}$S$_{x}$, where the mean free path is naturally suppressed by the chemical substitution, the low-field $\kappa/T$ at low temperatures exhibits $\sqrt{H}$ behavior, consistent with the presence of line nodes~\cite{Sato18}.
In the recent field dependence measurements of specific heat in FeSe, clear $\sqrt{H}$ dependence of $C_e/T$ was also observed~\cite{Hardy19}. 

Summarizing these bulk measurements, one can safely conclude that the superconducting gap structure of bulk FeSe has very strong $\bm{k}$ dependence with line nodes or deep gap minima.
The fact that some measurements suggest fully gapped behaviors, although most results are consistent with the presence of line nodes, implies that the nodes are unlikely protected by symmetry, but are accidental ones. 
The symmetry protected nodes are robust against impurity scattering, but the accidental nodes may be lifted by various perturbations such as disorder.  
This is most consistent with the $s$-wave $A_{1g}$ symmetry of the superconducting order parameter, either $s_{\pm}$ or $s_{++}$, with strong anisotropy in at least one of the multibands.
This strong anisotropy confirms the unconventional nature of superconductivity in FeSe.

\subsection{Angle-resolved photoemission spectroscopy (ARPES)}

\begin{figure}[t]
\centering
\includegraphics[width=\linewidth]{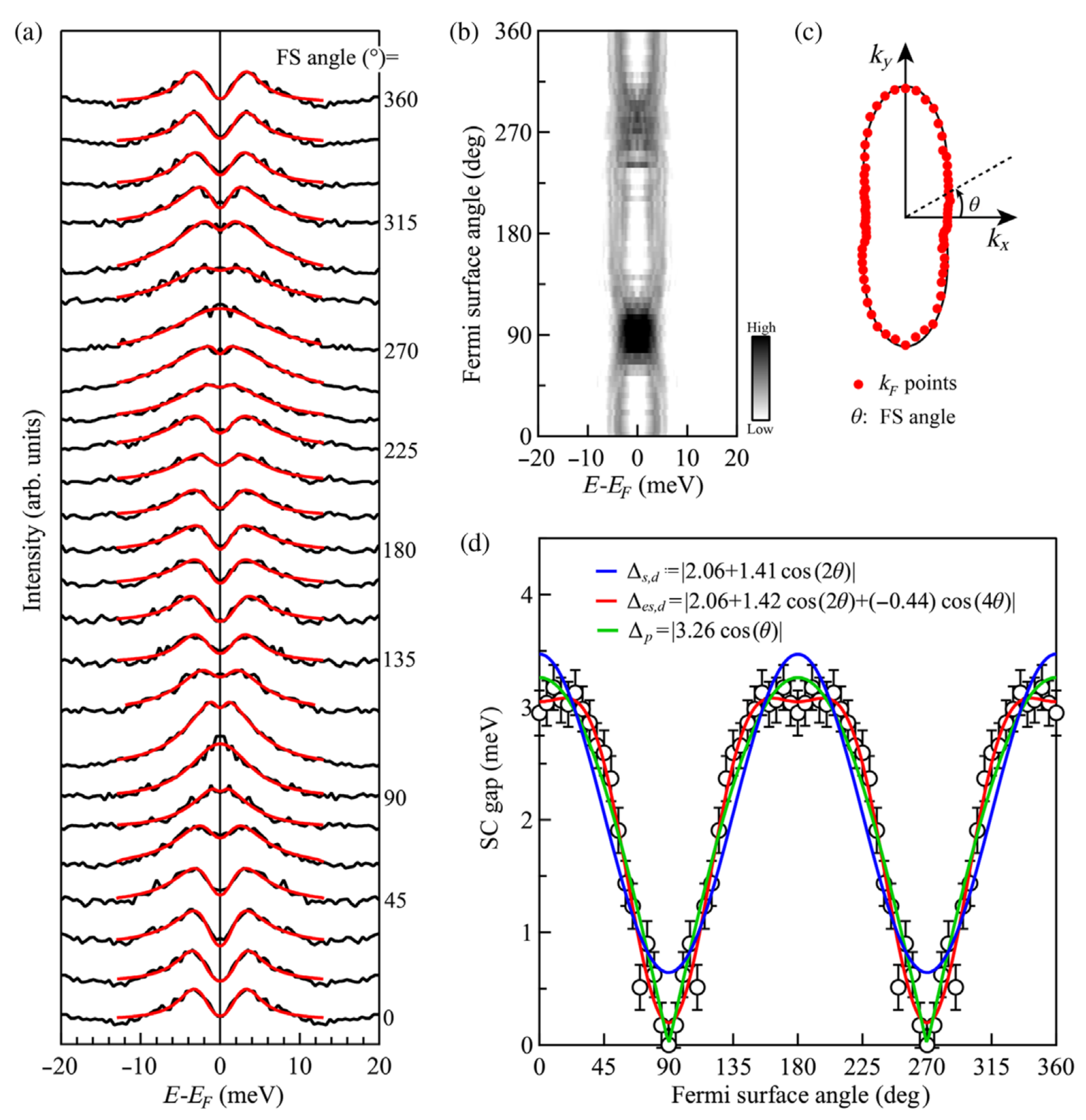}
\caption{(Color online) 
Momentum dependence of the superconducting gap on the hole Fermi surface of FeSe determined by ARPES measurement at 1.6~K.
(a) and (b) Symmetrized energy distribution curves at different Fermi momenta along the hole Fermi surface.
A phenomenological gap formula (red curves) is used to extract $\Delta(\bf{k})$.
(c) Location of the Fermi momentum defined by the Fermi surface (FS) angle $\theta$.
(d) Momentum dependence of the superconducting gap, with averaging over the four quadrants.
The measured gap (empty circles) is fitted by several gap forms.
Adopted from Ref.~\citen{Liu18a}.
}
\label{ARPESgap}
\end{figure}

ARPES measurements have a strong advantage over other techniques, namely, they can provide direct information on the momentum dependence of the energy spectrum.
In the superconducting state, the momentum dependence of the superconducting gap $\Delta(\bm{k})$ can be mapped out, and indeed the $d$-wave superconducting gap has been clearly found in cuprate superconductors.
In FeSe with a relatively low $T_{\rm c}$, however, a very high energy resolution is required to resolve the relatively small energy gap, and a high momentum resolution is also needed to resolve the $\bm{k}$ dependence around the very small Fermi surfaces.
Such high-resolution ARPES measurements have recently become available by using a laser light source with $\sim7$~eV energy.
However, in these laser-based ARPES measurements, the momentum space that can be accessed is limited to near the zone center ($\Gamma$ point), and the electron bands near the zone edge cannot be explored.
The laser ARPES results for single-domain samples of FeSe, independently obtained in the Institute for Solid State Physics, University of Tokyo~\cite{Hashimoto18}, and the Institute of Physics, Chinese Academy of Sciences~\cite{Liu18a}, show that $\Delta(\bm{k})$ in the hole band near the $\Gamma$ point is strongly anisotropic.
The hole Fermi surface has an ellipsoidal shape (see Fig.~\ref{FS_3D}), and $\Delta(\bm{k})$ is also two fold symmetric.
They found that along the long axis of the ellipsoid, the gap becomes almost zero, suggesting the presence of nodes near this direction, as shown in Figs.~\ref{ARPESgap}(a)-\ref{ARPESgap}(d).
In the two fold nematic phase, the $s$-wave and $d$-wave components can mix in the $A_{1g}$ symmetry, and the observed two fold anisotropic $\Delta(\bm{k})$ suggests that the $s$-wave and $d$-wave components are very close in magnitude. 
This can be called an anisotropic $A_{1g}$ pairing state with nascent nodes~\cite{Fernandes11}.
Although the gap structure of the electron band is not clear from ARPES measurements, this strong anisotropy in the hole band is consistent with the bulk measurements mentioned above.

\subsection{Scanning tunneling microscopy / spectroscopy (STM / STS)}

\begin{figure}[t]
\centering
\includegraphics[width=\linewidth]{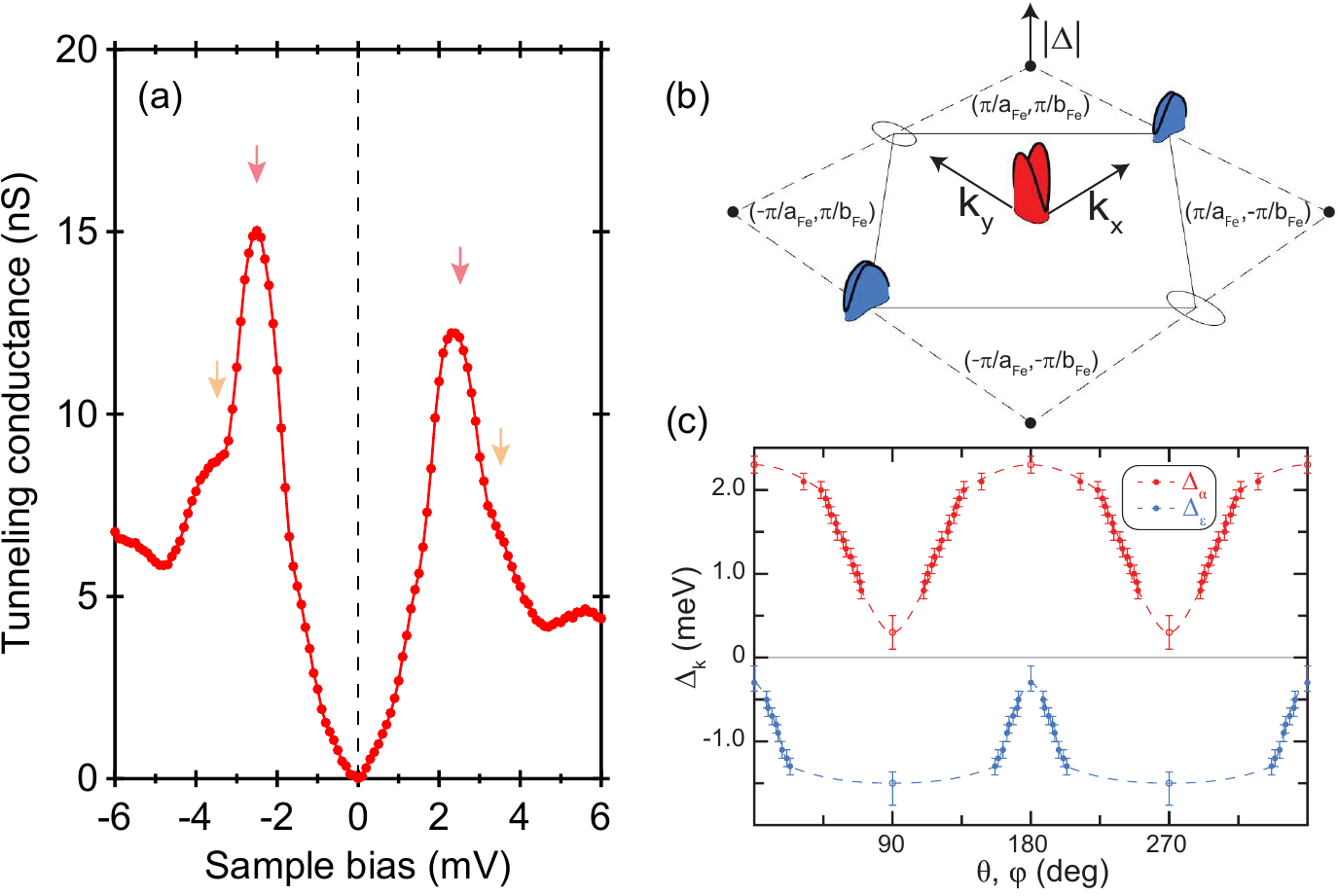}
\caption{(Color online) 
Superconducting gap structure of FeSe determined by STS measurements. 
(a) Typical conductance spectrum measured at 0.4~K at a cleaved surface of FeSe single crystal.
The bottom of the gap is V-shaped and there are at least two features at the gap edges as indicated by arrows.
Data adopted from Ref.~\citen{Hanaguri19}.
(b) and (c) Momentum dependence of the superconducting gap derived by QPI analysis.
The red and blue colors indicate the different signs of the two gap functions in the hole and electron bands.
One of the electron bands was not deduced by this analysis (thin ellipsoids).
Adopted from Ref.~\citen{Sprau17}.
}
\label{STSgap}
\end{figure}

The tunneling experiment is a powerful probe of the superconducting gap structure, because it can directly extract the quasiparticle DOS as a function of energy.
Owing to the recent advances of scanning tunneling microscopy / spectroscopy (STM / STS) techniques, one can obtain highly reliable DOS data with a very high energy resolution at very low temperatures.
The first evidence from STS for the presence of nodes  in thin films of FeSe was reported by Xue's group from Tsinghua~\cite{Song11}.
They found a V-shaped DOS in the energy dependence of the tunneling conductance near zero energy, consistent with the line nodes in the gap.
In bulk vapor-grown single crystals, similar V-shaped tunneling spectra are observed~\cite{Kasahara14,Watashige15,Hanaguri19}, again suggesting line nodes [Fig.~\ref{STSgap}(a)]. 
At higher energies, the spectra exhibit at least two distinct features of superconducting gaps at $\Delta_{\rm l}\approx3.5$~meV and $\Delta_{\rm s}\approx2.5$~meV.

This tunneling spectroscopy alone cannot resolve the positions of nodes in the momentum dependence of the gap $\Delta(\bm{k})$.
However, by analyzing the interference of standing waves induced by impurity scattering, which is called the Bogoliubov QPI imaging technique, the energy-dependent Fermi surface structures in the scattering wave vector plane can be mapped.
From these analyses, the superconducting gap structures $\Delta(\bm{k})$ of FeSe are extracted, revealing very strong anisotropies of the gap in the hole band near the zone center as well as in one of the electron bands near the zone edge [Fig.~\ref{STSgap}(b)]~\cite{Sprau17}.
In addition, by comparing the detailed energy dependence of the QPI mapping with theoretical calculations~\cite{Hirschfeld15}, it is concluded that the sign of the order parameter changes between the hole and electron bands in FeSe.
Although the extracted gap structure has no nodes but deep minima, the suggested pairing state is a strongly anisotropic $s_{\pm}$ state, which has been discussed in terms of orbital-dependent pairing.

Previously, QPI evidence for the sign-changing $s_{\pm}$ state was found in FeSe$_{1-x}$Te$_{x}$, where the effect of the magnetic field on QPI was analyzed~\cite{Hanaguri10}.
In FeSe$_{1-x}$Te$_{x}$, the tunneling spectra are more U-shaped and the gap is more isotropic than those in FeSe.
This nonuniversality of the superconducting gap structure is also found in iron-pnictides~\cite{Shibauchi14}; optimally doped Ba$_{1-x}$K$_x$Fe$_2$As$_2$ exhibits a fully gapped state, while BaFe$_2$(As$_{1-x}$P$_x$)$_2$ shows clear signatures of line nodes.
It has also been revealed that in the BaFe$_2$(As$_{1-x}$P$_x$)$_2$ system, in which the disorder can be controlled by electron irradiation, the nodes can be lifted by impurity scattering and that the observed nonmonotonic changes of low-energy excitations with disorder indicate a sign-changing $s_{\pm}$ state after node lifting~\cite{Mizukami14}.
Such a nonuniversal superconducting gap structure with $A_{1g}$ symmetry may indicate the presence and importance of multiple pairing mechanisms in iron-based superconductors, which are most likely based on spin fluctuations that favor the $s_{\pm}$ symmetry and orbital fluctuations that favor the $s_{++}$ state.

%%%%%%%%%%%%%%%%%%%%%%%%%%%%%%%%%%%%%%%%%%%%%%%%%%%%%%%%%%%%%%%%%%%%%%%%%%%%%%%
%%%%%%%%%%%%%%%%%%%%%%%%%%%%%%%%%%%%%%%%%%%%%%%%%%%%%%%%%%%%%%%%%%%%%%%%%%%%%%%

\section{BCS-BEC Crossover}
 
\subsection{BCS-BEC crossover}

An ideal gas consisting of noninteracting Bose particles can exhibit a phase transition called BEC; below some critical temperature $T_{\rm B}$, a macroscopic fraction of the bosons is condensed into a single ground state.
BEC occurs when the thermal de Broglie wavelength ($\propto 1/\sqrt{T}$) becomes comparable to the inter-particle distance $n_{\rm B}^{-1/3}$ at low temperatures, where $n_{\rm B}$ is the number of Bose particles.
Below $T_{\rm B}$, wavefunction interference becomes macroscopically apparent.
The superfluidity of the system is a consequence of BEC.

In Fermi systems, attractive interactions between fermions are needed to form bosonic  molecules (Cooper pairs), which are driven to BEC.
There are two limiting cases, weak-coupling BCS and strong-coupling BEC limits, where attractive interactions are weak and strong, respectively.
The physics of the crossover between the BCS and BEC limits  has been of considerable interest in the fields of condensed matter, ultracold atom and nuclear physics~\cite{BCSBEC}, giving a unified framework of quantum superfluid states of interacting fermions~\cite{Nozi85,Micnas90,Chen05,Randeria14}.
The crossover has hitherto been realized experimentally in ultracold atomic gases~\cite{BCSBEC}.
On the other hand,  almost all superconductors are in the BCS regime.  
In this section, a unique feature of the superconductivity of FeSe is covered.  
There is growing evidence that FeSe and FeSe$_{1-x}$S$_{x}$ are in the BCS-BEC crossover regime~\cite{Kasahara14,Hanaguri19,Hashimoto20,DHLee}.
FeSe-based superconductors may provide new insights into fundamental aspects of the physics of the crossover.

\begin{figure}[t]
\centering
\includegraphics[width=1\linewidth]{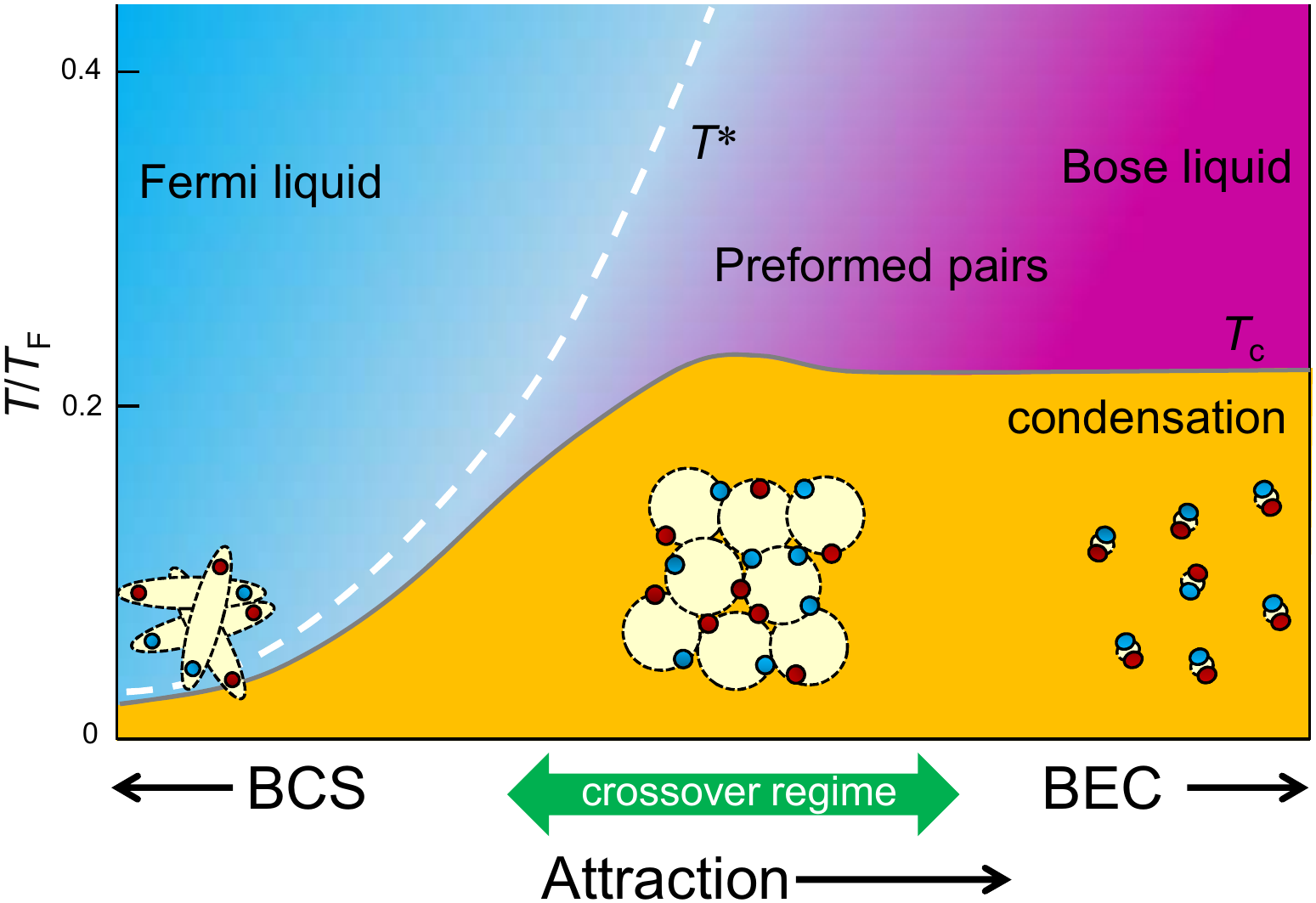}
\caption{(Color online) 
Canonical phase diagram of the BCS-BEC crossover.
With increasing attractive interaction, the superconducting condensation temperature $T_{\rm c}$ increases in the BCS regime and becomes independent of the interaction in the BEC regime.
The dashed white line represents the pairing temperature $T^{\ast}$, where the preformed pairs appear.
In the crossover regime, $T_{\rm c}$ exhibits a broad maximum.
As the pairing strength is increased, $T_{\rm c}$ and $T^{\ast}$ are separated. The pseudogap is expected at $T_{\rm c}<T<T^{\ast}$.
}
\label{crossover}
\end{figure}

In the BCS limit where the attraction is weak, the Cooper pairing is described as a momentum space pairing.
The pairwise occupation of states ($\bm{k}\uparrow, -\bm{k}\downarrow$) with zero center-of-mass momentum for bosonic  pairs leads to a profound rearrangement of the Fermi surface, leading to the formation of an energy gap $\Delta$, which corresponds to the pair condensation energy.
The size of the Cooper pairs, i.e.,  the coherence length $\xi$, is much larger than the average interelectron distance $\sim k_{\rm F}^{-1}$, where $k_{\rm F}$ is the Fermi momentum, indicating that Cooper pairs are strongly overlapped, $k_{\rm F}\xi \gg 1$. This corresponds to $\Delta/\varepsilon_{\rm F}\ll 1$, i.e., the pair condensation energy is much smaller than the Fermi energy.
In this regime, the condensation occurs simultaneously with the pair formation. 
In the BEC limit where the attraction is strong, two electrons are tightly bounded, forming a bound molecule, and the Cooper pairs behave as independent bosons.
In this limit, the size of the composite bosons is much smaller than the average interelectron distance, $k_{\rm F}\xi \ll1$; composite bosons are nonoverlapping, which can be regarded as a real-space pairing.
Even when the pair formation occurs at $T^{\ast}$, the thermal de Broglie wavelength is still much shorter than the interelectron distance.
As a result, the BEC transition occurs at $T_{\rm c}$, a temperature much lower than the pair formation temperature ($T^{\ast}\gg T_{\rm c}$), 
\begin{equation}
  T_{\rm c}=\frac{2\pi\hbar^2}{mk_{\rm B}}\left[\frac{(n/2)}{\zeta(3/2)}\right]^{2/3}=0.218 T_{\rm F},
\end{equation}
where $n$ is the density of fermion particles, $\zeta(z)$ is the Riemann zeta function ($\zeta(3/2)=2.612\cdots$), and $T_{\rm F}$ is the Fermi temperature.
Thus, in contrast to the BCS limit, the preformed Cooper pair regime extends over a wide range of temperatures (Fig.~\ref{crossover}).
The appearance of  preformed pairs has been suggested to lead to the pseudogap formation, which is the precursor of the well-developed superconducting gap.
The essential defining feature of the real-space pairing is that the chemical potential becomes negative ($\mu<0$), moving below the bottom of the band.

\begin{figure}[b]
\centering
\includegraphics[width=1\linewidth]{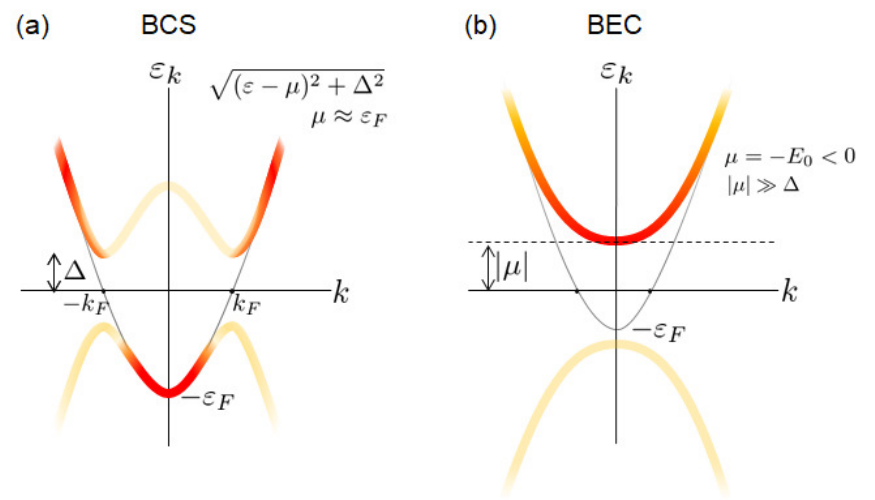}
\caption{(Color online) 
Dispersion of the Bogoliubov quasiparticle (a) in the BCS and (b) BEC regimes for the parabolic electron band in the normal state.
The intensity of the color represents the spectral weight of the Bogoliubov quasiparticle.
In the BCS regime, the minimum of the spectral gap $E_k=\Delta$ occurs at $|\bm{k}|=k_{\rm F}$.
In the BEC regime, the minimum spectral gap occurs at $\bm{k}=0$.
The minimum energy that breaks the Cooper pairs is determined to be $2\Delta$ and $2|\mu|$ for the BCS and BCE regimes, respectively.
}
\label{BQP_Dispersion}
\end{figure}

The bound energy of a Cooper pair is given by the superconducting gap in the BCS regime, while it is given by the chemical potential in the BEC regime.
The excitation energy of the quasiparticles (Bogoliubov quasiparticles) in the superfluid phase of the fermionic condensate is given by 
\begin{equation}
E_{k}=\pm \sqrt{(\varepsilon_k-\mu)^2+\Delta^2}, 
\end{equation}
where $\varepsilon_k=\hbar^2k^2/2m$ is the electron energy dispersion.
In the BCS regime, the chemical potential coincides with the Fermi energy at $T=0$, $\mu=\varepsilon_{\rm F}(=\hbar^2k_{\rm F}^2/2m)$.
The minimum of the spectral gap $E_k=\Delta$ opens at $\varepsilon=\mu$,  corresponding to $|\bm{k}|=k_{\rm F}$, as displayed in Fig.~\ref{BQP_Dispersion}(a).
In the BEC regime, where $\mu$ is negative, the minimum spectral gap is located at $\bm{k}=0$, as displayed in Fig.~\ref{BQP_Dispersion}(b).
In this regime where $|\mu|\gg\Delta$, $E_{k}=\sqrt{\mu^2+\Delta^2}\approx |\mu|$.
Thus, the minimum energy that breaks the Cooper pairs is determined to be $2\Delta$ and $2|\mu|$ for the BCS and BEC regimes, respectively.

The BCS state with cooperative Cooper pairing and the BEC state with composite bosons share the same type of spontaneous symmetry breaking.
The change between the two states is a continuous crossover at $T=0$~\cite{Legget80} and at a finite temperature,\cite{Nozi85} connected through a progressive reduction in the size of electron pairs involved as fundamental entities in both phenomena.
This crossover goes across the intermediate regime where the size of the pairs is comparable to  the average interparticle distance, $k_{\rm F}\xi\sim1$.
The BCS-BEC crossover has been extensively studied in  ultracold atomic systems, in which the attractive interaction can be controlled experimentally by a Feshbach resonance~\cite{BCSBEC}, but is extremely difficult to realize for electrons in solids.
Interest in the BCS-BEC crossover in high-$T_{\rm c}$ cuprate has also grown~\cite{Uemura89}, in which the size of the pairs appears to be comparable to the interparticle spacing.
In particular, the problem of the preformed pairs  in high-$T_{\rm c}$ cuprates has attracted considerable attention as an origin of the pseudogap formation in the underdoped regime~\cite{Chen05,Emery95,Franz07}.
However, the pseudogap and preformed pairs of cuprates remain highly controversial and unresolved issues~\cite{Keimer15}.

In FeSe, the values of $2\Delta_{\rm l}/k_{\rm B}T_{\rm c}\approx 9$ and $2\Delta_{\rm s}/k_{\rm B}T_{\rm c}\approx 6.5$ for the two gaps observed in spectroscopic measurements significantly increase  from the weak-coupling BCS value of 3.5, implying that the attractive interaction holding together the Cooper pairs takes an extremely strong coupling nature, as expected in the crossover regime.
Recently, it has been suggested that FeSe is deep inside the BCS-BEC crossover regime.
Below we discuss experimental evidence that FeSe provides a new platform for studying the electronic properties in the crossover regime.

\subsection{Evidence for the crossover in FeSe}

\subsubsection{Penetration depth}

High-quality single crystals of FeSe enable us to estimate the Fermi energies $\varepsilon_{\rm F}^{\rm e}$ and $\varepsilon_{\rm F}^{\rm h}$ from the band edges of electron and hole sheets, respectively, by using several techniques. 
All of them consistently point to extremely small Fermi energies.
First, we discuss the absolute value of the penetration depth in FeSe.

In 2D systems,  $\varepsilon_{\rm F}$ is related to the London penetration depth $\lambda(0)$ as $\varepsilon_{\rm F}=\frac{\pi \hbar^2 d}{\mu_0e^2}\lambda^{-2}(0)$, where $d$ is the interlayer distance~\cite{Prozorov06}.
For FeSe, $\lambda(0)\approx 400$~nm~\cite{Kasahara14}.
As the Fermi surface consists of one hole sheet and one (compensating) electron sheet, $\lambda$ can be written as  $1/\lambda^2_{\rm L}=1/(\lambda^{\rm e})^2+1/(\lambda^{\rm h})^2$, where $\lambda^{\rm e}$ and $\lambda^{\rm h}$ represent the contributions from the electron and hole sheets, respectively.
Assuming that the two sheets have similar effective masses, $\varepsilon_{\rm F}^{\rm h}\sim\varepsilon_{\rm F}^{\rm e}\approx 7$--8~meV is obtained~\cite{Kasahara14}.
The magnitude of the Fermi energy can also be inferred from the thermoelectric response in the normal state. From the Seebeck coefficient~\cite{Pourret11}, the upper limit of $\varepsilon_{\rm F}^{\rm e}$ is deduced to be $\sim 10$~meV.
These results indicate that the Fermi energies of the hole and electron pockets are both extremely small.
  
\begin{figure}[b]
\centering
\includegraphics[width=0.85\linewidth]{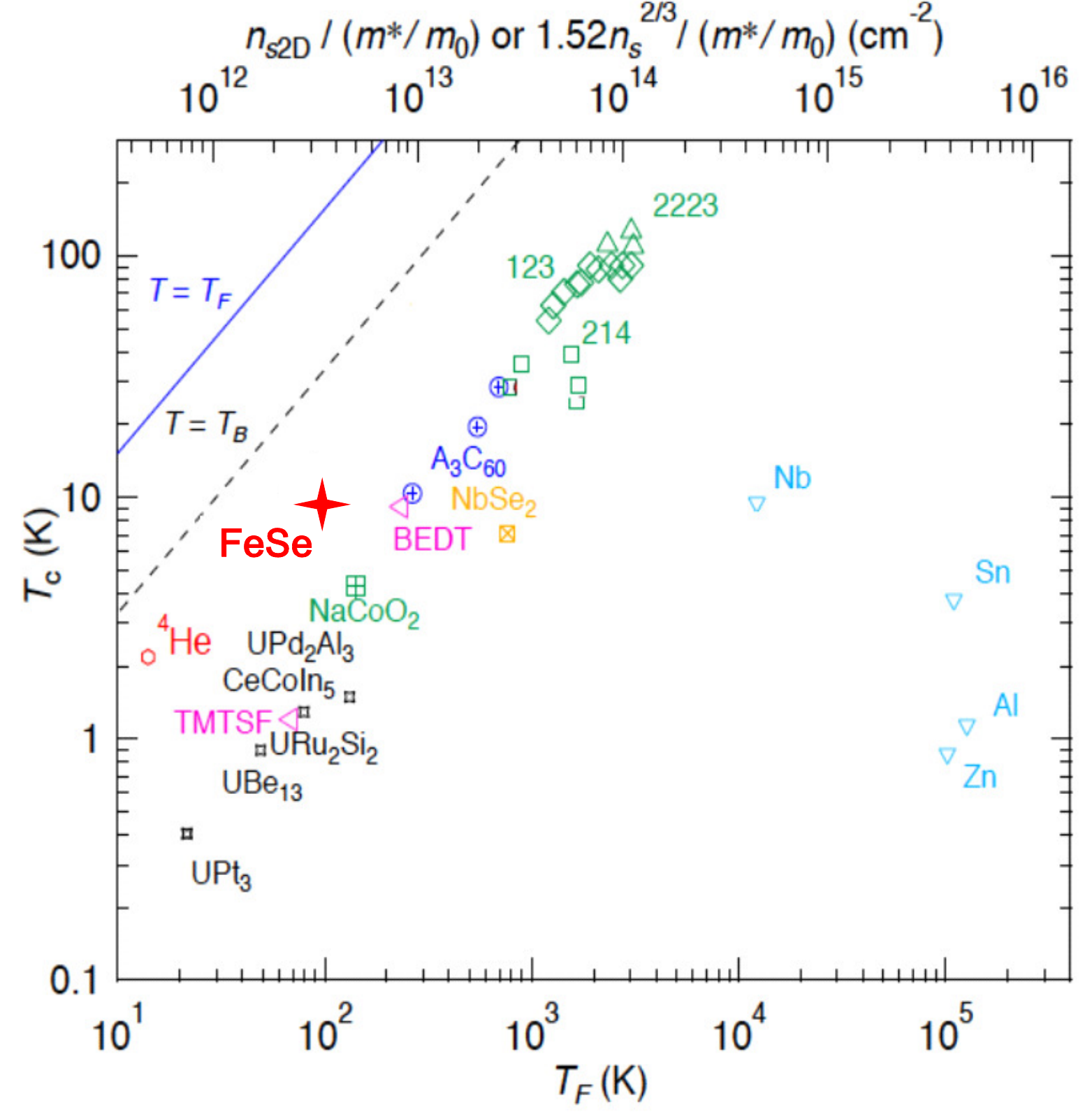}
\caption{(Color online) 
Uemura plot.
$T_{\rm c}$ plotted as a function of Fermi temperature $T_{\rm F}$ evaluated from $1/\lambda^2(0)$ for various 2D and 3D superconductors, including  conventional superconductors such as Nb;  high-$T_{\rm c}$ cuprates such as La$_{2-x}$Sr$_x$CuO$_4$ (214), YBa$_2$Cu$_3$O$_{7-\delta}$ (123), and Bi$_2$Sr$_2$Ca$_2$Cu$_3$O$_{y}$ (2223); and  organic and heavy fermion compounds.
The dashed line is the BEC temperature $T_{\rm B}$ for the ideal 3D Bose gas.
}
\label{Uemura}
\end{figure}

To place FeSe in the context of other superconductors, $T_{\rm c}$ is plotted as a function of Fermi temperature $T_{\rm F}\equiv\varepsilon_{\rm F}/k_{\rm B}$ or an equivalent critical temperature $T_{\rm B}$ for the BEC of electron pairs for several materials including FeSe (Uemura plot, Fig.\,\ref{Uemura})~\cite{Uemura89}.
Because the relevant Fermi surface sheets are nearly cylindrical, $T_{\rm F}$ for 2D systems may be estimated directly from $\lambda(0)$ via the relation $T_{\rm F}=\frac{\pi \hbar^2 n_{\rm 2D}}{k_{\rm B}m^{\ast}}\approx \left(\frac{\pi \hbar^2  d}{\mu_0k_{\rm B}e^2}\right) \lambda^{-2}(0)$, where $n_{\rm 2D}$ is the carrier concentration within the superconducting planes and $d$ is the interlayer distance.
For three-dimensional (3D) systems, $T_{\rm F}=(\hbar^2/2)(3\pi^2n)^{2/3}/k_{\rm B}m^{\ast}$.  
The dashed line corresponds to the BEC temperature for an ideal 3D Bose gas, $T_{\rm B}=\frac{2\pi\hbar^2}{m^{\ast}k_B}\left[\frac{(n/2)}{\zeta(3/2)}\right]^{2/3}$. 
In a quasi-2D system, this $T_{\rm B}$ provides an estimate of the maximum condensate temperature. Notably, the magnitude of $T_{\rm c}/T_{\rm F} \approx 0.10$ of FeSe exceeds that of cuprates and reaches nearly 50\% of the value of superfluid $^4$He.
It has been shown that $T_{\rm c}/T_{\rm F}$ in BaFe$_2$(As$_{1-x}$P$_x$)$_2$ markedly increases near an AFM QCP at $x_{\rm c}\approx 0.30$ owing to the increase in $m^{\ast}$~\cite{Hashimoto12}.
We note that the $T_{\rm c}/T_{\rm F}$ of FeSe is even larger than that of BaFe$_2$(As$_{1-x}$P$_x$)$_2$ at $x_{\rm c}$.
Thus, Fig.\,\ref{Uemura} indicates that FeSe is located closer to the BEC line than all other superconductors.

\subsubsection{ARPES}

As discussed above, in the BCS-BEC regime, the Bogoliubov quasiparticles exhibit a characteristic flat band dispersion near $\bm{k}=0$, which is distinctly different from the back-bending behavior at $k_{\rm F}$ expected in the BCS regime.
Such a quasiparticle dispersion can be directly observed by ARPES.   

A signature of the BCS-BEC crossover has been observed by ARPES measurements of  several iron-based superconductors.
For  Ba$_{1-x}$K$_x$Fe$_2$As$_2$~\cite{Shimojima17} and LiFe$_{1-x}$Co$_{x}$As~\cite{Miao15}, the crossover condition $\Delta/\varepsilon_{\rm F}\sim 1$ has been reported.
However, in these compounds, the crossover condition is satisfied only in a minor band with a small Fermi energy, and the Fermi energy is much larger than the gap energy in the main bands, $\Delta/\varepsilon_{\rm F}\ll1$.
In Fe$_{1+y}$Se$_{x}$Te$_{1-x}$,  $\Delta/\varepsilon_{\rm F}$ increases and the Bogoliubov quasiparticle band changes from BCS-like to flat-band-like by changing the concentration of excess Fe~\cite{Lubashevsky12,Rinott17,Okazaki15}.
However, in this system, owing to the crystal imperfection and excess Fe atoms, the superconducting gap is spatially inhomogeneous  compared with that in FeSe$_{1-x}$S$_{x}$.  
In addition, the Fermi energy of electron band is not well known. 

Very recently, flat energy dispersions, which are characteristic of the crossover regime, have been observed at the hole pocket in the superconducting state of FeSe.
As the Fermi energy of the electron pocket in FeSe is much smaller than that of the hole pocket, all the bands satisfy the crossover condition~\cite{Kasahara14,Hashimoto20}.
Moreover, it has been reported that the crossover signature is more pronounced with sulfur substitution.
In particular, in FeSe$_{1-x}$S$_{x}$ with $x=0.18$ in the tetragonal regime, an unusual quasiparticle dispersion, which is close to that expected in the BEC regime displayed in Fig.~\ref{BQP_Dispersion}(b), has been observed~\cite{Hashimoto20}.

\subsubsection{Quasiparticle interference (QPI)}

STM/STS can also be used to investigate  electronic dispersions through the QPI effect.
The QPI patterns are simply electronic standing waves scattered off defects and appear in the energy-dependent conductance images.
The Fourier transformation of the conductance images allows us to determine the energy-dependent scattering vectors $\bm{q}(E)$, from which one can infer the quasiparticle dispersions in momentum space.
Unlike ARPES, QPI can access not only the filled state but also the empty state above $\varepsilon_{\rm F}$, making it useful for exploring the electron bands in FeSe.

\begin{figure}[tb]
\centering
\includegraphics[width=0.9\linewidth]{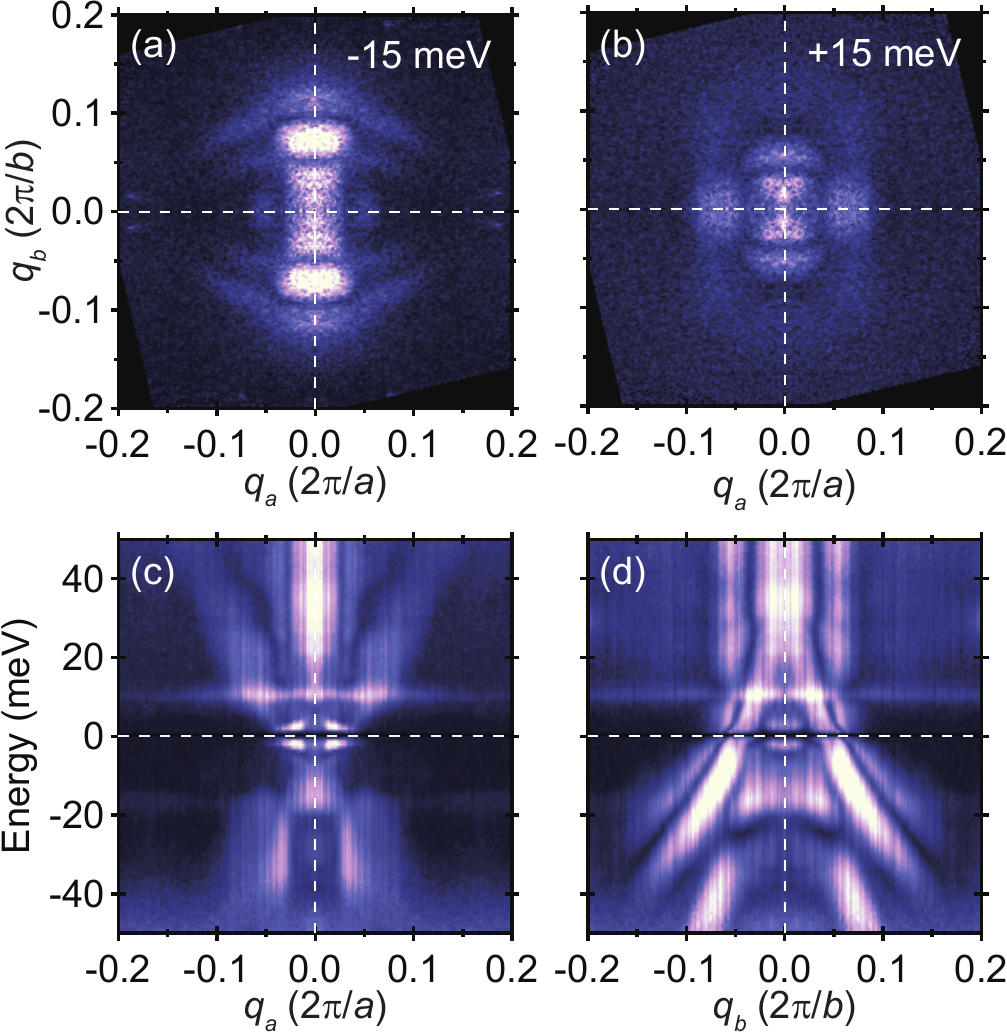}
\caption{(Color online) 
QPI patterns of FeSe at a low temperature (1.5~K).
A magnetic field $\mu_0H=12$~T was applied along the $c$-axis to suppress superconductivity.
(a) and (b) Fourier-transformed normalized conductance images at -15 and +15\,meV, respectively.
Uniaxial patterns are observed owing to nematicity.
Here, the axes of the orthorhombic unit cell are  defined as $a<b<c$.
(c) and (d) QPI dispersions obtained by taking line cuts from the energy-dependent Fourier-transformed QPI patterns along principle axes $\bm{q}_a$ and $\bm{q}_b$ in scattering space.
Electron-like and hole-like dispersions are identified along $\bm{q}_a$ and $\bm{q}_b$, respectively.
Adopted from Ref.\,\citen{Hanaguri18}.
}
\label{QPI}
\end{figure}

The QPI patterns of FeSe (Fig.~\ref{QPI}) are highly anisotropic owing to nematicity.
The obtained QPI dispersions consist of one electron branch and multiple hole branches (Fig.\,\ref{QPI}), and the electron and hole branches disperse along orthogonal axes~\cite{Kasahara14,Hanaguri18,Kostin18}.
There are at least two hole-like QPI branches that cross $\varepsilon_{\rm F}$, while there is only one hole band at $\varepsilon_{\rm F}$ (Sect.~2.1).
It has been argued that these two branches come from different $k_z$ states at $k_z=0$ and $k_z=\pi$~\cite{Hanaguri18,Rhodes19}.
These QPI signals may be associated with the scattering vectors that correspond to the minor axes of the cross sections of the nematicity-deformed hole and electron Fermi pockets~\cite{Kasahara14,Hanaguri18,Kostin18}.

In addition to these features, one can estimate Fermi energies and Fermi momenta from the QPI dispersions, which faithfully represent the band dispersions.
The top and bottom of the hole and electron branches correspond to $\varepsilon_{\rm F}^{\rm h}$ and $\varepsilon_{\rm F}^{\rm e}$, respectively, allowing us to estimate $\varepsilon_{\rm F}^{\rm h}\sim 10$--20~meV and $\varepsilon_{\rm F}^{\rm e}\sim 5$--10~meV (Fig.~\ref{QPI}).
The Fermi momenta can be obtained from the scattering vectors at zero energy and are estimated to be $k_{\rm F}^{\rm h}\approx 0.5-0.8$~nm$^{-1}$ and $k_{\rm F}^{\rm e}\approx 0.4$~nm$^{-1}$ for hole and electron cylinders, respectively.
These correspond to the minor axes of the deformed Fermi cylinder. The $k_{\rm F}$ along the major axes may be a few times larger depending on the Fermi surface pocket so as to satisfy the compensation condition. 
Such shallow bands are consistent with those observed by the quantum oscillations and ARPES measurements~\cite{Watson15,Terashima14}.

STM/STS can also directly evaluate the superconducting-gap size from the tunneling spectrum.
As discussed in Sect.~3.3, FeSe exhibits highly anisotropic superconducting gaps and there are at least two distinct superconducting gaps $\Delta_{\rm l} \approx3.5$~meV and $\Delta_{\rm s} \approx 2.5$~meV, which may represent the gaps opening on different Fermi surfaces~\cite{Kasahara14,Hanaguri18}.
It is still unclear which gap opens on which Fermi surface.
Nevertheless, since $\varepsilon_{\rm F}\lesssim 20$~meV and $\Delta \gtrsim 2.5$~meV, the ratio $\Delta/\varepsilon_{\rm F}$ must be larger than 0.1 for both bands, placing FeSe in the BCS-BEC crossover regime.
Additional strong evidence for the BCS-BEC crossover is provided by the extremely small $k_{\rm F}\xi$.
Since $\xi_{ab}$, which is an average value in a 2D plane, determined from the upper critical field ($\sim 17$~T) in  a perpendicular field ($\bm{H}\parallel c$),  is roughly 5\,nm, $k_{\rm F}\xi$ should be on the order of unity, again indicating BCS-BEC crossover superconductivity.

\subsubsection{Quantum-limit vortex core}

\begin{figure}[tb]
\centering
\includegraphics[width=0.9\linewidth]{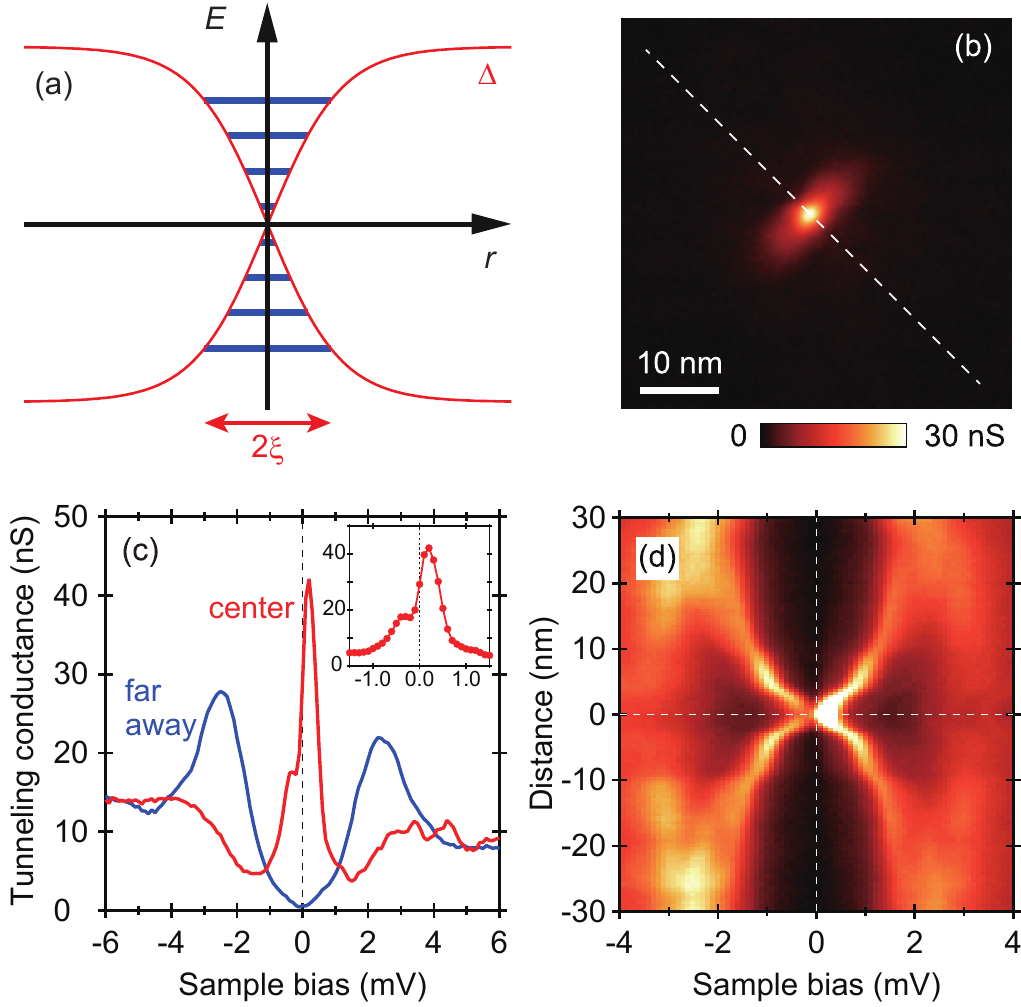}
\caption{(Color online) 
(a) Schematic energy $E$ diagram of an $s$-wave superconductor as a function of the distance $r$ from the center of the vortex core.
The superconducting gap $\Delta$ (red) recovers over the coherence length $\xi$ and the discrete Caroli-de~Gennes-Matricon states (blue) are formed in the core region.
(b) Zero-energy conductance map showing a single vortex at zero energy at 0.4\,K in a magnetic field of 0.25\,T along the $c$-axis of FeSe.
(c) Tunneling spectra taken at the vortex center (red) and away from vortices (blue).
The inset shows a magnified spectrum at the vortex center.
(d) Spatial evolution of tunneling spectra along the dashed line in (b).
Adopted from Ref.\,\citen{Hanaguri19}.
}
\label{vortex}
\end{figure}

The large $\Delta/\varepsilon_{\rm F}$ value should give rise to novel features in the vortex core.
Since the vortex core is a type of potential well, quantized bound states (Caroli-de~Gennes-Matricon states) should be formed inside, as schematically shown in Fig.~\ref{vortex}(a).
Such vortex-core states can be investigated by STM/STS, in principle.
The energies of these states are given by $\pm \mu_c\Delta^2/\varepsilon_{\rm F}$, where $\mu_c=1/2,3/2,5/2, \cdots$ is the quantum number that represents the angular momentum.
In the BCS limit, owing to the small ratio of $\Delta/\varepsilon_{\rm F}$, $\Delta^2/\varepsilon_{\rm F}$ is on the order of $\mu$eV for most known superconductors.
The number of   bound states is roughly $\varepsilon_{\rm F}/\Delta$, which is usually very large, more than 1000.
Therefore, because of inevitable smearing effects (e.g., thermal broadening), it is almost impossible to observe the individual Caroli-de~Gennes-Matricon states by STM/STS.
Instead, a large number of bound states overlap to form a broad particle-hole symmetric peak at zero energy in the tunneling spectrum at the vortex center.
With increasing distance from the center, this zero-energy peak splits and continuously approaches  $\pm\Delta$~\cite{Suderow14}.

By contrast, in the BCS-BEC-crossover regime, $\varepsilon_{\rm F}/\Delta$ should be on  the order of unity and thus the vortex core accommodates only a few levels, resulting in the so-called quantum-limit vortex~\cite{Hayashi98}.
Here, $\Delta^2/\varepsilon_{\rm F}$ can become large enough for each bound state to be resolved by STM/STS.
The spatial evolution of the bound states should  no longer be continuous and should show  Friedel-like oscillations~\cite{Hayashi98}.

Such characteristic signatures of BCS-BEC crossover have been reported for FeSe~\cite{Hanaguri19}.
Figure~\ref{vortex}(b) shows the vortex image of FeSe in a magnetic field of 0.25~T along the $c$-axis.
The vortex is elongated due to the nematicity.
As shown in Fig.\,\ref{vortex}(c), the lowest-energy local DOS (LDOS) peak of FeSe is not at zero energy, representing the lowest bound state in the quantum-limit vortex core.
The spatial evolution of the bound states exhibits   wiggling behavior~\cite{Hanaguri19}.
The Fourier analysis shows that the wavelength of such wiggles corresponds to $\pi/k_{\rm F}$,   consistent with the theoretical prediction of  Friedel-like oscillations~\cite{Hayashi98}.

\subsection{Superconducting fluctuations, preformed pairs, and pseudogap}

\subsubsection{Giant superconducting fluctuations}

It is well known that Cooper pairs can survive even above $T_{\rm c}$ as thermally fluctuating droplets.
These fluctuations arise from amplitude fluctuations of the superconducting order parameter and have been investigated for many decades~\cite{Larkin05}.
Their effects on thermodynamic, transport, and thermoelectric quantities in most superconductors are well understood in terms of the standard Gaussian fluctuation theories.
However, in the presence of preformed pairs associated with the BCS-BEC crossover, superconducting fluctuations are expected to be strikingly enhanced compared with the Gaussian theories owing to additional phase fluctuations~\cite{Emery95}.
Of particular interest is pseudogap formation, which is the central enigma of the underdoped cuprates~\cite{Keimer15}.
The origin of the pseudogap has been discussed in terms of the preformed pairs associated with the crossover phenomenon, which can lead to a partial reduction in DOS near the Fermi level~\cite{Chen05}.
However, it is still highly controversial.
Another important issue associated with the BCS-BEC crossover is the breakdown of Landau's Fermi liquid theory due to the strong interaction between fermions and fluctuating bosons.
In ultracold atomic systems,  Fermi liquid-like behavior has been observed in thermodynamics even in the crossover regime, but more recent photoemission experiments have suggested a sizable pseudogap opening and a breakdown of the Fermi liquid description~\cite{Gaebler10}.

Thus, the superconducting fluctuations and pseudogap formation in FeSe are highly intriguing. Superconducting fluctuations give rise to the reduction in normal-state resistivity.
In zero field, the ${\rm d}\rho_{xx}/{\rm d}T$ of FeSe shows a minimum at around $T^{\ast}\sim$20\,K, which can be attributed to the appearance of the additional conductivity due to the fluctuation of the order parameter (paraconductivity) below $T^{\ast}$~\cite{Kasahara16}. However, a quantitative analysis of the paraconductivity is difficult to achieve because its evaluation strongly depends on the extrapolation of the normal-state resistivity above $T^{\ast}$ to lower $T$.
The fluctuation-induced magnetoresistance of FeSe is also difficult to analyze owing to a large and complicated magnetoresistance, which is characteristic of compensated semimetals (see Fig.~\ref{rho_T}).

The superconducting fluctuations in FeSe have been examined through  magnetic measurements by several research groups~\cite{Kasahara16,Yang17,Takahashi19}.
The diamagnetic response due to superconducting fluctuations is clearly observed in the magnetization $M(H)$ for $\bm{H}\parallel c$, which exhibits a pronounced decrease below $T^{\ast}$.
A crossing point in the diamagnetic response to magnetization, where $M_{\rm dia}(T, H)$ exhibits a field-independent value, is observed near $T_{\rm c}$~\cite{Kasahara16}.
Such crossing behavior,  which has been pointed out to be a signature of large superconducting fluctuations,  is also observed in cuprates~\cite{Welp91}.
The fluctuation-induced diamagnetic susceptibility of most superconductors including multiband systems can be well described by the standard Gaussian-type (Aslamasov--Larkin) fluctuation susceptibility $\chi_{\rm AL}$, which is given by
\begin{equation}
 \chi_{\rm AL}=-\frac{2\pi^2}{3}\frac{k_{\rm B}T_{\rm c}}{\Phi_0^2}\frac{\xi_{ab}^2}{\xi_{c}}\sqrt{\frac{T_{\rm c}}{T-T_{\rm c}}},
 \label{Eq:torque}
\end{equation}
in the zero-field limit, where $\Phi_0$ is the flux quantum  and $\xi_{ab}$ and $\xi_{c}$ are the coherence lengths parallel and perpendicular to the $ab$ plane at zero temperature, respectively~\cite{Larkin05}.
In the multiband case, the behavior of $\chi_{\rm AL}$ is determined by the smallest  coherence length of the main band, which governs the orbital upper critical field.

\begin{figure}[t]
\centering
\includegraphics[width=0.75\linewidth]{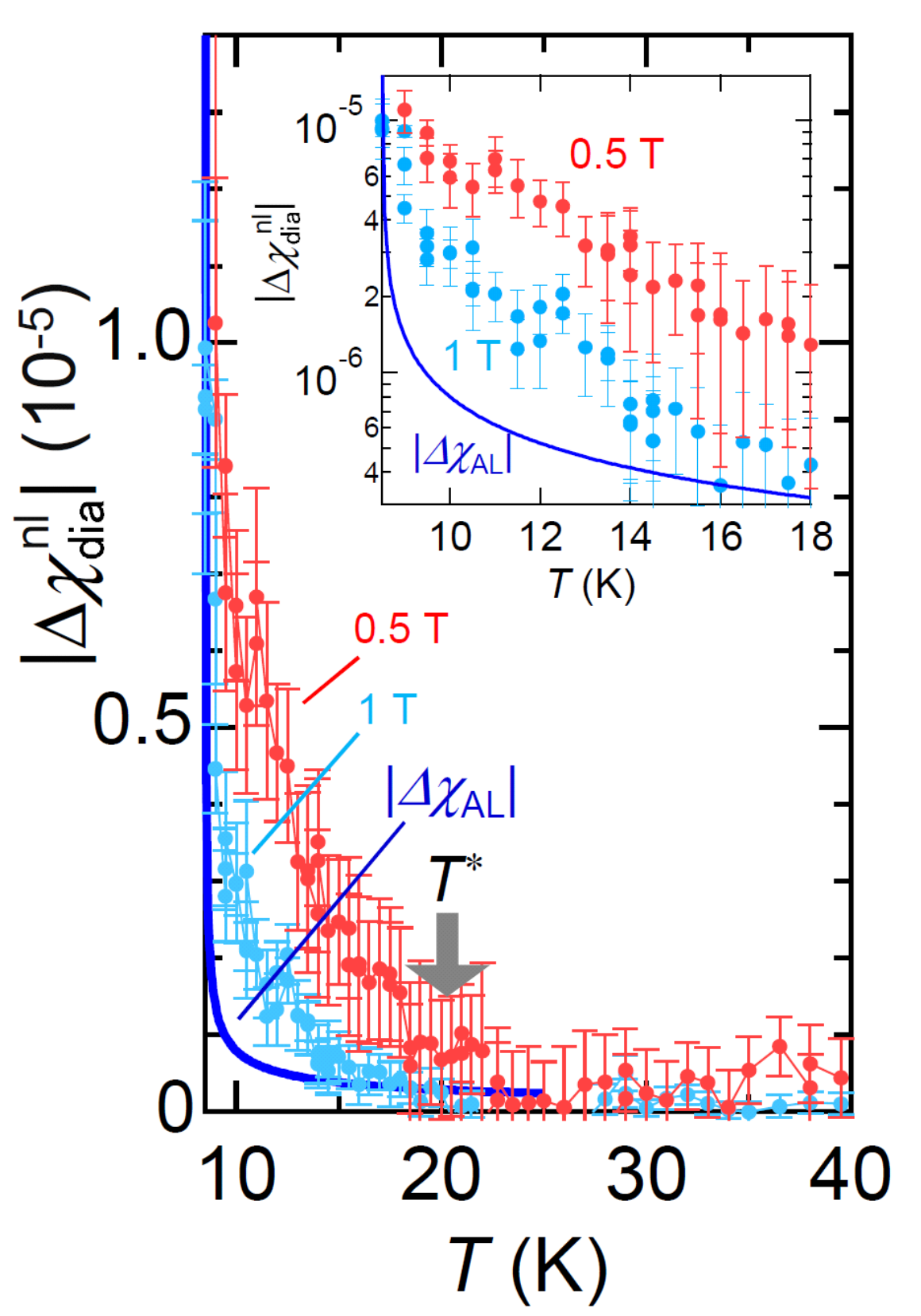}
\caption{(Color online) 
Temperature dependence of the nonlinear diamagnetic response of FeSe at $\mu_0H=0.5$~T (red) and 1~T (blue) obtained by  magnetic torque measurements.
Below $T^{\ast}$,  this diamagnetic response is largely enhanced.
The blue line represents $|\Delta\chi_{\rm AL}|$ in the standard Gaussian fluctuation theory calculated from Eq.\,(\ref{Eq:torque}).
The inset shows $|\Delta\chi_{\rm dia}^{\rm nl}|$ plotted in a semi-log scale at low temperatures.
Adopted from Ref.~\citen{Kasahara16}.
}
\label{Fluctuations}
\end{figure}

The low-field diamagnetic response in FeSe, which is measured by the magnetic torque $\bm{\tau}=\mu_0V\bm{M}\times \bm{H}$ (where $V$ is the sample volume), has been reported to exhibit the superconducting fluctuations with a highly unusual nature~\cite{Kasahara16}.
From the torque measurements, the difference between the $c$-axis and $ab$-plane susceptibilities, $\Delta\chi=\chi_c-\chi_{ab}$, can be determined.
It has been reported that $\Delta\chi(T)$ markedly increases with decreasing $T$ and exhibits divergent behavior near $T_{\rm c}$, indicating the presence of the superconducting fluctuation-induced diamagnetic contribution~\cite{Kasahara16}.
The Gaussian-type fluctuation susceptibility given by Eq.\,(\ref{Eq:torque}) indicates that the diamagnetic response of the magnetization is $H$-linear.
In contrast, the observed diamagnetic response of FeSe contains both $H$-linear and nonlinear contributions of the magnetization~\cite{Kasahara16}. 
Figure\,\ref{Fluctuations} and its inset show the temperature dependence of the nonlinear part $\Delta\chi_{\rm dia}^{\rm nl}$, which is estimated by subtracting the $H$-linear contribution obtained at the highest field as $\Delta\chi_{\rm dia}^{\rm nl}(H)\approx \Delta \chi(H)-\Delta \chi(7~{\rm T})$.
In Fig.\,\ref{Fluctuations} and its inset, the contribution expected from the Gaussian fluctuation theory given by $\Delta\chi_{\rm AL}\approx -\frac{2\pi^2}{3}\frac{k_{\rm B}T_{\rm c}}{\Phi_0^2}\left(\frac{\xi_{ab}^2}{\xi_{c}}-\xi_c\right)\sqrt{\frac{T_{\rm c}}{T-T_{\rm c}}}$ is also plotted, where we use $\xi_{ab}=5.5$~nm and $\xi_{c}=1.5$~nm.
Near $T_{\rm c}$, $\Delta \chi_{\rm dia}^{\rm nl}$ at 0.5~T is nearly 10 times larger than $\Delta \chi_{\rm AL}$.  

The above results provide evidence that the amplitude of the diamagnetic fluctuations of FeSe by far exceeds that expected in the standard Gaussian theory, implying that the superconducting fluctuations in FeSe are distinctly different from those in conventional superconductors.
The superconducting fluctuations in FeSe have been further examined by several research groups.
The NMR relaxation rate divided by the temperature $(T_1T)^{-1}$, which increases below $T_{\rm s}$, starts to be suppressed below $T^{\ast}$.~\cite{Shi18}
The NMR results indicate the presence of  superconducting fluctuations that deviate from the standard Gaussian theory.
On the other hand,  magnetization measurements using a superconducting quantum interference device (SQUID) did not reveal such a large superconducting fluctuation signal~\cite{Yang17}.
Moreover,  magnetic torque measurements using an optical detection technique indicated a considerably smaller fluctuation signal originating from the vortex liquid~\cite{Takahashi19}.
The origin of the discrepancy between these measurements is not clear.  
It may be due to the sample quality. 
%The discrepancy between these measurements may be due to the sample quality.
In fact, the crystals in which a giant superconducting fluctuation is observed exhibit very large magnetoresistance and distinct quantum oscillations at high magnetic fields.

Although giant superconducting fluctuations are observed in the diamagnetic response in FeSe, the jump of the heat capacity at $T_{\rm c}$ is still  ordinary mean-field BCS-like as shown in Fig.~\ref{BulkT}(c).
However, a recent detailed analysis of the heat capacity near $T_{\rm c}$ suggests the presence of superconducting fluctuations  that substantially exceed Gaussian fluctuations~\cite{Mizukami20}.
The heat capacity measurement revealed significant fluctuation effects not only in the zero field but also in the vortex state with a magnetic field applied both parallel and perpendicular to the $ab$ plane.

Thermal fluctuations have a marked effect on the vortex system in type-II superconductors~\cite{Blatter94}.
One of the most prominent effects is  vortex lattice melting.
A vortex lattice melts when the thermal displacement of the vortices is an appreciable fraction of the distance between vortices.
In quasi-2D high-$T_{\rm c}$ cuprates, the magnetic field at which the melting transition occurs is much lower than the mean upper critical field $H_{\rm c2}$.
The strength of the thermal fluctuations is quantified by the dimensionless Ginzburg number, $G_{\rm i} = [\epsilon k_{\rm B}T_{\rm c}/H_{\rm c}^2(0)\xi_{ab}^3]^2/2$, which measures the relative size of the thermal energy $k_{\rm B}T_{\rm c}$ and the condensation energy within the coherence volume.
Here, $\epsilon\equiv \lambda_{c}/\lambda_{ab}$ ($\lambda_{c}$ is the penetration depth for screening current perpendicular to the $ab$ plane) is the anisotropy ratio and $H_{\rm c}=\Phi_0/2\sqrt{2}\lambda_{ab}\xi_{ab}$ is the thermodynamic critical field.
The large $G_{\rm i}$ leads to the reduction in the vortex lattice melting temperature $T_{\rm melt}$.
$G_{\rm i}$ is roughly proportional to $(\Delta/\varepsilon_{\rm F})^4$.
In conventional low-$T_{\rm c}$ superconductors, $G_{\rm i}$ ranges from $10^{-11}$ to $10^{-7}$, while in FeSe with a large $\Delta/\varepsilon_{\rm F}$, $G_{\rm i}$ is estimated to be as large as $10^{-2}$, which is comparable to or even larger than that of YBa$_2$Cu$_3$O$_7$.
We therefore expect a sizable separation of the melting transition below $T_{\rm c}(H)$ over a large portion of the phase diagram, as  is the case for the high-$T_{\rm c}$ cuprates.
Very recently, the vortex lattice melting transition has been observed  by heat capacity measurements~\cite{Hardy20}.
It has been reported that the melting line merges with $H_{\rm c2}$ at a finite temperature, which is consistent with the theory of the vortex lattice melting in strongly Pauli limited superconductors~\cite{Adachi03}.

\subsubsection{Pseudogap}

An important question related to the preformed pairs is the formation of the pseudogap, which is a characteristic signature of BCS-BEC crossover other than the giant superconducting fluctuations.
The pseudogap formation in FeSe is a nontrivial issue because FeSe is a compensated semimetal with hole and electron pockets, which may give rise to more complex phenomena than in the single-band case~\cite{Chubukov16b}.
Below $T^{\ast}$, the NMR $(T_1T)^{-1}$ is suppressed and exhibits a broad maximum at $T_{\rm p}(H)$, which bears a resemblance to the pseudogap behavior in optimally doped cuprate superconductors~\cite{Shi18}.
It has been reported that  $T^{\ast}$  and $T_{\rm p}(H)$ decrease in the same manner as $T_{\rm c}(H)$ with increasing $H$.
This suggests that the pseudogap behavior in FeSe is due to superconducting fluctuations, which presumably originate from the theoretically predicted preformed pairs~\cite{Shi18}.

\begin{figure}[t]
\centering
\includegraphics[width=0.7\linewidth]{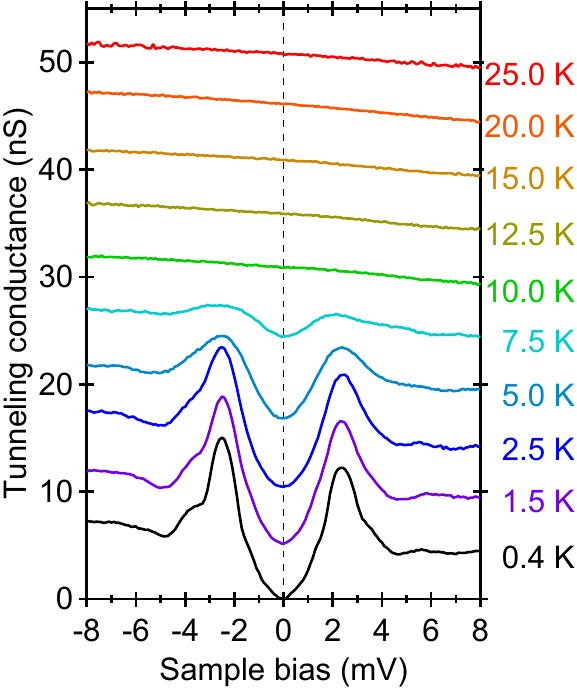}
\caption{(Color online) 
Tunneling spectra of FeSe obtained at different temperatures.
Each curve is shifted by 5~nS for clarity.
Adopted from Ref.~\citen{Hanaguri19}.
}
\label{STS_PG}
\end{figure}

A spectroscopic signature for the pseudogap formation above $T_{\rm c}$ was first reported by the STM/STS measurements~\cite{Rossler15}.
However, as shown in Fig.~\ref{STS_PG}, the subsequent STM/STS measurements indicated the absence of the pseudogap~\cite{Hanaguri19}.
The absence of the spectroscopic pseudogap despite a large $\Delta/\varepsilon_{\rm F}$ has been discussed in terms of the multiband character  of FeSe and its compensated semimetal nature.
In the BCS-BEC crossover superconductivity of a single-band system, the chemical potential is shifted outside of the band edge. 
However, if there are hole and electron bands, which nearly compensate each other and have strong interband interactions, the chemical potential should be pinned at the original energy position.
In the case of perfectly symmetrical electron and hole bands, the chemical potential should always be pinned at zero energy because $\mu_{\rm e}+\mu_{\rm h}=0$  in both the normal and superconducting states, where $\mu_{\rm e}$ and $\mu_{\rm h}$ are the chemical potentials of the electron and hole bands, respectively.
It has been pointed out that in such a case, the splitting between $T^{\ast}$ and $T_{\rm c}$ is largely reduced~\cite{Chubukov16b}, leading to the suppression of the pseudogap formation even in the BCS-BEC crossover regime.

In a real material, however, there is a certain asymmetry between the electron and hole bands, which should cause a shift of the chemical potential as a function of temperature.
The multiband character brings about another effect that suppresses the pseudogap formation~\cite{Hanaguri19}.
If there are electron and hole pockets,  two pairing channels associated with the interband and intraband interactions should appear.
In the case of FeSe, superconductivity occurs in the nematic phase and the superconducting gap exhibits strong anisotropy as discussed in Sect.~3.
A possible scenario is that the anisotropy is caused by the mixture of $s$-wave and $d$-wave symmetries, where the former and the latter are caused by interband and intraband interactions, respectively.
Theoretical calculations based on such a model have been performed to estimate the pair formation and  superconducting transition temperatures~\cite{Hanaguri19}.
It has been shown that if the interband pairing is stronger than the intraband pairing, which may be the case for FeSe, the pair formation and superconducting transition temperatures do not split despite the large $\Delta/\varepsilon_{\rm F}$~\cite{Hanaguri19}.
The BCS-BEC crossover in the multiband system may be difficult to realize in ultracold atomic systems.
Therefore, FeSe is unique and can open a new research field for the BCS-BEC crossover.

\subsubsection{Evolution of BCS-BEC crossover in FeSe$_{1-x}$S$_{x}$}

Very recently, the evolution of the BCS-BEC crossover properties with sulfur substitution in FeSe$_{1-x}$S$_{x}$ has been investigated.
ARPES measurements revealed that $\Delta/\varepsilon_{\rm F}$ decreases with increasing $x$, but it still remains large~\cite{Hashimoto20}.
Therefore, it is expected that the crossover nature will be less pronounced in FeSe$_{1-x}$S$_{x}$ with increasing $x$.
Contrary to this expectation, it has been found that the crossover nature becomes more significant in FeSe$_{1-x}$S$_{x}$~\cite{Mizukami20,Hashimoto20}.
ARPES experiments found that the flat dispersion of the Bogoliubov quasiparticles is more pronounced with increasing $x$ in the nematic regime.
Surprisingly, on entering the tetragonal regime beyond the nematic QCP,  the flat dispersion changes to a BEC-like one, which shows a minimum gap at $\bm{k}=0$~\cite{Hashimoto20}.
Heat capacity measurements reveal a highly unusual BEC-like transition with strong fluctuations in tetragonal FeSe$_{1-x}$S$_{x}$~\cite{Mizukami20}, which appears to be consistent with the ARPES measurements.

However, the formation of the spectroscopic pseudogap in  tetragonal FeSe$_{1-x}$S$_{x}$ is controversial.
Although ARPES measurements indicated a the distinct pseudogap-like reduction in DOS in the hole pocket above $T_{\rm c}$~\cite{Hashimoto20}, no discernible reduction in DOS was observed in STS measurements~\cite{Mizukami20}.
The evolutions of the BCS-BEC crossover behavior in FeSe$_{1-x}$S$_{x}$ again suggest that a multiband system exhibits a unique feature that is absent in a single-band system.
Thus,  FeSe$_{1-x}$S$_{x}$ offers a unique playground to search for as-yet-unknown novel phenomena in strongly interacting fermions and deserves future attention.

%%%%%%%%%%%%%%%%%%%%%%%%%%%%%%%%%%%%%%%%%%%%%%%%%%%%%%%%%%%%%%%%%%%%%%%%%%%%%%%
%%%%%%%%%%%%%%%%%%%%%%%%%%%%%%%%%%%%%%%%%%%%%%%%%%%%%%%%%%%%%%%%%%%%%%%%%%%%%%%

\section{Exotic Superconducting State Induced by Magnetic Field}

\subsection{Field-induced superconducting phase}

The emergence of a novel superconducting phase at high magnetic fields, whose pairing state is distinctly different from that of the low-field phase, has been a long-standing issue in the study of superconductivity.
One intriguing issue related to this field-induced phase concerns whether the spin imbalance or spin polarization will lead to a strong modification of the properties of the electron systems.
This problem has been of considerable interest not only for superconductors in solid state physics, but also in the studies of the neutral fermion superfluid in the field of ultracold atomic systems and for the color superconductivity in high-energy physics.
Among several possible exotic states associated with the spin imbalance, a spatially nonuniform superconducting state caused by the paramagnetism of conduction electrons has been one of the most intensively studied topics over the past half-century after the pioneering work by Fulde and Ferrell (FF) as well as Larkin and Ovchinnikov (LO)~\cite{Fulde64,Larkin65,Matsuda07,Zwicknagl10,Wosnitza18}.
In the FFLO state, an inhomogeneous superconducting state with a modulated superconducting order parameter is formed.

\begin{figure*}[t]
	\centering
	\includegraphics[width=0.7\linewidth]{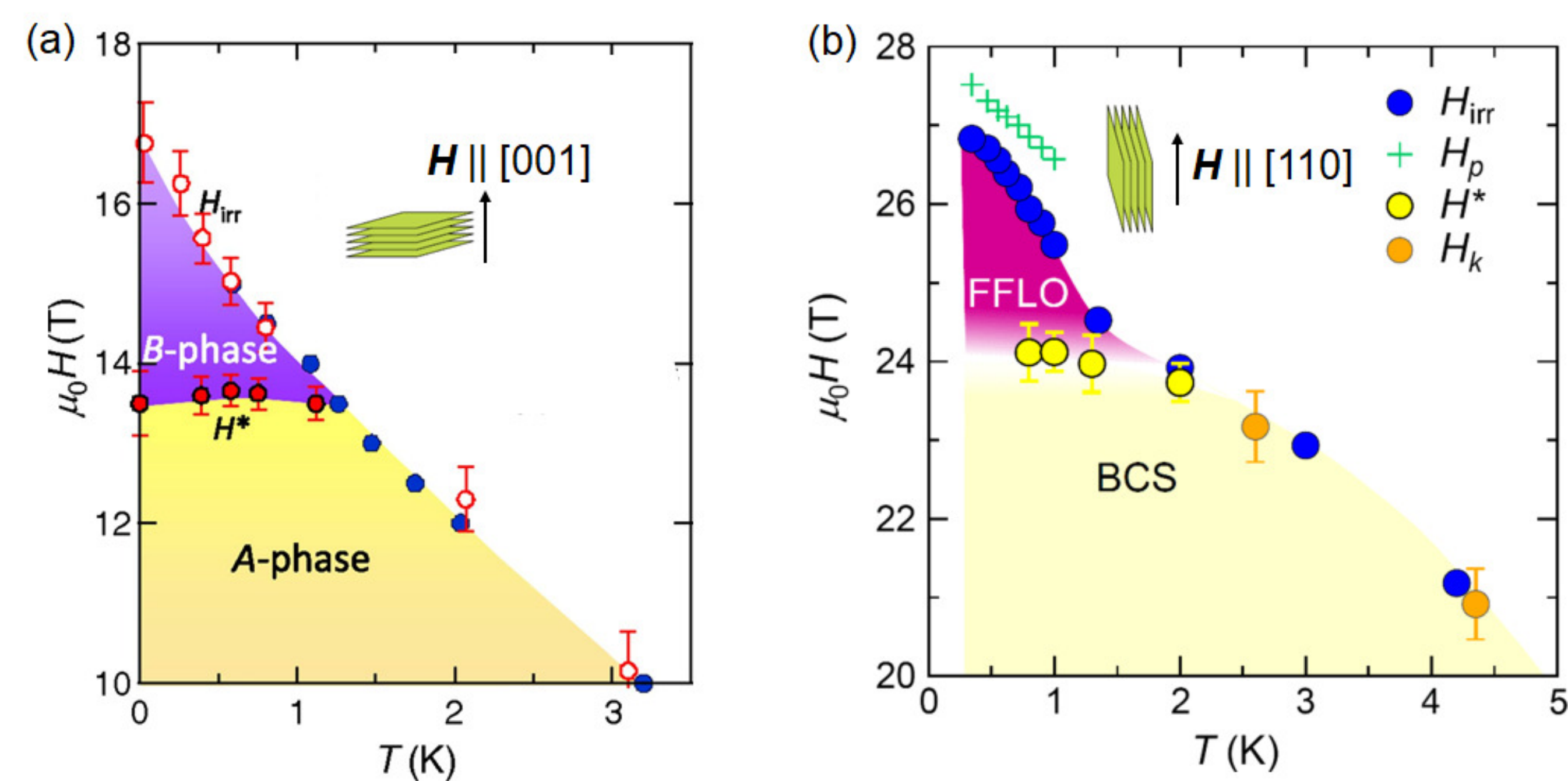}
	\caption{(Color online) 
	(a) High-field phase diagram of FeSe for $\bm{H}\parallel c$.
	Solid blue and open red circles represent the irreversible field $H_{\rm irr}$ determined by the resistivity and the magnetic torque, respectively.
	The mean-field upper critical field is above $H_{\rm irr}$.
	Solid red circles represent $H^{\ast}$ determined by the cusp of the field dependence of the thermal conductivity.
	A high-field superconducting $B$-phase separated from the low-field $A$-phase, which is the BCS pairing ($\bm{k}\uparrow, -\bm{k}\downarrow$) phase, has been proposed.
	Adopted from Ref.~\citen{Kasahara14}.
	(b) Phase diagram for $\bm{H}\parallel ab$ plane.
	Blue circles and green crosses show $H_{\rm irr}$ and $H_{\rm p}$ determined by resistivity measurements, respectively.
	Orange and yellow circles show $H_{\rm k}$ and $H^{\ast}$ determined by thermal conductivity measurements, respectively.
	Above the first-order phase transition field $H^{\ast}$,  a distinct field-induced superconducting phase emerges at low temperatures.
	The high-field phase has been attributed to the FFLO pairing ($\bm{k}\uparrow, -\bm{k}+\bm{q}\downarrow$) state.
	Adopted from Ref.~\citen{Kasahara20}.
	}
	\label{HFSC}
\end{figure*} 

It has been reported that in FeSe, a new superconducting phase appears at the low-temperature/high-field corner in the superconducting state of the $H$-$T$ phase diagram for both the $\bm{H}\parallel ab$ plane~\cite{Ok20,Kasahara20} and the $\bm{H}\parallel c$-axis~\cite{Kasahara14}.
It has been argued that the field-induced phase for $\bm{H}\parallel ab$ is at least consistent with the FFLO phase.
However, even if the FFLO state is realized in FeSe, its physical properties are expected to be very different from those of the originally predicted FFLO state in several aspects, such as the extremely highly spin-polarized state~\cite{Kasahara14}, the coexistence of the FFLO and Abrikosov vortex states~\cite{Kasahara20}, the strongly orbital-dependent pairing interaction~\cite{Sprau17}, the nontrivial Zeeman effect due to spin-orbit coupling~\cite{Kasahara20}, and the multiband electronic structure.
In particular, the magnetic-field-induced superconducting phase in FeSe provides insights into previously poorly understood aspects of the highly spin-polarized Fermi liquid in the BCS-BEC crossover regime.

\subsection{FeSe in strong magnetic field} 
 
The presence of the high-field superconducting phase separated from the low-field one has been observed by several measurements, including resistivity, magnetic torque, heat capacity, and thermal transport measurements.
The phase diagram in a magnetic field applied parallel to the $c$-axis ($\bm{H}\parallel c$) is shown in Fig.\,\ref{HFSC}(a)~\cite{Kasahara14}.
At a field $H^{\ast}$, the thermal conductivity exhibits a cusp-like feature. 
As the Cooper pair condensate does not contribute to heat transport, the thermal conductivity can probe quasiparticle excitations out of the superconducting condensate~\cite{Matsuda06}.
Moreover, as the thermal conductivity has no fluctuation corrections, the cusp of $\kappa/T$ usually corresponds to a mean-field phase transition.
The presence of $H^{\ast}$ has also been indicated by a distinct kink anomaly of the thermal Hall conductivity $\kappa_{xy}$.
The analysis of the thermal Hall angle $\kappa_{xy}/\kappa$ indicates a change in quasiparticle scattering rate at $H^{\ast}$~\cite{Watashige17}.
Very recently, the anomaly at $H^{\ast}$ has also been confirmed by heat capacity measurements.

The irreversibility field $H_{\rm irr}$ caused by the vortex pinning is determined by the magnetic torque, which is used to measure the bulk properties, and by the resistivity.
The irreversibility line at low temperatures extends to high fields well above $H^{\ast}$, demonstrating that $H^{\ast}$ is located inside the superconducting state.
These results suggest the presence of a field-induced superconducting phase [$B$-phase in Fig.~\ref{HFSC}(a)].
However, the presence of the $B$-phase is controversy among different research groups~\cite{Kasahara14,Watashige17,Hardy20}.

\begin{figure}[t]
	\centering
	\includegraphics[width=0.5\linewidth]{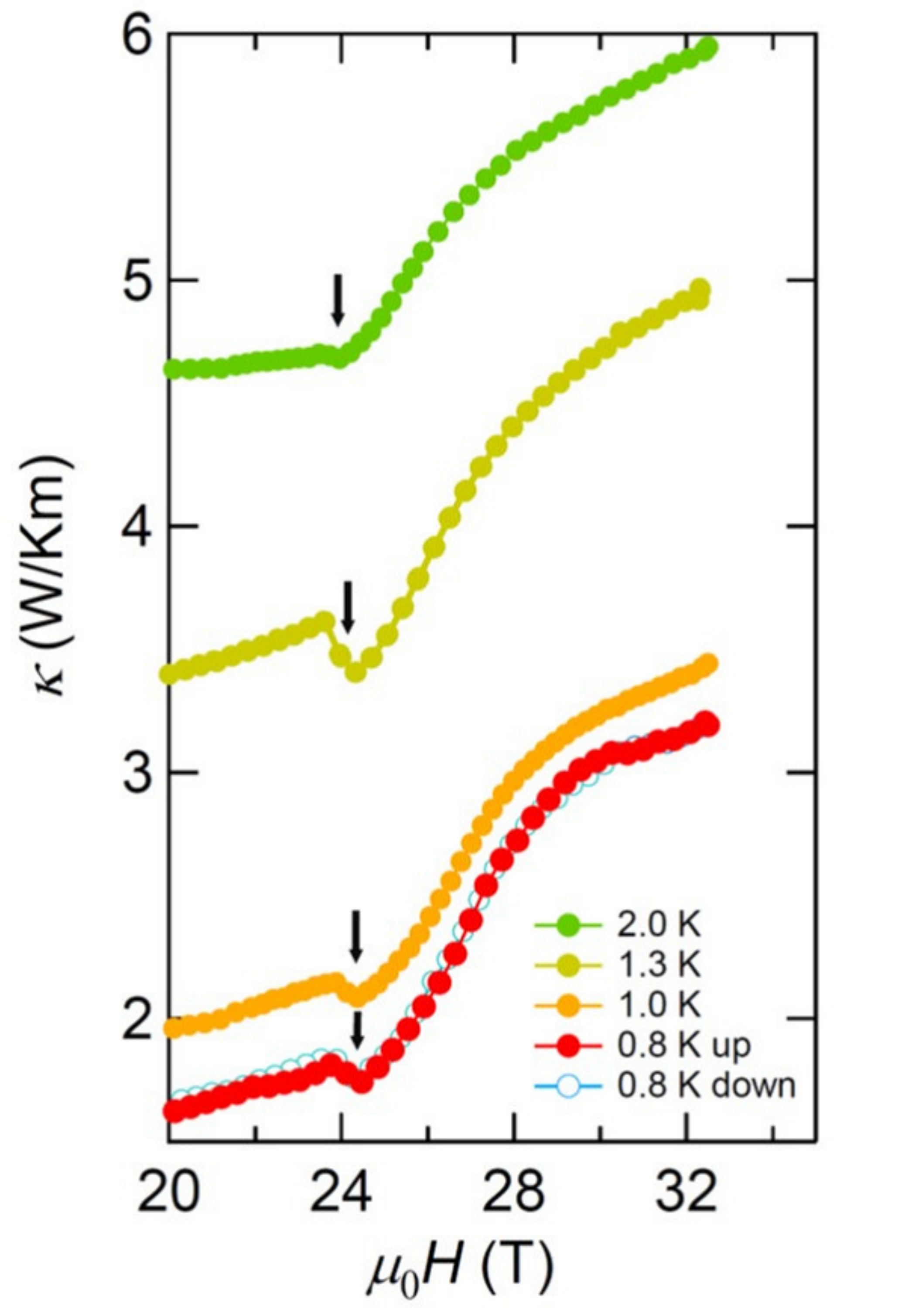}
	\caption{(Color online) 
	Magnetic field dependence of thermal conductivity in the high-field regime at low temperatures in FeSe for $\bm{H}\parallel ab$.
	A discontinuous downward jump appears at $\mu_0H^{\ast}\approx 24$~T inside the superconducting state as indicated by black arrows.
	Adopted from Ref.~\citen{Kasahara20}.
	}
	\label{FFLOTC}
\end{figure}

The $H$-$T$ phase diagram of FeSe for $\bm{H}\parallel ab$ has also been studied recently by several research groups via the measurements of in-plane electrical resistivity, thermal conductivity~\cite{Kasahara20}, magnetocaloric effect~\cite{Ok20},  and heat capacity~\cite{Hardy20} up to 35~T.
All measurements appear to consistently show the presence of a field-induced superconducting phase in this geometry.
Figure~\ref{HFSC}(b) displays the $H$-$T$ phase diagram.
The anomaly in the superconducting state was first detected from the magnetocaloric effect~\cite{Ok20}.
The most marked anomaly was observed by  thermal conductivity measurements on a twinned crystal in $\bm{H}$ applied along the diagonal direction in the $ab$ plane ($\bm{H}\parallel [110]_{\rm O}$  in orthorhombic notation).
As displayed in Fig.~\ref{FFLOTC}, $\kappa(H)$ exhibits a discontinuous downward jump at $\mu_0H^{\ast}\approx 24$~T inside the superconducting state.
At $H^{\ast}$, $\kappa(H)$ shows a large change in field slope and increases steeply with $H$ above $H^{\ast}$.
It should be stressed that the jump of $\kappa(H)$, which is caused by a jump in entropy, is a strong indication of a first-order phase transition, as reported for CeCoIn$_5$~\cite{Izawa01} and URu$_2$Si$_2$~\cite{KasaharaY07}.

In the $H$-$T$ phase diagram, the irreversibility field $H_{\rm irr}$ determined by the onset field of nonzero resistivity is also shown.
Below $\sim 2$\,K, $H_{\rm irr}$ exhibits an anomalous upturn.
It should be stressed that $H^{\ast}$ is deep inside the superconducting state at low temperatures, as evidenced by the fact that $H^{\ast}$ is well below $H_{\rm irr}$.  
No discernible anomaly of $\kappa(H)$ is observed above 2\,K, indicating that the first-order transition occurs only within the superconducting state.
It has been reported that the resistive transition under magnetic fields exhibits   significant broadening at high temperatures.
This is attributed to a strongly fluctuating superconducting order parameter, which gives rise to the drift motion of vortices in the liquid state.
On the other hand, below $\sim 1$~K, the resistive transition becomes very sharp~\cite{Kasahara20}.
Thus, there is also a distinct high-field superconducting phase for $\bm{H}\parallel ab$, which is well separated by a first-order phase transition from the low-field phase.
The quantum oscillation measurements exclude the possibility that the high-field superconducting phases observed for both $\bm{H}\parallel c$ and $\bm{H}\parallel ab$ are AFM ordered phases.

\subsection{Fulde--Ferrell--Larkin--Ovchinnikov (FFLO) state}

Superconductivity is destroyed by an external magnetic field through the orbital and Pauli pair-breaking effects.
The former effect is associated with the Lorentz force acting on electrons, which results in the formation of vortices.
This orbital pair-breaking field is given as $H_{\rm orb}=\Phi_0/2\pi\xi^2$.
The latter effect is associated with the spin paramagnetic effect that tries to align the spin of the original singlet Cooper pairs through the Zeeman effect.
This Pauli pair-breaking limit takes place when the paramagnetic energy in the normal state $E_{\rm P}=\frac{1}{2}\chi_{\rm n} H^2$, where $\chi_{\rm n}=g\mu_{\rm B}^2N(\varepsilon_{\rm F})$ is the normal-state spin susceptibility ($g$ is the $g$-factor and $\mu_{\rm B}$ the Bohr magneton),  coincides with the superconducting condensation energy $E_{\rm s}=\frac{1}{2}N(\varepsilon_{\rm F})\Delta^2$, which yields $H_{\rm P}=\Delta/\sqrt{g}\mu_{\rm B}$.
The ratio of $H_{\rm orb}$ to $H_{\rm P}$, which is called the Maki parameter, is given by
\begin{equation}
\alpha_{\rm M}\equiv \sqrt{2}\frac{H_{\rm orb}}{H_{\rm P}}\approx \frac{m^{\ast}}{m_0}\frac{\Delta}{\varepsilon_F},  
\label{Eq:Maki}
\end{equation} 
where $m^{\ast}$ is the effective mass of the conduction electron and $m_0$ is the free electron mass. 
The Maki parameter is usually much less than unity, indicating that the impact of the paramagnetic effect is negligibly small in most superconductors.
However, in quasi-2D layered superconductors (for parallel fields) and heavy fermion superconductors, $\alpha_{\rm M}$ markedly increases owing to large $m^{\ast}/m_0$ values, and thus the superconductivity may be limited by the Pauli paramagnetic effect.
It should be stressed that in superconductors in the BCS-BEC crossover regime, a large $\Delta/\varepsilon_{\rm F}$ leads to the increase in $\alpha_{\rm M}$.   

\begin{figure}[t]
\centering
\includegraphics[width=0.9\linewidth]{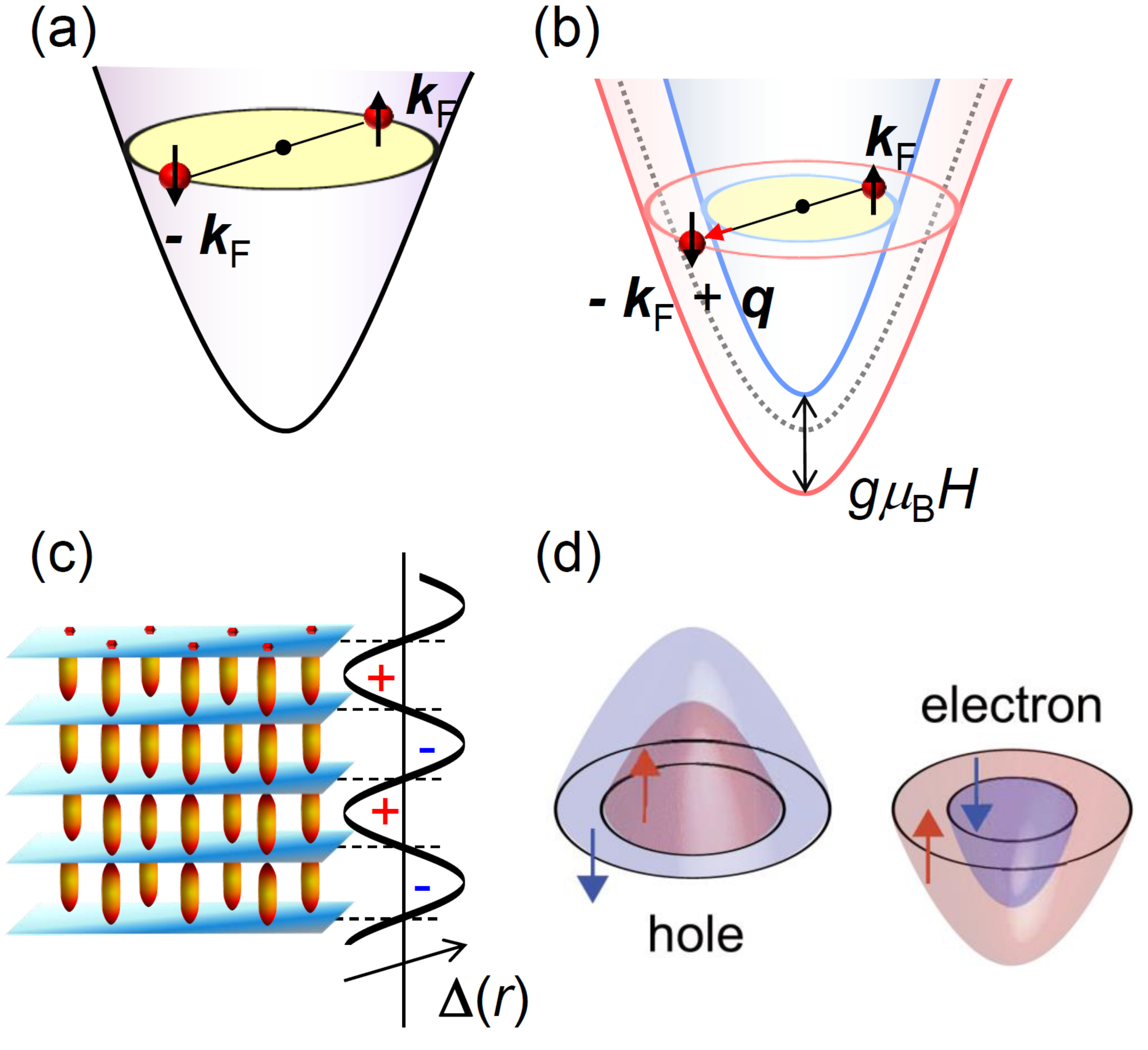}
\caption{(Color online) 
(a) Schematic illustration of Cooper pairing ($\bm{k}\uparrow, -\bm{k}\downarrow$) in the BCS state.
(b) Pairing state with ($\bm{k}\uparrow, -\bm{k}+\bm{q}\downarrow$) in the FFLO state.
(c) Schematic illustration of the superconducting order parameter $\Delta$ in real space and segmentation of the magnetic flux lines by planar nodes.
(d) Schematic electronic structures of hole and electron pockets at fields around $H^{\ast}$ in FeSe.
Both Fermi surfaces are highly spin-imbalanced.
Adopted from Ref.~\citen{Kasahara20}.
}
\label{FFLO}
\end{figure}

FFLO proposed that when the superconductivity is limited by the Pauli paramagnetic effect, the upper critical field can be enhanced by forming an exotic pairing state~\cite{Fulde64,Larkin65}.
In contrast with the ($\bm{k}\uparrow, -\bm{k}\downarrow$) pairing in the traditional BCS state, as shown in Fig.~\ref{FFLO}(a), the Cooper pair formation in the FFLO state occurs between Zeeman split  parts of the Fermi surface,  leading to a new type of ($\bm{k}\uparrow, -\bm{k}+\bm{q}\downarrow$) pairing with $|\bm{q}|\sim g\mu_{\rm B} H/\hbar \upsilon_{\rm F}$ ($\upsilon_{\rm F}$ is the Fermi velocity), as shown in Fig.~\ref{FFLO}(b); the Cooper pairs have finite center-of-mass momenta.
Because of the finite $q\equiv |\bm{q}|$, the superconducting order parameter $\Delta(\bm{r}) \propto \langle \psi^{\dagger}_{\downarrow}(\bm{r})\psi^{\dagger}_{\uparrow}(\bm{r})\rangle$ has an oscillating component $\exp ({\rm i}\bm{q}\cdot\bm{r})$.  
The FF superconducting state has a spontaneous modulation in the phase of the order parameter~\cite{Fulde64}, while the LO state has a spatial modulation of Cooper pair density~\cite{Larkin65}.
It is generally found that the LO states are favored over the FF states, but henceforth both states are simply referred to as the FFLO state.
In the FFLO state, spatial symmetry breaking originating from the appearance of the $\bm{q}$-vector appears, in addition to gauge symmetry breaking.
A fascinating aspect of the FFLO state is that it exhibits inhomogeneous superconducting phases with a spatially oscillating order parameter.
In its simplest form, the order parameter is modulated as $\Delta(\bm{r}) \propto \sin(\bm{q}\cdot \bm{r})$, and periodic planar nodes appear perpendicular to the magnetic field, leading to a segmentation of the vortices into pieces of length $\Lambda=\pi/q$, as illustrated in Fig.~\ref{FFLO}(c).

Despite considerable research efforts in the search for the FFLO states in the past half-century, the FFLO state still constitutes a challenge for   researchers.
Very stringent conditions are required for the realization of the FFLO state.   
In real bulk type-II superconductors, the orbital effect is invariably present, which is detrimental to the formation of the FFLO state.
The FFLO state can exist at finite temperatures if $\alpha_{\rm M}$ is larger than 1.8, but the FFLO region shrinks considerably from that in the absence of the orbital effect.
Moreover, the FFLO state is highly sensitive to disorder. 
Despite a tremendous amount of study of the FFLO state, its firm experimental confirmation is still lacking.
Some signatures of the FFLO state have been reported in only a few candidate materials, including heavy fermion~\cite{Matsuda07} and quasi-2D organic superconductors~\cite{Zwicknagl10,Wosnitza18}.
Among them, the organic $\kappa$-(ET)$_2$Cu(NCS)$_2$ and heavy fermion CeCoIn$_5$ have been studied most extensively.
In both systems, a thermodynamic phase transition occurs below the upper critical fields and a high-field superconducting phase emerges at low temperatures~\cite{Radvan03,Bianchi03,Lortz07,Agosta17}.
In the former, each superconducting layer is very weakly coupled via the Josephson effect.
A possible FFLO state has been reported in a magnetic field applied parallel to the layers, where the magnetic flux is concentrated in the regions between the layers forming coreless Josephson vortices.
However, the position of the first-order transition in $H$-$T$ phase diagram has been controversial, depending on the measurement method used.
Moreover, it has been pointed out that vortex phase transitions have given rise to considerable ambiguity in the interpretation of the experimental data.
The presence of the FFLO phase in CeCoIn$_5$ remains a controversial issue.
In fact, the $H$-$T$ phase diagram of CeCoIn$_5$ is very different from that expected for the original FFLO state.
Moreover, the magnetic order occurs simultaneously at the putative FFLO transition~\cite{Kenzelmann08}, indicating that this phase is not a simple FFLO phase~\cite{Agterberg09, Yanase09, Hatakeyama15}.
Possible FFLO states have also been discussed recently for other systems, including the heavy fermion CeCu$_2$Si$_2$~\cite{Kitagawa18} and iron-pnictide KFe$_2$As$_2$~\cite{Cho17}.

FeSe may satisfy some of the prerequisites for the realization of the FFLO state.
In FeSe in the vicinity of the BCS-BEC crossover regime, an estimate gives $\Delta/\varepsilon_{\rm F} \sim 0.1$--0.3 for the hole band and an even larger $\Delta/\varepsilon_{\rm F}$ for the electron band~\cite{Kasahara14,Hanaguri18}.
By using $m^{\ast}\approx 7m_0$ ($4m_0$) for the hole (electron) pocket determined by SdH oscillation experiments~\cite{Terashima14}, $\alpha_{\rm M}$ is found to be as large as $\sim 5$ ($\sim 2.5$) for the electron (hole) pocket.
This fulfills a requirement for the formation of the FFLO state.
Moreover, the analyses of magnetoresistance and quantum oscillations show that high-quality single crystals of FeSe, obtained through flux/vapor-transport growth techniques~\cite{Song19}, are in the ultraclean limit with an extraordinary large mean free path $\ell$.

On the other hand, there are several unique aspects in FeSe that have not been taken into account in the original idea of the FFLO state.
They arise from the extremely shallow pockets and multiband character.

\begin{itemize}

	\item Extremely large spin imbalance.
The BCS-BEC crossover nature in FeSe gives rise to the large spin imbalance near the upper critical field, which will be discussed in the next subsection.

	\item  Strong spin-orbit coupling, $\lambda_{\rm so} \sim \varepsilon_{\rm F}$, which yields a highly orbital-dependent Zeeman effect.
It has been suggested that this seriously modifies the Pauli limiting field through the $g$-factor.

	\item  Orbital-dependent pairing interaction, which is expected to seriously  affect the $\bm{q}$-vector.
The FFLO pairing may also be orbital-dependent.  

\end{itemize}

It has been argued that the high-field phase for $\bm{H}\parallel ab$ can be associated with an FFLO phase for the following reasons.
First, in the $H$-$T$ phase diagram shown in Fig.~\ref{HFSC}(b), the steep enhancement of $H_{\rm c2}^{ab}$ at low temperatures and the first-order phase transition at  an almost $T$-independent $H^{\ast}$ are consistent with the original prediction of the FFLO state.
Second, planar nodes perpendicular to $\bm{H}$ are expected to be the most optimal solution for the lowest Landau level.
In the present geometry with the thermal current density $\bm{j}_{\rm T} \parallel \bm{H}$, quasiparticles that conduct heat are expected to be scattered by the periodic planar nodes upon entering the FFLO phase.
This leads to a reduction in  $\kappa(H)$ just above $H^{\ast}$, which is consistent with the present results.
Third, as the $c$-axis coherence length ($\xi_{c} \approx 1.5$~nm) well exceeds the interlayer distance (0.55~nm)~\cite{Hsu08,Terashima14}, one-dimensional (1D) tube-like Abrikosov vortices are formed even in a parallel field.
In this case, the planar node formation leads to a segmentation of the vortices into pieces of length $\Lambda$.
The pieces are largely decoupled and, hence, better able than conventional vortices to position themselves at pinning centers, leading to increases in the pinning forces of the flux lines in the FFLO phase.
This is consistent with the observed sharp resistive transition above $H^{\ast}$.
Fourth, as will be discussed in the next subsection, the electron pocket is extremely spin-polarized near the upper critical fields.
Therefore, it is questionable that superconducting pairing is induced in the electron pocket in such a strongly spin-imbalanced state.
It has been shown that the FFLO instability is sensitive to the nesting properties of the Fermi surface.
When the Fermi surfaces have flat parts, the FFLO state is more stabilized through nesting.
As the portion of the hole pocket derived from the $d_{yz}$ orbital forms a Fermi surface sheet that is more flattened than the other portion of the Fermi surface, this 1D-like Fermi sheet is likely to be responsible for the FFLO state.

In contrast to $\bm{H}\parallel ab$, the high field phase for $\bm{H}\parallel c$ remains elusive and its identification is a challenging issue.
It cannot be simply explained by the FFLO state, although the $H$-$T$ phase diagram has some common features with that for $\bm{H}\parallel ab$.
The $\bm{q}$-vector, which is always in the $ab$ plane, does not stabilize the FFLO state for $\bm{H}\parallel c$.
Therefore, an FFLO state may be difficult to form owing to the lack of a $\bm{q}$-vector for $\bm{H}$ applied perpendicular to the quasi-2D Fermi surface of FeSe.
According to the calculation of the effective $g$-factor obtained by the orbitally projected model~\cite{Kang18a}, the formation of the FFLO state is more favored for $\bm{H}\parallel ab$ than for $\bm{H}\parallel c$~\cite{Kasahara20}.
Recently,  a possible FFLO state has been proposed even for $\bm{H}\parallel c$~\cite{Song19}.

\subsection{Highly spin-polarized field-induced state in the BCS-BEC crossover regime}
 
The field-induced superconducting phase provides insights into previously poorly understood aspects of the highly spin-polarized Fermi liquid in the BCS-BEC crossover regime.
In the standard BCS theory of spin-singlet pairing, the pairing occurs between fermions with opposite spins.
The question of what happens if a large fraction of the spin-up fermions cannot find spin-down partners has been widely discussed by researchers from different aspects.
In conventional superconductors, however, a highly unequal population of spin-up and spin-down electrons is very difficult to realize, essentially because superconductivity is usually destroyed by the orbital pair-breaking effects.
Even when the superconductivity is destroyed by the Pauli paramagnetic effect, such a spin imbalance is usually negligibly small.

In paramagnetic metals, the spin imbalance is caused by the Zeeman splitting in a magnetic field, as shown in Fig.~\ref{FFLO}(d).
The magnitude of the spin imbalance $P_{\rm spin}=(N_\uparrow-N_\downarrow)/(N_\uparrow-N_\downarrow)$, where $N_\uparrow$ and $N_\downarrow$ are the numbers of up and down spins, respectively, is roughly estimated as $P_{\rm spin}\approx \mu_{\rm B}H/\varepsilon_{\rm F}$.
Therefore, $P_{\rm spin}$ is estimated to be $P_{\rm spin}\approx \Delta/\varepsilon_{\rm F}$ at $H_{\rm P}$ in Pauli-limited superconductors.
In orbital-limited superconductors, where $H_{\rm orb}<H_{\rm P}$, $P_{\rm spin}$ at $H_{\rm orb}$ is smaller than that expected in Pauli-limited superconductors.
Therefore, in almost all superconductors, $P_{\rm spin}$ is usually negligibly small, $P_{\rm spin}< 10^{-2}$, near the upper critical field, i.e.,  the effect of the spin imbalance is not taken into account in the original FFLO proposal.

\begin{figure*}[t]
	\centering
	\includegraphics[width=0.8\linewidth]{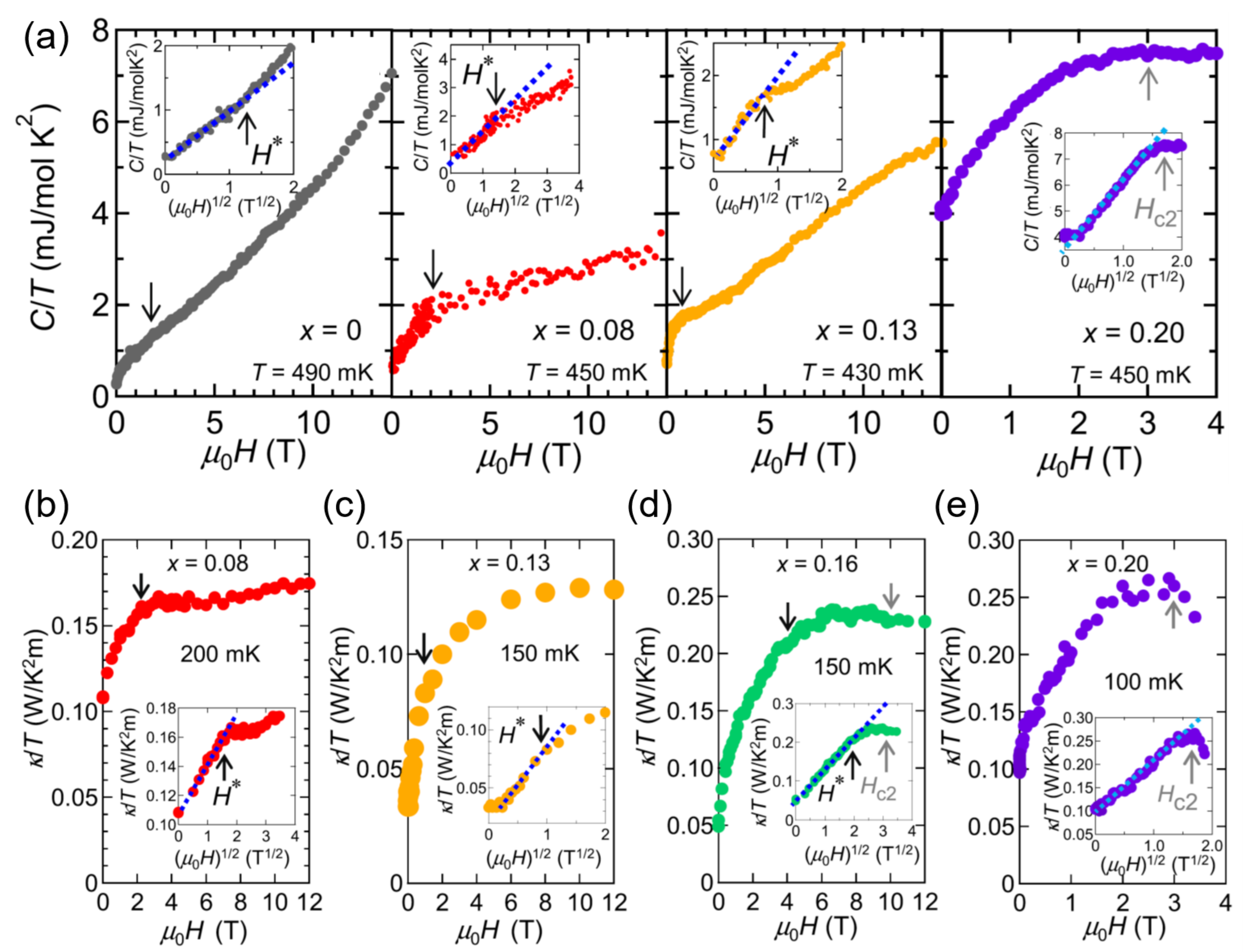}
	\caption{(Color online) 
	%Superconducting gap structure inferred from field dependence of specific heat and thermal conductivity~\cite{Sato18}.
	(a) Field dependence of specific heat at low temperatures below $\sim0.5$~K in FeSe$_{1-x}$S$_{x}$ covering orthorhombic ($x=0, 0.08, 0.13$) and tetragonal ($x=0.20$) phases. 
	(b-e) Field dependence of thermal conductivity at low temperatures below $\sim0.2$~K for orthorhombic [$x=0.08$ (b), 0.13 (c), 0.16 (d)], and tetragonal [$x=0.20$ (e)] phases.
	The insets are the same data plotted against $H^{1/2}$. The dashed lines are the fits to $\sqrt{H}$ dependence.
	Adopted from Ref.~\citen{Sato18}.
	}
	\label{CkappaH}
\end{figure*}

One intriguing issue concerns whether a large spin imbalance will lead to a strong modification of the properties of the correlated Fermi systems.
Although highly spin-imbalanced Fermi systems have been realized in ultracold atomic gases, the nature of the spin-imbalanced superfluid remains unexploited experimentally owing to the difficulty in cooling the systems to sufficiently low temperatures.
In FeSe in the BCS-BEC crossover regime, the Zeeman effect is particularly effective in shrinking the Fermi volume associated with the spin minority, giving rise to a highly spin-imbalanced phase where $\varepsilon_{\rm F}\sim \Delta\sim \mu_{\rm B}H_{\rm c2}$ near the upper critical fields.
For $\bm{H}\parallel ab$, an estimate yields $P_{\rm spin}\sim 0.5$ and 0.2 for electron and hole pockets, respectively, assuming $g=2$~\cite{Kasahara20}.
This indicates that electron pockets are extremely highly polarized.
Therefore,  in the high-field phase of FeSe, a large fraction of the spin-up fermions cannot find spin-down partners.

The presence of a possible FFLO phase in FeSe should stimulate considerable further work in understanding and exploiting strongly interacting Fermi liquids near the BCS-BEC crossover regime, which remains largely unexplored and might bridge the areas of condensed-matter and ultracold-atomic systems.

%%%%%%%%%%%%%%%%%%%%%%%%%%%%%%%%%%%%%%%%%%%%%%%%%%%%%%%%%%%%%%%%%%%%%%%%%%%%%%%
%%%%%%%%%%%%%%%%%%%%%%%%%%%%%%%%%%%%%%%%%%%%%%%%%%%%%%%%%%%%%%%%%%%%%%%%%%%%%%%

\section{Superconductivity near the Nematic Critical Point}

\subsection{Abrupt change in superconducting gap}

As discussed in Sect.~2.3, the electronic nematic phase in FeSe can be suppressed by isovalent S substitution for the Se site. 
Near the nematic QCP ($x_{\rm c}\approx 0.17$), the nematic fluctuations are strongly enhanced~\cite{Hosoi16}, and the transport properties show non-Fermi liquid properties~\cite{Licciardello19,Huang20}.
The impact of such nematic quantum criticality on superconductivity is an important subject in the field of condensed matter physics~\cite{Lederer16}.
Inside the nematic phase ($x<x_{\rm c}$), the superconducting transition temperature $T_{\rm c}$ shows a broad peak at $x\sim 0.08$ (see Fig.~\ref{Nematic_Sus}).
At $x=x_{\rm c}$, $T_{\rm c}$ jumps from $\sim 8$~K inside the nematic phase to $\sim 4$~K outside the nematic phase~\cite{Sato18,Mizukami20}.
As can be seen in Fig.~\ref{CkappaH}, the upper critical field for $H \parallel c$ is also strongly suppressed from $\sim 10$~T ($x=0.16<x_{\rm c}$) to $\sim 3$~T ($x=0.20>x_{\rm c}$)~\cite{Sato18}.
These significant changes in superconducting properties at the nematic critical concentration imply that the nematicity, or rotational symmetry breaking, strongly affects superconductivity.

\begin{figure*}[t]
	\centering
	\includegraphics[width=0.8\linewidth]{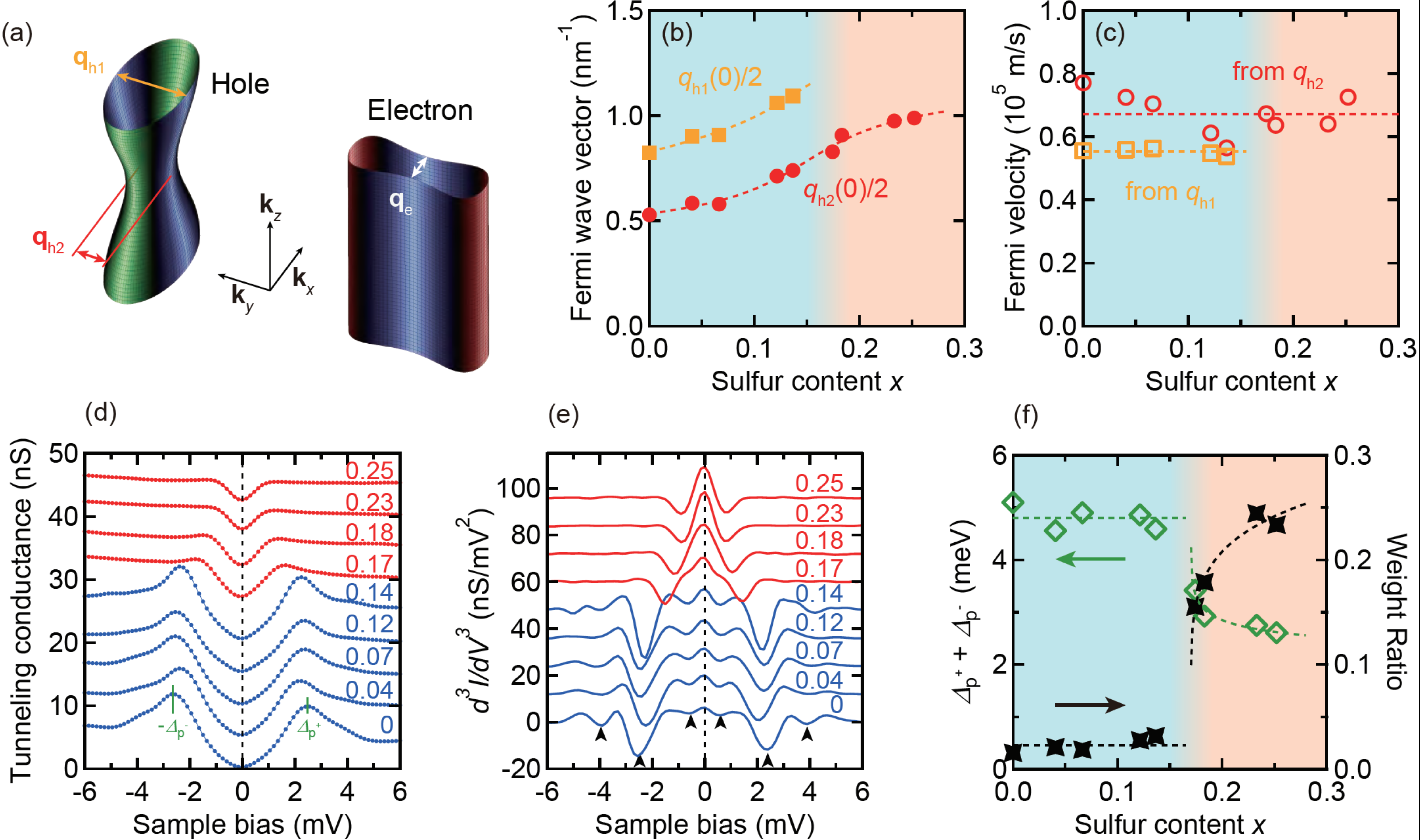}
	\caption{(Color online) 
	Evolutions of electronic structure and low-energy excitations with S substitution determined by QPI and STS measurements of FeSe$_{1-x}$S$_{x}$. 
	(a) Schematics of the 3D constant energy surface of hole and electron bands, and definitions of characteristic scattering wave vectors $\bm{q}_{\rm h1}$ and $\bm{q}_{\rm h2}$.
	(b) Evolutions of the scattering wave vectors as functions of sulfur content $x$.
	Lines are the guides to the eyes.
	(c) Evolution of the Fermi velocity with $x$.
	(d) Averaged tunneling conductance spectra of FeSe$_{1-x}$S$_{x}$ for $0\le x \le 0.25$.
	Each curve is shifted vertically for clarity. Blue (red) curves are for the orthorhombic (tetragonal) phase.
	(e) Energy second derivative of averaged tunneling spectra.
	Each curve is shifted vertically for clarity.
	(f) Evolutions of the apparent gap amplitude (green diamond) and the zero-energy spectral weight normalized by the weights at the gap-edge energies (black stars).
	Adopted from Ref.~\citen{Hanaguri18}.
	}
	\label{STSxdep}
\end{figure*}

The superconducting gap structure is changed abruptly at the nematic QCP, which is evidenced by the field dependence studies of the specific heat $C$ and thermal conductivity $\kappa$~\cite{Sato18}.
Inside the nematic phase, the field dependence of both $C/T$ and $\kappa/T$ shows $\sqrt{H}$ behavior at low fields, as expected in the nodal superconductors, while it deviates from $\sqrt{H}$ at fields much lower than the upper critical field $H_{\rm c2}$, as shown in Figs.~\ref{CkappaH}(a)-\ref{CkappaH}(d).
This deviation can be explained by the multigap effect, and the deviation field $H^{\ast}$ has been attributed to the virtual upper critical field of the smaller gap.
In contrast, in the tetragonal samples outside the nematic phase, such multigap behavior is not observed and, as shown in Figs.~\ref{CkappaH}(a) and \ref{CkappaH}(e), the field dependence of $C/T$ and $\kappa/T$ can be fitted to the $\sqrt{H}$ dependence in the entire field range up to $H_{\rm c2}$.
Near the nematic QCP, charge fluctuations of $d_{xz}$ and $d_{yz}$ orbitals are enhanced equally in the tetragonal phase, while they develop differently in the nematic phase.
From these results, it has been suggested that the orbital-dependent nature of the nematic fluctuations has a strong impact on the superconducting gap structure and hence on the pairing interaction~\cite{Sato18}. 

This marked change in superconducting gap has been corroborated by systematic STS studies~\cite{Hanaguri18}, which are summarized in Fig.~\ref{STSxdep}.
The tunneling conductance remains essentially unchanged with increasing sulfur content $x$ inside the nematic phase, but once the nematicity vanishes at $x>x_{\rm c}\approx 0.17$, the superconducting-gap spectrum shows a drastic change [Figs.~\ref{STSxdep}(d)-\ref{STSxdep}(f)].
Below $x_{\rm c}$, clear quasiparticle peaks are observed at energies $\sim\pm 2.5$~meV and the zero-energy conductance is very small.
Above $x_{\rm c}$, however, quasiparticle peaks are strongly damped with a reduced gap size below $\sim 1.5$~meV, and at the same time, the zero-energy state is much more enhanced than that of the orthorhombic samples.
The marked difference in zero-bias conductance is also consistent with the thermodynamic properties, and the specific heat data shows a large residual DOS only for $x>x_{\rm c}$~\cite{Sato18,Mizukami20}.
By using the QPI technique, the evolution of the normal-state electronic structure with $x$ has also been studied with the same specimens [Figs.~\ref{STSxdep}(a)-\ref{STSxdep}(c)], which revealed that the Fermi surface structure changes smoothly across the nematic QCP.
This implies that the abrupt changes in superconducting properties are not linked to the strength of nematicity, but the presence or absence of nematicity results in two distinct pairing states separated by the nematic QCP~\cite{Hanaguri18}.

\subsection{Possible ultranodal pair state with Bogoliubov Fermi surface}

An intriguing theoretical proposal that may account for these anomalous superconducting states in FeSe$_{1-x}$S$_{x}$ has been recently made by Setty {\it et al.}~\cite{Setty20a,Setty20b}.
This is based on the recently developed notion of a novel superconducting state, dubbed the Bogoliubov Fermi surface~\cite{Agterberg17}.

\begin{figure*}[t]
	\centering
	\includegraphics[width=\linewidth]{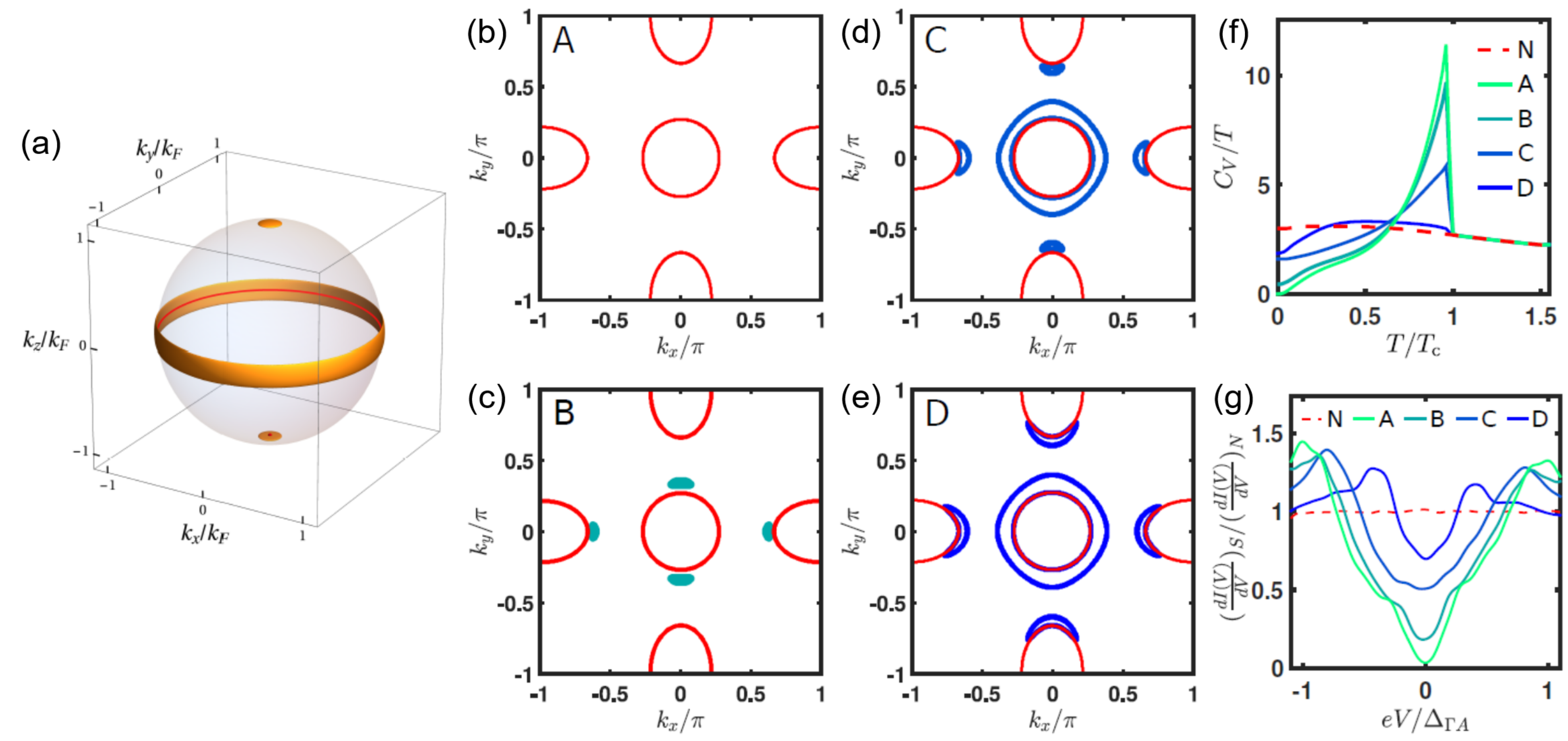}
	\caption{(Color online) Exotic superconducting state with Bogoliubov Fermi surfaces. 
	(a) Schematics of the Bogoliubov Fermi surfaces (yellow region), which may appear near the point and line nodes (red points and lines) when TRS is broken.
	Adopted from Ref.~\citen{Agterberg17}.
	(b)-(e) Schematic Fermi surfaces of FeSe-based superconductors in the normal (red) and superconducting states (green and blue patches) for different interband and intraband gap anisotropy parameters.
	(f) Corresponding temperature dependence of specific heat divided by temperature $C/T$.
	(g) Corresponding tunneling conductance spectra at low energies.
	Panels (b)-(g) are adopted from Ref.~\citen{Setty20a}.
	}
	\label{BogoliubovFS}
\end{figure*}

In unconventional superconductors, the superconducting gap is anisotropic in the momentum space and often exhibits nodes at certain $\bm{k}$ points.
Thus, there are three possible types of superconducting gap: the gap is nodeless, it has point nodes, or it has line nodes. 
It has been shown theoretically that when even-parity superconductors break time reversal symmetry (TRS), there is a possibility of a fourth type having a surface of nodes in some circumstances~\cite{Agterberg17}.
Such a novel state with Bogoliubov Fermi surface [see Fig.~\ref{BogoliubovFS}(a) for an example], which may also be called a topological ultranodal pair state, can be realized when the Pfaffian of the Bogoliubov--de Gennes Hamiltonian, which is non-negative for TRS preserved states, becomes negative.

In FeSe, as discussed in Sect.~3, the gap structure has line nodes or deep minima, and the pairing is likely to be spin-singlet.
Theoretical calculations show that when the relative strength of intraband to interband pairing interactions is altered as a function of sulfur substitution, the Pfaffian may change sign to negative, indicating that the TRS is broken, which gives rise to exotic superconducting states with Bogoliubov Fermi surfaces as shown in Figs.~\ref{BogoliubovFS}(b)-\ref{BogoliubovFS}(e)~\cite{Setty20a}.
If this condition is realized in the tetragonal phase of FeSe$_{1-x}$S$_{x}$, then the presence of the Bogoliubov Fermi surface markedly changes the zero-energy DOS, which is consistent with the experimental observations of the substantially large residual low-energy states in tunneling spectra as well as in specific heat [Figs.~\ref{BogoliubovFS}(f) and \ref{BogoliubovFS}(g)]. 

\begin{figure}[b]
	\centering
	\includegraphics[width=0.7\linewidth]{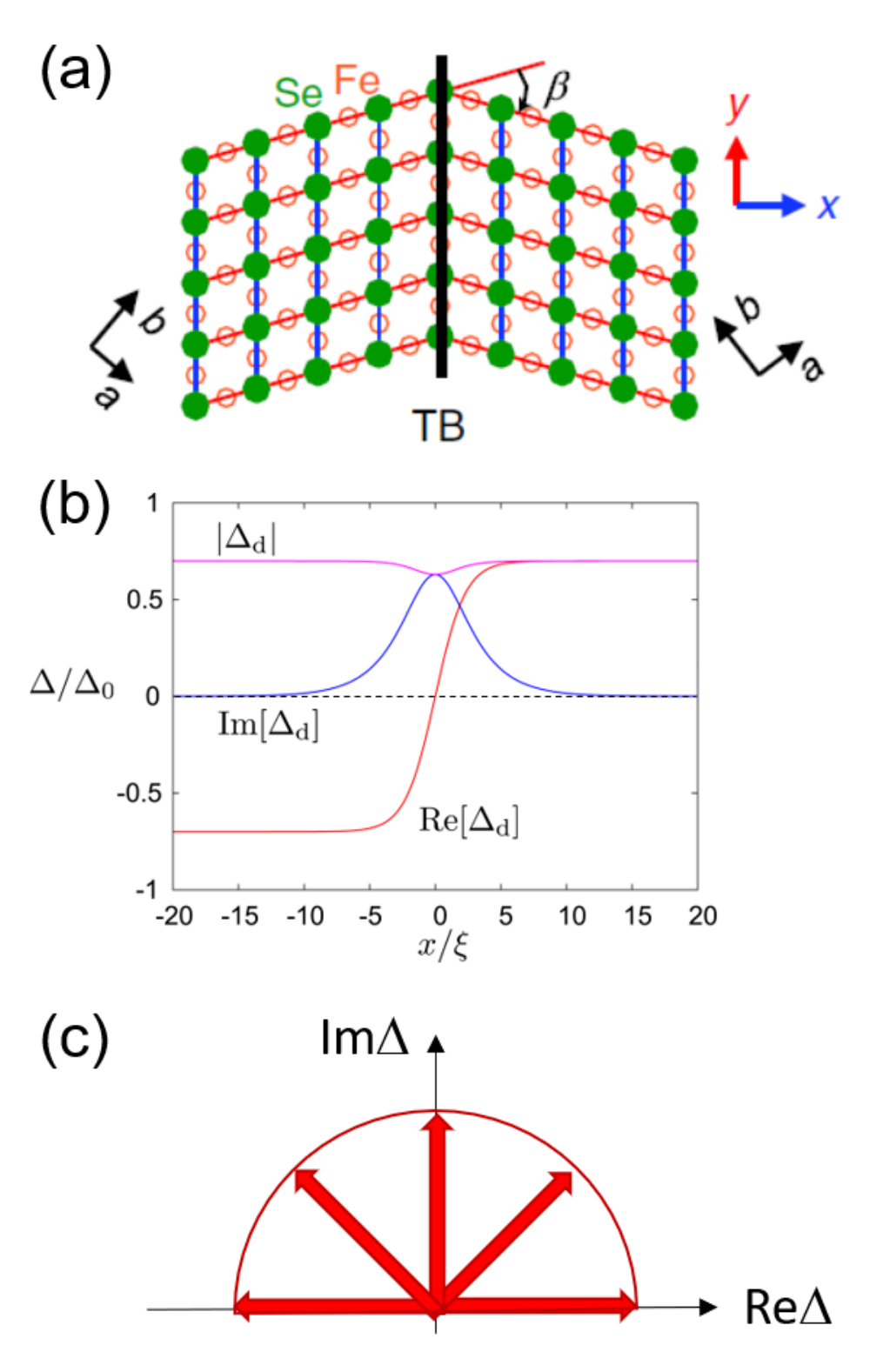}
	\caption{(Color online) Possible complex order parameter induced near nematic twin boundaries~\cite{Sigrist96,Watashige15}.
	(a) Schematic view of the crystal structure of FeSe near a twin boundary.
	Adopted from Ref.~\citen{Watashige15}.
	(b) Possible position dependence of $d$-wave component near a twin boundary with a finite imaginary part.
	The real part should change sign across the boundary, while the magnitude of the gap remains finite owing to the induced imaginary part.
	(c) Schematic trajectory of the gap function as a function of position in the complex plane.
	}
	\label{TRSB_TB}
\end{figure}

To realize such a state, the superconducting state must break TRS, and thus the question is whether the FeSe-based superconductors have TRS breaking states or not. The current experimental situation on TRS breaking is reviewed in the next section.

%%%%%%%%%%%%%%%%%%%%%%%%%%%%%%%%%%%%%%%%%%%%%%%%%%%%%%%%%%%%%%%%%%%%%%%%%%%%%%%
%%%%%%%%%%%%%%%%%%%%%%%%%%%%%%%%%%%%%%%%%%%%%%%%%%%%%%%%%%%%%%%%%%%%%%%%%%%%%%%

\section{Time-Reversal Symmetry (TRS) Breaking}

\subsection{Effect of nematic twin boundary}

\begin{figure*}[t]
	\centering
	\includegraphics[width=0.7\linewidth]{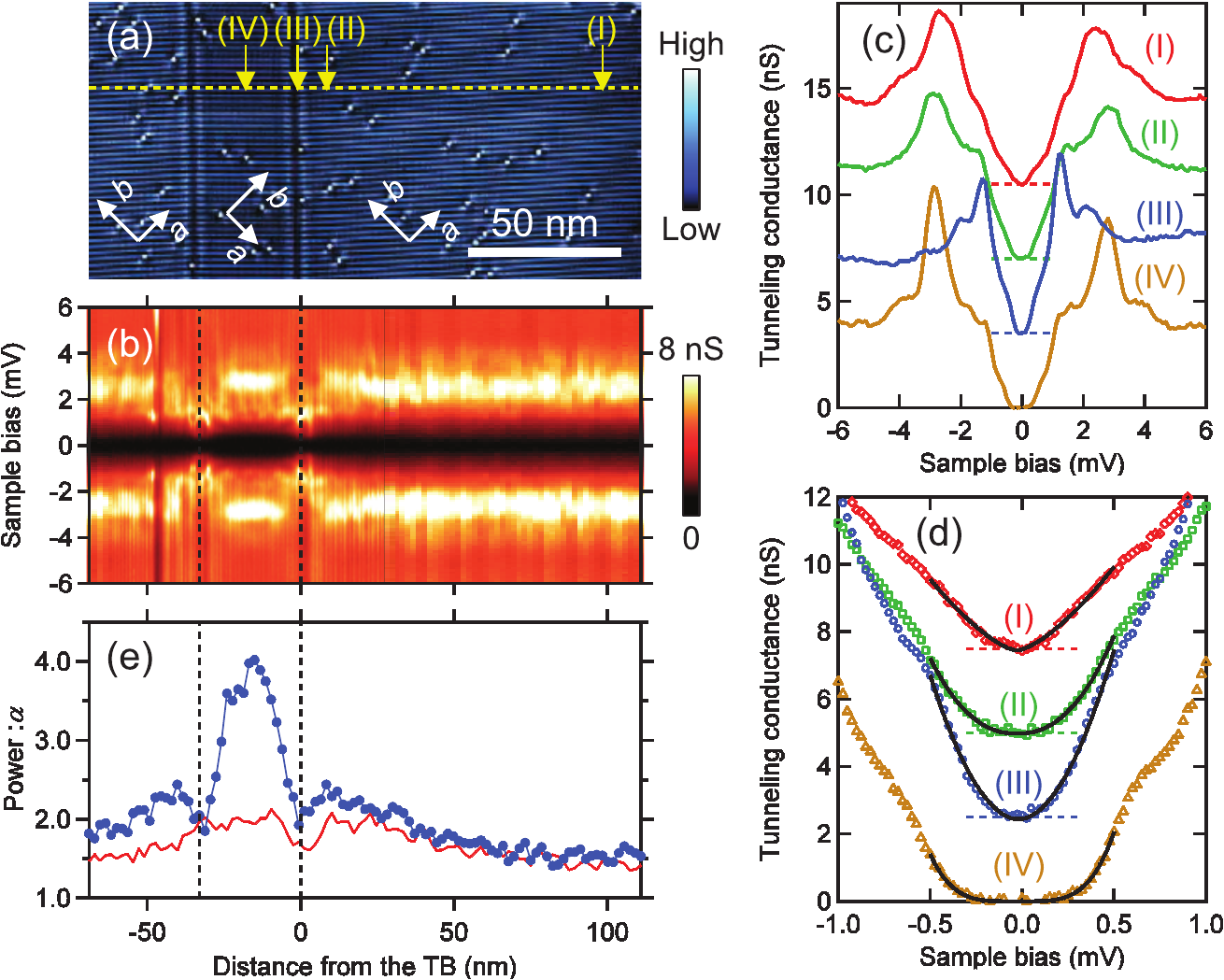}
	\caption{(Color online) 
	Evolution of STM conductance spectra across nematic twin boundaries in FeSe. 
	(a) STM topographic image in an area containing two vertical twin boundaries separated by $\sim33$~nm.
	(b) Intensity plot for the position dependence of conductance spectra along the yellow dashed line in (a).
	(c) Tunneling spectra at four positions indicated in (a).
	Each curve is shifted vertically for clarity.
	(d) Expanded view of spectra at low energies.
	The solid lines are the fits to the power-law energy dependence $|E|^\alpha$.
	(e) Position dependence of the exponent $\alpha$ obtained by the power-law fitting (blue circles), compared with the result for single twin-boundary case.
	Adopted from Ref.~\citen{Watashige15}.
	}
	\label{TRSB_STS}
\end{figure*}

Time reversal is simply equivalent to the complex conjugation of the wave functions for a spinless system. 
Thus, in spin-singlet superconductors, a TRS breaking state can be described by a complex order parameter $\Delta=\Delta_1 +{\rm i}\Delta_2$, whose time reversal $\Delta^{\ast}=\Delta_1 -{\rm i}\Delta_2$ is not identical to $\Delta$.
In a tetragonal $D_{4h}$ system, the $s$-wave and $d$-wave even-parity pairing states belong to different irreducible representations, and thus these states generally have different transition temperatures.
Usually one of the transition temperatures wins over the other, but when the pairing interactions that drive these different states are comparable, these two states may mix in the form of an $s+{\rm i}d$ state whose onset is at a temperature lower than the actual $T_{\rm c}$.

In the nematic phase with orthorhombic symmetry, the $s$-wave and $d$-wave states no longer belong to different irreducible representations, and they can mix in the real form $s+d$~\cite{Kang18b}.
This can be easily understood by the fact that the nematicity is characterized by the two fold symmetry in the plane, and thus the superconducting order parameter should also be two fold symmetric, as evidenced by the observation of the elongated ellipsoidal shape of vortices~\cite{Song11,Watashige15,Hanaguri19}, which can be described by the sum of fourfold $s$-wave and twofold $d$-wave components.

In the nematic phase, another important aspect is the formation of domains with different nematic directions.
Across a boundary of the two domains, namely, the nematic twin boundary, the crystal structure is rotated by 90 degrees as schematically shown in Fig.~\ref{TRSB_TB}(a).
Thus, the twofold $d$-wave component of the superconducting order parameter must change sign across a twin boundary, i.e., one domain has an $s+d$ state and the other domain has an $s-d$ state.
Then the question is how to reverse the sign near the twin boundary.
One possibility is to change sign while keeping the order parameter real, and the magnitude of the $d$-wave component shrinks when approaching the boundary and becomes zero at the twin boundary.
Another possibility is to have an imaginary component to avoid a vanishing order parameter as shown in Fig.~\ref{TRSB_TB}(b); as a function of the position across the boundary, the order parameter follows an arch trajectory in the complex plane from $s-d$ to $s+d$ through the $s+{\rm i}d$ state [Fig.~\ref{TRSB_TB}(c)].
Such a problem was first considered by Sigrist {\it et al.}, who developed a theory for a $d+s$ order parameter in the orthorhombic high-$T_{\rm c}$ superconductor YBa$_2$Cu$_3$O$_{7-\delta}$ with twin boundaries~\cite{Sigrist96}.
They found that the imaginary component may appear near twin boundaries in a length scale much larger than the coherence length $\xi$. 

\begin{figure}[t]
	\centering
	\includegraphics[width=\linewidth]{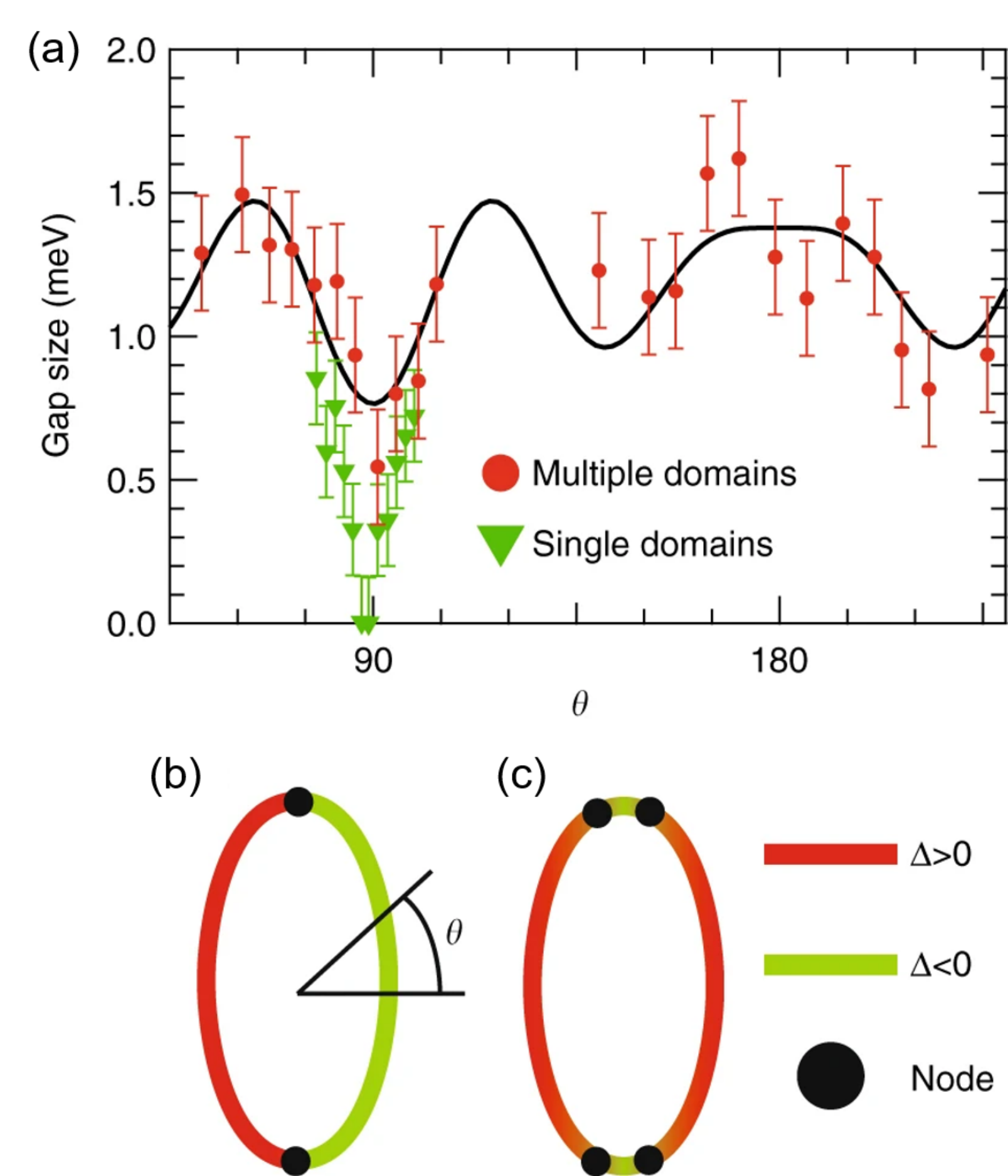}
	\caption{(Color online) 
	ARPES measurements of the superconducting gap for the hole band in single-domain and multiple-domain samples of FeSe.
	(a) Momentum dependence of the gap along the hole Fermi surface ellipsoid in single-domain (green triangles) and multiple-domain (red circles) samples.
	The solid line is a fit for $\Delta(\bm{k})$ in the multiple-domain sample.
	(b) Possible positions of nodes for a spin-triplet $p$-wave state.
	(c) Possible positions of nodes for a spin-singlet $s+d$ state.
	Adopted from Ref.~\citen{Hashimoto18}.
	}
	\label{TRSB_ARPES}
\end{figure}

\subsection{Evidence from gap structure}

One of the consequences of the presence of an imaginary component in the superconducting order parameter $\Delta(\bm{k})=\Delta_1(\bm{k})+{\rm i}\Delta_2(\bm{k})$ is that the low-energy quasiparticle excitations given by 
\begin{equation}
E_k=\sqrt{(\varepsilon_k-\mu)^2+|\Delta(\bm{k})|^2}=\sqrt{(\varepsilon_k-\mu)^2+\Delta_1(\bm{k})^2+\Delta_2(\bm{k})^2}
\end{equation}
are strongly modified. 
In general, the momentum dependence of the imaginary part $\Delta_2(\bm{k})$ is different from that of the real part, so the low-lying excitations that are determined by nodes in $\Delta_1(\bm{k})$ are expected to be gapped out.

Detailed position-dependent STM/STS studies on FeSe clean crystals have shown that the nematic twin boundaries can be considered as ideal interfaces with no notable structural distortion in an atomic scale~\cite{Watashige15}.
The low-energy conductance spectra at positions far away from the boundaries have a V shape, indicating the presence of low-lying quasiparticle excitations.
In contrast, the STS conductance curves near a twin boundary show flattening behavior at low energies, without exhibiting a zero-energy conductance peak that may be expected when the order parameters at neighboring domains change sign without having an imaginary part.
As shown in Fig.~\ref{TRSB_STS}, this flattening behavior is more pronounced at the positions between two twin boundaries, and the presence of a finite excitation gap is clearly resolved.
These observations of the twin-boundary-induced gap opening are consistent with the complex order parameter near the twin boundaries, and the essential features of conductance spectra have been reproduced by theoretical calculations assuming the presence of an imaginary component near boundaries~\cite{Watashige15}.

The low-energy flattening behavior is found over an extended length scale of $\gtrsim 50$~nm (see Fig.~\ref{TRSB_STS}), an order of magnitude larger than the averaged in-plane coherence length $\xi_{ab}\approx 5$~nm.
This can also be explained by the above theory of complex order parameter, where the characteristic length scale $\bar{\xi}$ can be much longer than the coherence length and diverges when approaching the phase boundary between the time-reversal symmetric $s+d$ state and the TRS-broken $s+{\rm i}d$ state in the bulk.
This phase boundary is determined by the closeness of the transition temperatures of the $s$ and $d$ states and by the amount of orthorhombicity.
In YBa$_2$Cu$_3$O$_{7-\delta}$, the superconducting order parameter is dominated by the $d$-wave component and the orthorhombicity-induced $s$-wave component is much smaller.
Thus, the onset temperature of the TRS breaking state may be much lower than the actual transition temperature, and there have been no reports showing clear evidence for such a TRS breaking state near twin boundaries.
This may also be related to the fact that STM/STS measurements of YBa$_2$Cu$_3$O$_{7-\delta}$ single crystals are challenging owing to the difficulty in cleavage.
In FeSe, in contrast, the superconducting order parameter has comparable $s$ and $d$ components as discussed in Sect.~3, which may lead to the observation of such a state.

In a laser-ARPES study, the angle dependence of the superconducting gap in the hole band near the $\Gamma$ point has been compared between almost single-domain and multiple-domain samples~\cite{Hashimoto18}.
In the former, the anisotropic gap reaches almost zero along the long axis of the underlying ellipsoidal Fermi surface, but in the latter the nodes are lifted and gap minima are found as shown in Fig.~\ref{TRSB_ARPES}.
This difference in gap structure between samples with different domain structures has been interpreted as another piece of experimental evidence that a finite gap opens near the twin boundary.
This observation is consistent with the STS results, although another laser-ARPES study from a different research group did not confirm such node lifting behavior in samples with multiple domains~\cite{Liu18a}.
The results for multiple-domain samples may depend on the density of twin boundaries, and further studies are required to fully understand how $\Delta(\bm{k})$ evolves as a function of position near the boundaries.

\subsection{Evidence from muon spin rotation ($\mu$SR)}

Studies of the gap structure can provide only indirect information on the TRS breaking in the superconducting state. 
A more direct consequence of the TRS breaking is that a finite magnetic field is induced inside the superconducting sample. 
This is related to the fact that the $s+{\rm i}d$ and $s-{\rm i}d$ states are energetically degenerate, forming chiral domains, which is analogous to the magnetic domain formation at zero field in ferromagnets.
Near the boundaries and impurities, a small but finite magnetic field is induced, which can be detected by experimental probes.
The $\mu$SR at zero external field is one of the very sensitive magnetic probes that have been used as direct probes of TRS breaking states in unconventional superconductors.
Very recent zero-field $\mu$SR measurements in vapor-grown single crystals of FeSe have shown that while the muon relaxation rate is almost independent of temperature above $T_{\rm c}$, which is consistent with the absence of magnetic order in FeSe, it starts to develop immediately below $T_{\rm c}\approx 9$~K and continues to increase down to the lowest measured temperature of $\sim2$~K~\cite{Matsuura20}. 

These results also provide strong evidence for a TRS breaking state in FeSe.
The onset temperature of the magnetic induction is very close to $T_{\rm c}$, which may be explained by the comparable magnitudes of $s$- and $d$-wave components of the order parameter.
The magnitude of the magnetic induction at low temperatures is estimated to be as small as $\sim 0.15$~G.
Whether or not this TRS breaking state occurs only in the vicinity of nematic twin boundaries cannot be concluded by this $\mu$SR result for FeSe alone.
This motivates similar experiments on tetragonal FeSe$_{1-x}$S$_{x}$, which we discuss below.

\begin{figure}[t]
	\centering
	\includegraphics[width=0.85\linewidth]{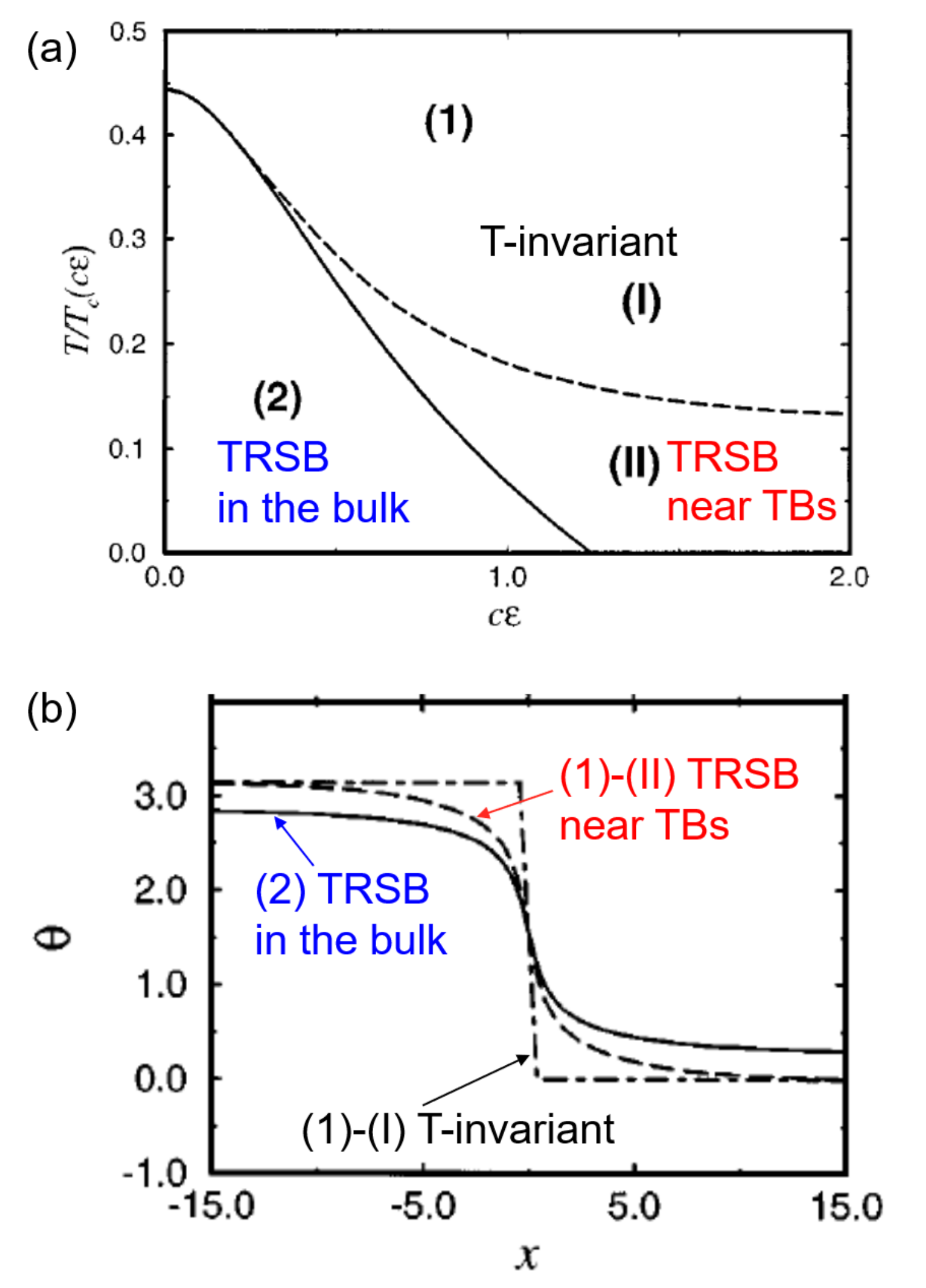}
	\caption{(Color online) 
	Phase diagram of TRS breaking (TRSB) state and corresponding phase difference between $d$ and $s$ components.
	(a) Temperature versus orthorhombic distortion ($\varepsilon$) phase diagram of TRSB states calculated for a $d+s$ state in which the $s$-wave transition temperature is assumed to be $0.5T_{\rm c}$.
	(b) Phase difference $\theta$ between $d$ and $s$ components in a $d+e^{{\rm i}\theta}s$ state as a function of position near a twin boundary.
	The three curves correspond to the three phases in (a).
	In the TRSB state in the bulk, phase (2) in (a), the characteristic length scale $\tilde{\xi}$ diverges and a finite phase difference is present deep in the bulk, but the dependence near the boundary is similar to that in phase (1)-(II).
	Adopted from Ref.~\citen{Sigrist96}.
	}
	\label{TRSB_phase}
\end{figure}

\subsection{TRS breaking in the bulk}

In the tetragonal phase of FeSe$_{1-x}$S$_{x}$ ($x>0.17$), the nematic twin boundaries do not exist, and thus $\mu$SR measurements in tetragonal FeSe$_{1-x}$S$_{x}$ can test TRS breaking inside the bulk.
Most recent data of zero-field $\mu$SR for $x\gtrsim 0.20$ show a similar increase in relaxation rate below $T_{\rm c}\approx4$~K, as found in orthorhombic FeSe~\cite{Matsuura20}.
This immediately implies that tetragonal FeSe$_{1-x}$S$_{x}$ has a superconducting state with broken TRS in the bulk.
In addition, the observations of finite induced magnetic fields with similar magnitudes in orthorhombic and tetragonal samples suggest that, in orthorhombic FeSe, TRS breaking occurs not only near the nematic twin boundaries but also deep in the single domains.
Indeed, the phase diagram in Fig.~\ref{TRSB_phase}(a) studied by Sigrist {\it et al.} for orthorhombic superconductors indicates that the bulk TRS breaking state exists in a wide range of parameters at temperatures below the state of broken TRS only near the twin boundaries~\cite{Sigrist96}.
We note that the difference between these two TRS breaking states in the orthorhombic phase is characterized by the presence or absence of a small phase difference between $s$ and $d$ components away from the boundaries, and in both cases, the position dependence of the phase difference is similarly significant near the twin boundaries as shown in Fig.~\ref{TRSB_phase}(b).
The presence of a small phase difference deep in the single domain implies that a tiny gap opening may be present in low-energy quasiparticle excitations.
The observation of such a tiny gap requires measurement techniques having very high energy resolutions at very low temperatures.

The TRS breaking in FeSe$_{1-x}$S$_{x}$ superconductors fulfills one of the strong requirements of realizing a novel ultranodal superconducting state with the Bogoliubov Fermi surface introduced in the previous section.
The $\mu$SR experiments under magnetic fields provide further support for this.
Transverse-field $\mu$SR measurements in type-II superconductors can provide quantitative information on the magnetic penetration depth that characterizes the field distributions around superconducting vortices.
The magnitude of the penetration depth $\lambda(0)$ is directly related to the density of superconducting electrons, which participate in the supercurrent flows that screen the magnetic field.
It is found that the magnitude of $\lambda(0)$ is larger in tetragonal FeSe$_{1-x}$S$_{x}$ than in orthorhombic FeSe.
This shows the opposite trend to that expected from the increase in the Fermi surface volume with S substitution found in quantum oscillations~\cite{Coldea19}.
This immediately indicates that the density of normal electrons is larger on the tetragonal side, whereas the density of superconducting electrons is smaller.
This surprising result can be consistently explained by the presence of a Bogoliubov Fermi surface in the superconducting ground state for the tetragonal side of FeSe$_{1-x}$S$_{x}$, which reduces the density of superconducting electrons from the normal-state value. 
This  is also consistent with the large residual DOS observed in STS~\cite{Hanaguri18} and thermodynamic measurements~\cite{Mizukami20}.

It has been widely established that in unconventional superconductors with line nodes, the application of a magnetic field gives rise to a rapid enhancement of the low-energy DOS as discussed in Sect.~3.1.2.
%This phenomenon, called as Volovik effect, is explained by the Doppler shift of quasiparticle excitation energy due to the supercurrent around vortices.
To some extent, the Bogoliubov Fermi surface is related to the induced zero-energy states in momentum regions close to the original line nodes, owing to the virtual ``magnetic field'' associated with the spontaneous TRS breaking, which may be called a spontaneous Volovik effect.
The reason why such a state can appear only on the tetragonal side of FeSe$_{1-x}$S$_{x}$ deserves further theoretical and experimental investigations.
It is also intriguing to study how this is related to the BEC-like superconducting state found in tetragonal FeSe$_{1-x}$S$_{x}$ as discussed in Sect.~4.

%%%%%%%%%%%%%%%%%%%%%%%%%%%%%%%%%%%%%%%%%%%%%%%%%%%%%%%%%%%%%%%%%%%%%%%%%%%%%%%
%%%%%%%%%%%%%%%%%%%%%%%%%%%%%%%%%%%%%%%%%%%%%%%%%%%%%%%%%%%%%%%%%%%%%%%%%%%%%%%

\section{Topological Superconducting States}

\subsection{Topological quantum phenomena}

This section reviews a rather different aspect of the FeSe-family of compounds, namely, the topological nature.
Topological quantum physics is one of the most active areas in recent condensed matter research, which was ignited by the theoretical prediction of a quantum spin-Hall insulator~\cite{Kane05a,Kane05b,Bernevig06} followed by the experimental verification~\cite{Konig07}.
A quantum spin-Hall insulator or a 2D topological insulator (TI) is a 2D insulator with a band gap across which the order of the energy bands is inverted from that expected from the energies of the corresponding atomic orbitals.
Such a situation can be caused by the spin-orbit interaction.
Unlike ordinary insulators, quantum spin-Hall insulators always host spin-polarized helical edge states with a linear energy-momentum dispersion, which can be described by a massless Dirac equation.
Such edge states are not related to the chemical and structural properties of the edge but are robust and nontrivial in the sense that they are associated with the topological nature of the Bloch wavefunctions due to the band inversion.

The topological nature can be classified by the topological invariant $Z_2=\{0,1\}$~\cite{Kane05a}.
The $Z_2$ invariants in ordinary and quantum spin-Hall insulators are 0 and 1, respectively.
The concept of the quantum spin-Hall insulator can be extended to three dimensions, where the topological nature is classified by a set of four $Z_2$ invariants~\cite{Moore07,Fu07}.
There are two types of 3D TI, namely, weak and strong TIs.
Strong TIs are an insulator in the bulk but exhibit spin-polarized Dirac surface states all over the surface, irrespective of the chemical and structural properties of the surface, corresponding to a quantum spin-Hall insulator with the Dirac edge states.
In weak TIs, such a Dirac surface state appears only on particular surfaces.

The experimental realization of a quantum spin-Hall insulator was achieved in a HgTe quantum well in 2007~\cite{Konig07}, and Bi-Sb alloy was identified as a 3D strong TI in 2008~\cite{Hsieh08}.
This was around the time when iron-based superconductivity was discovered.
Since then, iron-based superconductivity and topological quantum physics have been investigated actively but in parallel, and thus there has been little interaction between them.
Nevertheless, in principle, if the concept of topology is applied to superconductors, unique and useful phenomena may emerge.
Here, the prerequisite is that the superconducting state should be topologically nontrivial.
There is growing evidence that Te substitution for Se in FeSe gives rise to topological superconductivity at the surface.
In the following, we describe the topological phenomena in FeSe$_{1-x}$Te$_{x}$, paying particular attention to the Majorana quasiparticles in a vortex core.

\subsection{Topological superconductivity and Majorana quasiparticles}

We start by briefly introducing topological superconductivity.
For details, textbooks~\cite{Bernevig13,Nomura16} and a review article~\cite{Sato17b} are available.
A topological superconductor can be viewed as a superconducting counterpart of a TI, where the Cooper pair wavefunction has a topologically nontrivial nature.
Here, the superconducting gap corresponds to the band gap in a TI.
Similar to the case of TIs, gapless boundary states with a linear quasiparticle dispersion appear in topological superconductors.
Such quasiparticles can take exactly zero energy, whereas all the quasiparticle states in an ordinary superconductor appear at finite energies.
Irrespective of the topological nature, quasiparticle states in the superconducting state are coherent superpositions of electron and hole states.
The zero-energy state in the topological superconductor is unique because the electron and hole weights are exactly equal.
Such a half-electron half-hole state can be regarded as a quasiparticle that is its own antiparticle, which is known as a Majorana quasiparticle.
Majorana quasiparticles can be used as a fundamental building block for future fault-tolerant quantum computing~\cite{Sato17b,Nayak08,Alicea12,Ivanov01,Fu08} and thus are attracting much attention.

\subsection{Potential platforms for topological superconductivity}

Unfortunately, topological superconductivity and therefore Majorana quasiparticles have been elusive.
The early proposals mostly focused on chiral $p$-wave superconductors~\cite{Read00}, but such superconductors have yet to be established experimentally.
To overcome this situation, several ways to realize effective chiral $p$-wave superconductivity in artificial systems have been proposed .
These include the use of 1D Rashba semiconductor nanowires~\cite{Mourik12} and magnetic-atom chains~\cite{NadjPerge14,Kim18} in which superconductivity is induced via the proximity effect from an attached ordinary $s$-wave superconductor.

In 2008, Fu and Kane proposed a novel way to realize 2D topological superconductivity using a heterostructure that consists of an ordinary $s$-wave superconductor and a 3D strong TI~\cite{Fu08}.
If superconductivity is induced in the spin-polarized Dirac surface state of a TI via the proximity effect, it can be effectively regarded as a chiral $p$-wave superconducting state owing to the spin polarization of the Dirac surface state~\cite{Fu08}.
A magnetic field applied perpendicular to the interface generates quantized superconducting vortices, in which Majorana quasiparticles would be localized to form a Majorana bound state (MBS)~\cite{Fu08}.
Since the MBS has exactly zero energy, it may show up as a zero-bias peak (ZBP) in the LDOS spectra measured by STM/STS, in principle.

However, there are two immediate issues to be addressed to implement actual experiments.
First, in the Fu-Kane proposal, topological superconductivity emerges at the interface, which is buried beneath either TI or $s$-wave superconductor films.
Therefore, surface-sensitive probes such as STM/STS cannot directly access the MBS, even if it exists.
Second, even though the topologically trivial vortex bound states are expected to appear only at finite energies, their lowest energy $\sim \Delta^2/\varepsilon_{\rm F}$ is generally very small $\sim \mu$eV (see Sect.~4).
This is below the energy resolutions of the conventional electron spectroscopy techniques.

Experimental attempts using STM/STS were made on heterostructures where \ce{Bi2Te3} (TI) thin films were epitaxially grown on a \ce{NbSe2} ($s$-wave superconductor) substrate by molecular-beam epitaxy~\cite{Sun17}.
The superconducting gap size observed at the \ce{Bi2Te3} surface decayed exponentially with increasing film thickness, consistent with proximity-induced superconductivity.
This suggests that the observed LDOS spectra include information at the interface, although it is indirect.
The above-mentioned energy resolution issue remains, but other characteristics, such as the spatial dependence of the LDOS spectrum and the spin polarization, have been investigated to determine the features that may signify the MBS~\cite{Sun17}.

Obviously, it is desirable to investigate a system where the MBS is exposed at the surface and is energetically well separated from other trivial vortex bound states.
To have topological superconductivity at the surface, one can think of a superconductor that has TI-like character in its normal-state bulk band structure.
This is possible if the superconductor is a certain semimetal.
As in the case of insulators, semimetals exhibit a band gap, although it meanders in the Brillouin zone and turns out to place the highest energy of the valence band above the lowest energy of the conduction band.
If such a meandering band gap is topologically nontrivial, a spin-polarized Dirac surface state that can cross $\varepsilon_{\rm F}$ should emerge.
Furthermore, if the bulk of such a semimetal is an $s$-wave superconductor, the spin-polarized Dirac surface state may turn out to exhibit topological superconductivity due to the self-proximity effect from the bulk.
This is a natural realization of the Fu-Kane proposal at the exposed surface, providing a platform to investigate the MBS with surface-sensitive probes.
There are several candidate materials for such {\it connate topological superconductors}~\cite{Hao19}, such as $\beta$-\ce{PdBi2}~\cite{Sakano15,Iwaya17} and \ce{PbTaSe2}~\cite{Guan16}.
FeSe$_{1-x}$Te$_{x}$ can also be categorized as a connate topological superconductor.

\subsection{Basic properties of FeSe$_{1-x}$Te$_{x}$}

Before discussing its topological nature, we briefly summarize the basic properties of FeSe$_{1-x}$Te$_{x}$.
As in the case of S substitution, Te substitution retains the compensation condition because Te is isovalent to Se.
Sample preparation in the low-Te-concentration regime has been challenging because of the possible miscibility gap in $0.1 \lesssim x \lesssim 0.3$~\cite{Mizuguchi10}.
Very few attempts using pulsed laser deposition~\cite{Imai15} and flux growth~\cite{Terao19} have been reported in this regime.
Single crystals with higher $x$ can be obtained by the melt-growth technique but they tend to contain excess iron atoms at the interstitial sites, which affect various properties~\cite{Rinott17,Liu09}.
Subsequent annealing is generally required to remove the excess irons~\cite{Sun14}.
In short, high-quality single crystals are more difficult to prepare in FeSe$_{1-x}$Te$_{x}$ than in FeSe$_{1-x}$S$_{x}$ and the samples thus far available inevitably contain certain amounts of disorders~\cite{Machida19}.

As in the case of of S substitution, nematicity tends to be suppressed upon Te substitution and diminishes above $x \sim 0.4$~\cite{Mizuguchi10}.
$T_{\rm c}$ is rather insensitive to $x$ in the region $0.3 \lesssim x \lesssim 0.7$ and reaches a value as high as 14.5~K, which is higher than that of FeSe.
Unlike FeSe$_{1-x}$S$_{x}$ with highly anisotropic superconducting gaps, a rather isotropic superconducting gap of $\Delta \sim 1.5$~meV has been observed in FeSe$_{1-x}$Te$_{x}$~\cite{Hanaguri10}.
A double-stripe-type long-range magnetic order appears at $x \gtrsim 0.8$, where the superconductivity diminishes~\cite{Mizuguchi10}.

In addition to the topological nature discussed in the next subsection, FeSe$_{1-x}$Te$_{x}$ is advantageous for the MBS search because it is in a BCS-BEC crossover regime (see Sect.~4).
ARPES experiments suggest that $\varepsilon_{\rm F}^{\rm h} \sim \varepsilon_{\rm F}^{\rm e} \sim 10$~meV~\cite{Lubashevsky12,Okazaki15,Rinott17}.
Considering that $\Delta \sim 1.5$~meV~\cite{Hanaguri10}, we can estimate the lowest trivial bound-state energy $\Delta^2/2\varepsilon_{\rm F}$ to be as large as 100~$\mu$eV.
This is still small but enough to distinguish the MBS from the trivial states if the highest-energy-resolution STM/STS technology is employed.
Therefore, except for the issue associated with the disorders, FeSe$_{1-x}$Te$_{x}$ provides an excellent platform to search for the MBS.

\subsection{Topological phenomena in FeSe$_{1-x}$Te$_{x}$}

\subsubsection{Band structure and ARPES experiments}

Early experimental evidence that suggested the topological nature of FeSe$_{1-x}$Te$_{x}$ was the ZBP found in the tunneling spectra at the interstitial excess irons~\cite{Yin15}.
The ZBP was robust in the sense that it did not split or shift to finite energies, even if the STM tip was moved away from the excess iron site and even if magnetic fields were applied~\cite{Yin15}.
Such an apparent robustness triggered theoretical analyses of the topological nature of FeSe$_{1-x}$Te$_{x}$, including a proposal of a quantum anomalous vortex generated by the magnetic excess iron atom and the spin-orbit interaction~\cite{Jiang19,Fan20}.

\begin{figure}[tb]
\centering
\includegraphics[width=\linewidth]{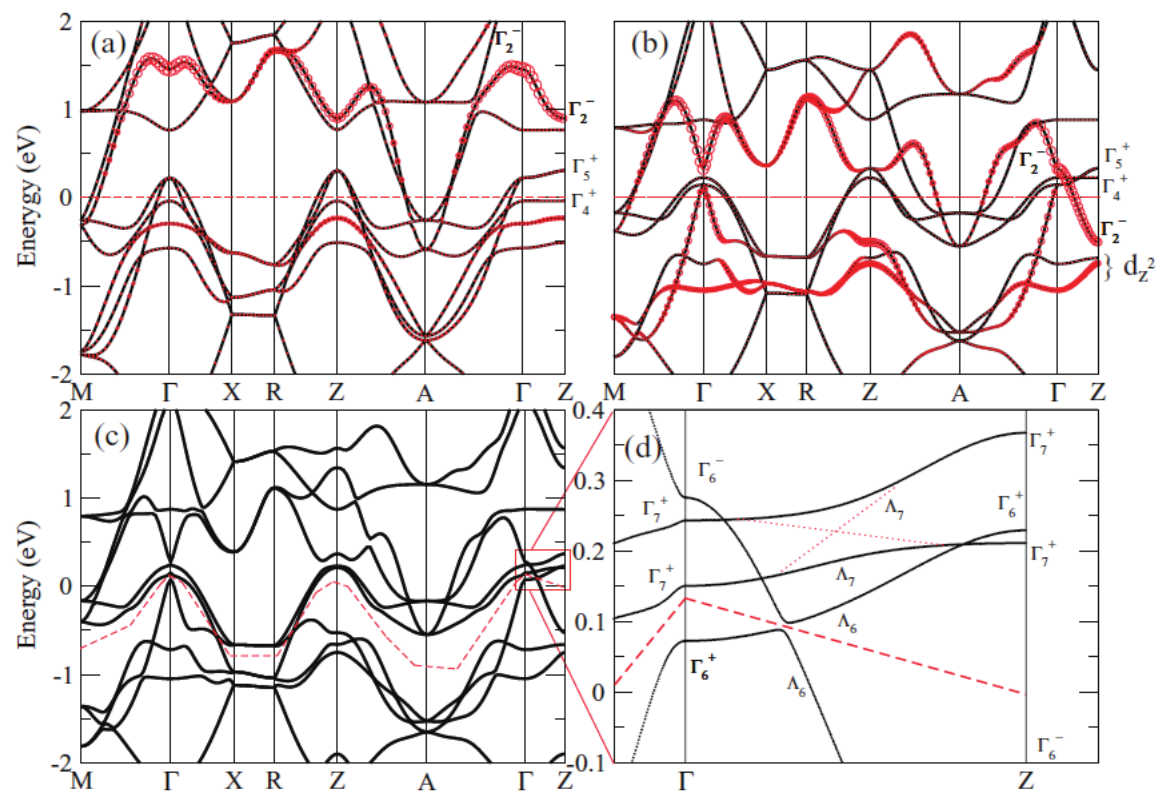}
\caption{(Color online) 
(a) and (b) Calculated band structures without the spin-orbit interaction for FeSe~and for FeSe$_{0.5}$Te$_{0.5}$, respectively.
The size of the red circles denotes the weight of the chalcogen $p_z$ orbitals.
(c) and (d) Calculated band structure for FeSe$_{0.5}$Te$_{0.5}$ with the spin-orbit interaction.
A gap opens at one of the band crossing points along the $\Gamma-Z$ direction, giving rise to the meandering band gap indicated by the red dashed line.
Adopted from Ref.~\citen{Wang15b}.
}
\label{FeSeTe_band}
\end{figure}

First-principles band-structure calculations for FeSe and FeSe$_{0.5}$Te$_{0.5}$ revealed that topological nature indeed emerges upon Te substitution~\cite{Wang15b}.
As discussed in Sect.~2, the band structure of FeSe is complex but trivial from a band topology point of view.
Te substitution alters this original band structure through the following two effects.
First, the Te 5$p$ orbital is more extended than the Se 4$p$ orbital.
This brings about stronger coupling between chalcogen $p_z$ orbitals and thus a larger band dispersion along the $\Gamma-Z$ direction for the associated band [Fig.~\ref{FeSeTe_band}(b)].
This gives rise to additional band crossings along the $\Gamma-Z$ direction, resulting in the band inversion at the $Z$ point.
Second, because Te is heavier than Se, a stronger spin-orbit interaction is expected.
Indeed, the spin-orbit interaction opens a gap at one of the additional band crossing points along the $\Gamma-Z$ direction and brings about a meandering band gap that is topologically nontrivial [Figs.~\ref{FeSeTe_band}(c) and \ref{FeSeTe_band}(d)].
The topologically nontrivial nature of FeSe$_{1-x}$Te$_{x}$ has been pointed out from different points of view~\cite{Wu16,Xu16} and also discussed for other iron-based superconductors~\cite{Hao19,Zhang19}.

\begin{figure}[tb]
\centering
\includegraphics[width=\linewidth]{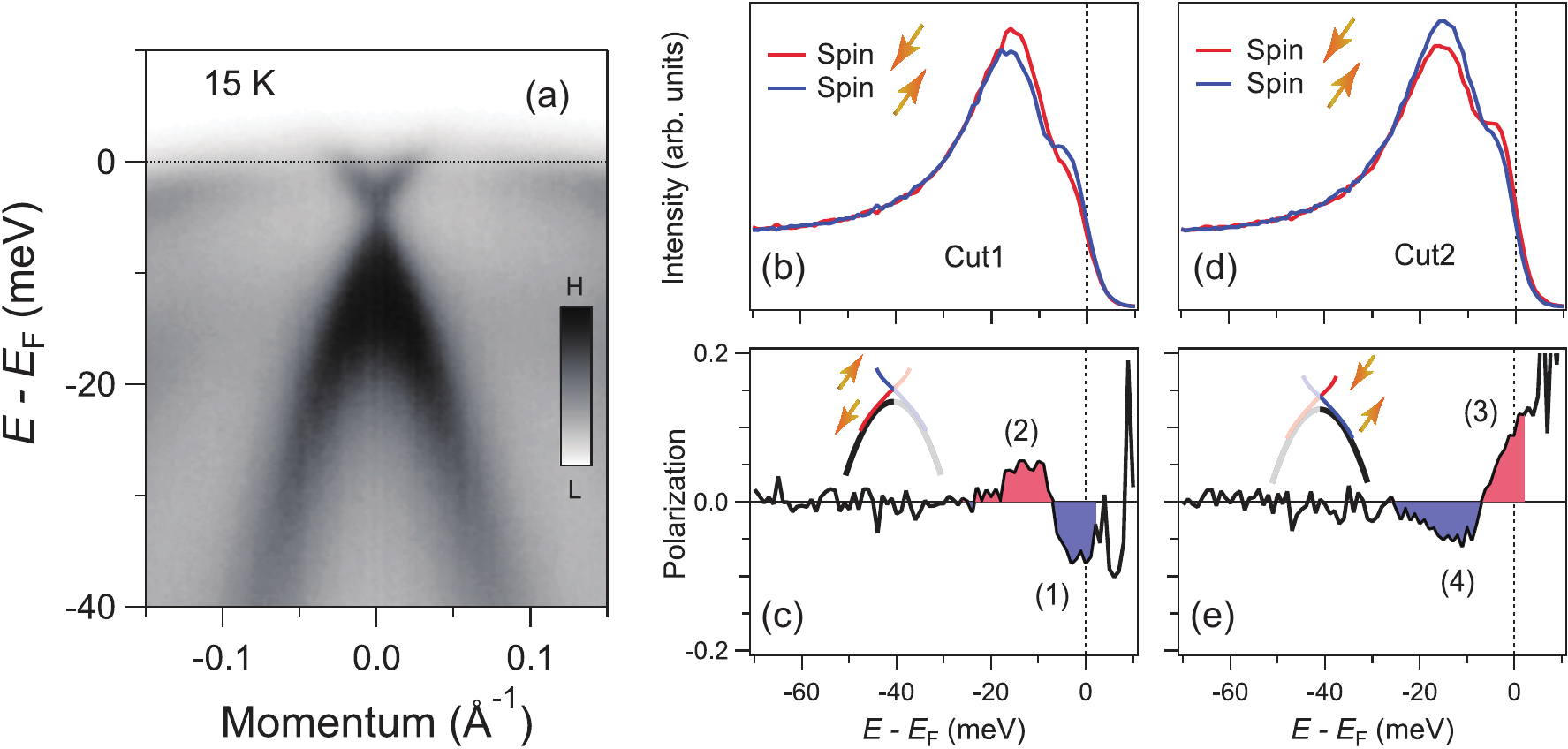}
\caption{(Color online) 
(a) ARPES intensity map showing the Dirac dispersion at the $(001)$ surface of FeSe$_{0.45}$Te$_{0.55}$.
(b) and (c) Spin-resolved energy distribution curves and their difference, respectively, taken on one side of the Dirac cone.
Spin polarizations are illustrated in the inset of (c).
(d) and (e) Same as (b) and (c) but taken on the other side of the cone.
Spin polarizations are reversed, consistent with the helical spin structure.
Adopted from Ref.~\citen{Zhang18a}.
}
\label{FeSeTe_ARPES}
\end{figure}

The above band structure results in the spin-polarized Dirac surface states, which are a hallmark of the topological nature and provide an important platform to host topological superconductivity.
The experimental observation of such surface states was challenging because they are close to $\varepsilon_{\rm F}$ and a bulk band, requiring a high energy resolution for ARPES along with a spin sensitivity.
Later on, ultrahigh-resolution laser-based spin ARPES was utilized and the spin-polarized Dirac surface state was successfully observed at the $(001)$ surface of an FeSe$_{0.45}$Te$_{0.55}$ single crystal (Fig.~\ref{FeSeTe_ARPES})~\cite{Zhang18a}.

\subsubsection{Majorana bound state (MBS) in the vortex core}

\begin{figure}[tb]
\centering
\includegraphics[width=0.9\linewidth]{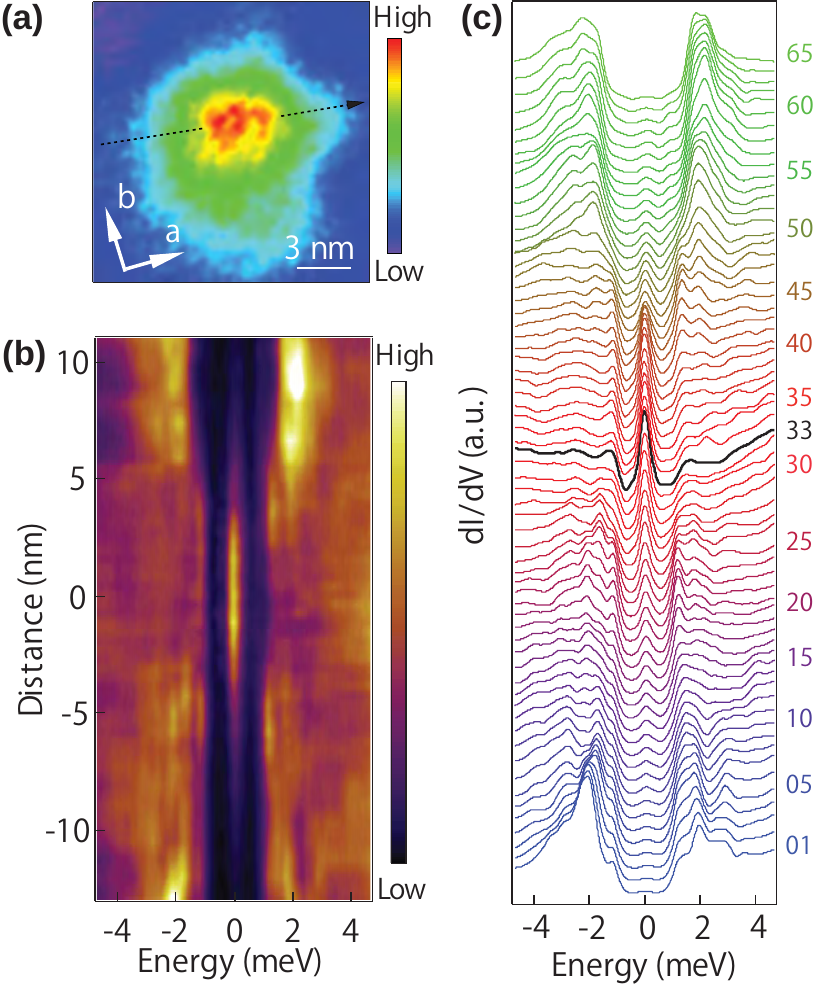}
\caption{(Color online) (a) Image of the vortex in FeSe$_{0.45}$Te$_{0.55}$ obtained by mapping the zero-bias tunneling conductance.
A magnetic field of 0.5~T was applied perpendicular to the observed $(001)$ surface.
(b) Line profile of the tunneling conductance taken along the black dashed arrow in (a), showing a nonsplitting ZBP.
(c) Waterfall plot of the data shown in (b).
The spectrum taken at the vortex center is shown in black.
Adopted from Ref.~\citen{Wang18}.
}
\label{FeSeTe_Vortex}
\end{figure}
\begin{figure*}[t!]
	\centering
	\includegraphics[width=0.7\linewidth]{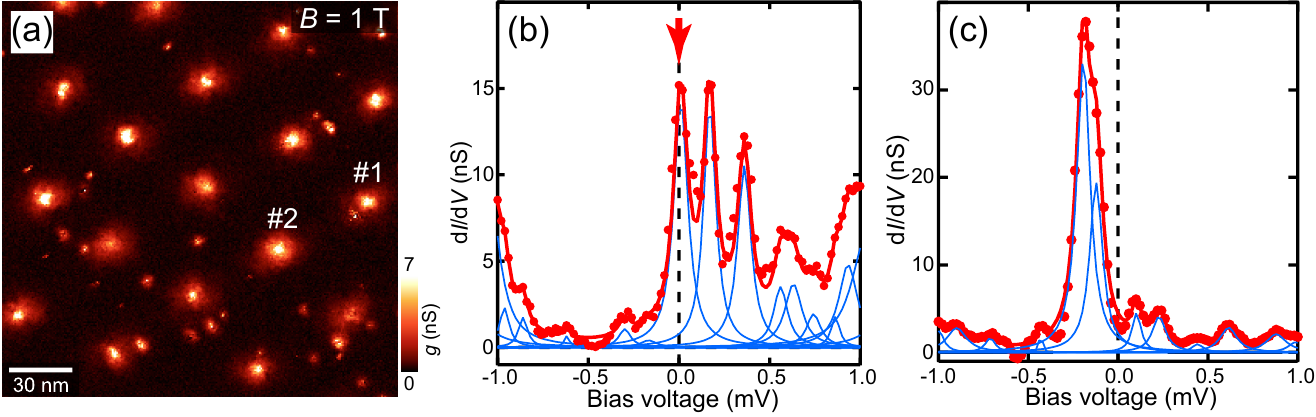}
	\caption{(Color online) 
		(a) Zero-energy conductance map $g$ showing the disordered vortex lattice in  FeSe$_{0.4}$Te$_{0.6}$ under a magnetic field of 1~T applied perpendicular to the observed $(001)$ surface.
		(b) High-energy-resolution tunneling spectrum taken at the center of the vortex labeled as 1 in (a).
		A ZBP is observed as indicated by the red arrow.
		(c) High-energy-resolution tunneling spectrum taken at the center of the vortex labeled as 2 in (a).
		No ZBP is observed.
		Adopted from Ref.~\citen{Machida19}.
	}
	\label{FeSeTe_ZVBS}
\end{figure*}

\begin{figure*}[tb]
	\centering
	\includegraphics[width=0.8\linewidth]{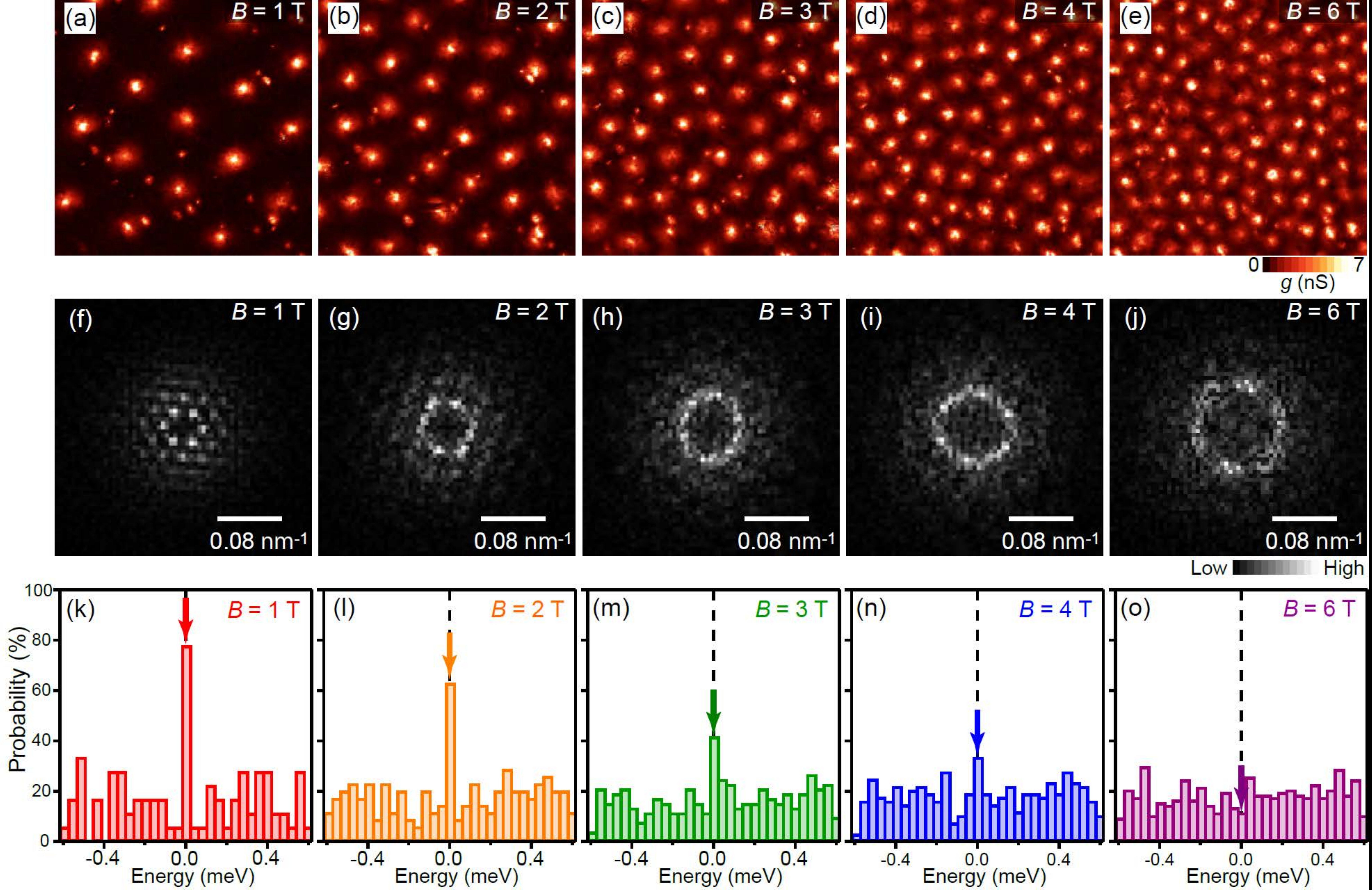}
	\caption{(Color online) (a)--(e) Series of zero-energy conductance $g$ maps showing the vortex lattice in FeSe$_{0.4}$Te$_{0.6}$ under magnetic fields of
		(a) 1~T, (b) 2~T, (c) 3~T, (d) 4~T, and (e) 6~T.
		(f)--(j) Fourier-transformed images from (a)--(e), respectively.
		Ring-like features mean that there is a distance correlation while the long-range orientation order is lost.
		(k)--(o) Respective histograms of the appearance frequency of the conductance peaks at given energies.
		The probability of finding the ZBP decreases with increasing magnetic field.
		Adopted from Ref.~\citen{Machida19}.
	}
	\label{FeSeTe_ZVBS_Hdep}
\end{figure*}

Given the observation of the spin-polarized Dirac surface state in FeSe$_{1-x}$Te$_{x}$, there is hope that the MBS would be formed in its vortex cores.
The first STM/STS experiment in this context succeeded in detecting the ZBP in the vortex cores of FeSe$_{0.45}$Te$_{0.55}$ (Fig.~\ref{FeSeTe_Vortex})~\cite{Wang18}.
Unlike trivial vortex bound states, the observed ZBP remains at zero energy over a certain distance from the vortex center and its intensity evolution agrees with that expected from the theoretical spatial profile of the MBS~\cite{Wang18}.
A similar ZBP has also been observed in other iron-based superconductors such as (Li$_{0.84}$Fe$_{0.16}$)OHFeSe~\cite{Liu18b} and CaKFe$_4$As$_4$~\cite{Liu19}.

The early experiments were carried out at about 0.5~K with an energy resolution of $\sim 250~\mu$eV~\cite{Wang18,Chen18}, which is larger than the estimated energy of the lowest trivial bound states in the vortex core ($\sim 100~\mu$eV).
Therefore, there still remains an ambiguity whether the observed ZBP indeed represents the MBS or whether it is a bundle of trivial vortex bound states that are thermally broadened to form an apparent peak at zero energy.
In addition, there is a puzzle that the ZBP has been observed in only a fraction of vortices and the rest of the vortices do not host the ZBP~\cite{Wang18,Chen18}.
The problem is that FeSe$_{1-x}$Te$_{x}$ samples inevitably contain various chemical and electronic disorders as mentioned above.
It is important to clarify what type of disorder governs the ZBP.

Subsequent STM/STS experiments addressed these issues~\cite{Machida19}.
Dilution-fridge-based STM~\cite{Machida18} was employed to reach ultralow temperatures below 90~mK.
As a result, an energy resolution as high as $\sim 20~\mu$eV was achieved.
This is enough to distinguish the ZBP from the finite-energy trivial bound states and gave a strong constraint that the origin of the ZBP is the MBS~\cite{Machida19} (Fig.~\ref{FeSeTe_ZVBS}).
The correlations between vortices with and without the ZBP and various quenched disorders were also investigated systematically~\cite{Machida19}.
Interestingly, any chemical and electronic disorders preexisting in the sample do not affect the presence or absence of the ZBP.
It was found that the fraction of vortices with the ZBP decreases with increasing applied magnetic field, namely, increasing vortex density (Fig.~\ref{FeSeTe_ZVBS_Hdep})~\cite{Machida19}.
This suggests that interactions among the MBSs in different vortices are responsible for the diminishing ZBP at higher fields.
Moreover, since there are two types of vortex with and without the ZBP, and the quenched disorders do not play any role in this distinction, one can infer that the disorder in the vortex-lattice structure affects the ZBP.
Large-scale theoretical simulations have been performed to confirm this idea~\cite{Chiu20}.
The employed model includes the Majorana-Majorana interaction and disorder in the vortex lattice.
The results reproduced the basic features of the experimental observations~\cite{Chiu20}.
Very recently, an alternative theoretical model based on the spatially inhomogeneous Zeeman effect has been proposed~\cite{Ghazaryan20,Wu20}.

\subsubsection{Search for other Majorana features}

Strictly, the ZBP is only one of the necessary conditions for the MBS.
There have been further attempts to detect features that are unique to the MBS.
The detailed energy spectrum in the vortex core should provide an important clue to revealing such features.
As described in Sect.~4, the quantized energies of the low-lying vortex bound states are given by $\pm\mu_c\Delta^2/\varepsilon_{\rm F}$, where $\mu_c$ is a half-odd integer.
This is actually the case for the vortices in topologically trivial superconductors.
In the case of the topological vortex with the MBS, the quantized sequence becomes $\pm\mu_t\Delta^2/\varepsilon_{\rm F}$, where $\mu_t$ is an integer~\cite{Kong19}.
There is a 1/2 shift between the two cases and the MBS corresponds to the $\mu_t=0$ state.
Experiments have been performed on two types of vortex in FeSe$_{1-x}$Te$_{x}$: those with and without the ZBP~\cite{Kong19}.
Half-odd integer and integer sequences were observed for the former and latter vortices, respectively, suggesting that FeSe$_{1-x}$Te$_{x}$ hosts both topologically trivial and nontrivial vortices depending on the location~\cite{Kong19}.
This appears to be incompatible with the observation that the quenched disorders are unrelated to the ZBP~\cite{Machida19}.
The LDOS spectrum and its spatial evolution are different from vortex to vortex in FeSe$_{1-x}$Te$_{x}$~\cite{Wang18,Chen18,Machida19,Kong19}, making it difficult to reach clear conclusions.
Experiments on more homogeneous samples are highly desired.

Another signature that is expected to be unique for the MBS is the quantization of the tunneling conductance.
It has been theoretically predicted that if Majorana quasiparticles are involved in the tunneling process, induced resonant Andreev reflections may quantize the tunneling conductance to be $2e^2/h$~\cite{Law09}.
This has been experimentally tested using the Majorana state formed in a superconducting InSb nanowire covered with a superconductor (Al) shell~\cite{Zhang18b}.
In the case of the MBS in the vortices, STM/STS is a powerful tool for observation, but the challenge is that a very high tunneling conductance must be achieved to see the expected quantization. 
Such experiments have been performed on FeSe$_{1-x}$Te$_{x}$~\cite{Zhu20} and the related compound (Li$_{0.84}$Fe$_{0.16}$)OHFeSe~\cite{Chen19}.
Plateau-like behaviors in the tunneling conductance have indeed been observed.
However, the quantization behavior is not yet clear.
Moreover, multiple tunneling paths can exist because of the contributions from the multiple bulk bands.
This should affect the quantization condition.
Further experimental and theoretical efforts are anticipated.

The vortex core at the surface of FeSe$_{1-x}$Te$_{x}$ can be regarded as a zero-dimensional boundary in the 2D topological superconductor.
An extended 1D boundary, namely, the edge, may also host Majorana quasiparticles that can move along the edge.
STM/STS has been utilized to detect such dispersing Majorana quasiparticles in FeSe$_{1-x}$Te$_{x}$~\cite{Wang20}.
The platform was a novel naturally formed domain boundary, across which the crystal lattice exhibits a half-unit-cell shift in its structure~\cite{Wang20}.
The LDOS spectrum observed at the domain boundary is constant as a function of energy, similar to an LDOS spectrum of normal metals.
It has been argued that this behavior is consistent with the linear energy-momentum dispersion expected for Majorana quasiparticles moving along a 1D channel.
Another 1D Majorana mode has been suggested to exist at the hinge between the facets of the FeSe$_{1-x}$Te$_x$ crystal, in relation to the possible higher-order topological character~\cite{Gray19}. 
Edge and/or hinge states prepared in more controlled ways may provide a more direct opportunity to investigate the dispersing Majorana quasiparticles in detail.
The development of a sample fabrication technique is awaited.

%%%%%%%%%%%%%%%%%%%%%%%%%%%%%%%%%%%%%%%%%%%%%%%%%%%%%%%%%%%%%%%%%%%%%%%%%%%%%%%
%%%%%%%%%%%%%%%%%%%%%%%%%%%%%%%%%%%%%%%%%%%%%%%%%%%%%%%%%%%%%%%%%%%%%%%%%%%%%%%

\section{Conclusion}

In this review, we have discussed a wide variety of exotic superconducting states observed in bulk FeSe-based superconductors.
What makes this system unique from other superconductors lies in its peculiar electronic properties, in particular, the extremely small Fermi energy, multiband nature, and orbital-dependent electron correlations.
Because of these properties, spin and orbital degrees of freedom, i.e., magnetism and nematicity, both of which are intimately related to the electron pairing, can be largely tuned by nonthermal parameters, such as pressure, chemical substitution, and magnetic field.
Therefore, even though that the pairing mechanism of FeSe-based superconductors is still unknown, it is natural to consider that many different pairing states emerge as a result of this large tunability.

Although more experimental and theoretical works are clearly needed to arrive at a more quantitative description of the data, we feel confident that FeSe-based materials serve as a novel platform for many exotic pairing states, some of which have never been realized in other superconductors. However, we believe that the following questions regarding the superconducting states in FeSe-based materials remain to be answered.

\begin{itemize}

	\item The electron Fermi surface still requires thorough and comprehensive investigations.
The precise shape of the Fermi surface and the detailed superconducting gap properties are yet to be determined.

  	\item  The most fundamental question is whether the prevailing $s_{\pm}$ pairing state with the sign reversal between electron and hole Fermi surfaces is realized even in FeSe$_{1-x}$S$_{x}$ near the nematic QCP, where no sizable spin fluctuations are observed and nematic fluctuations are strongly enhanced.

  	\item Closely related to the above issue, it is an open question why the superconducting gap function markedly changes at the nematic QCP.
In addition, it will also be intriguing to clarify whether the highly unusual superconducting gap function in the tetragonal phase of FeSe$_{1-x}$S$_{x}$ at $x\geq 0.17$ is related to the Bogoliubov Fermi surface.

  	\item It appears that the BCS-BEC crossover properties are largely modified by the multiband character and the orbital-dependent nature.
However, we still lack a quantitative explanation of why the pseudogap is hardly observed despite giant superconducting fluctuations.

  	\item The field-induced superconducting phase can be attributed to the FFLO state.
However, it is not a conventional FFLO state because of a large spin polarization and orbital-dependent pairing.

  	\item  To confirm the time-reversal symmetry breaking, more direct measurements, such as the observation of chiral domains in the superconducting state, are desired.

  	\item The correspondence between the zero-energy conductance peak in the vortex core of FeSe$_{1-x}$Te$_x$ and the Majorana bound state should be examined further to find the features that represent ``Majorananess''.
The obvious goal is to manipulate the Majorana state.
More homogeneous samples are indispensable for this purpose.

\end{itemize}

Lines of evidence for exotic superconducting states of FeSe-based materials have continued to motivate researchers to further investigate and develop novel superconducting states, which are at the forefront of modern research.  We hope that this overview presented here is helpful.

\section*{Acknowledgments}
The authors acknowledge the collaboration of S.~Arsenijevi\'{c}, A.~E.~B\"{o}hmer, J.-G.~Chen, A.~I.~Coldea, I.~Eremin, N.~Fujiwara, T.~Fukuda,  A.~Furusaki, Y.~Gallais, S.~K.~Goh, T.~Hashimoto, S.~Hosoi, N.~E.~Hussey, K.~Ishida, K.~Ishizaka, K.~Iwaya, Y.~Kasahara, W.~Knafo, Y.~Kohsaka, M.~Konczykowski, S.~Licciardello, H.~v.~L\"{o}hneysen, T.~Machida, K.~Matsuura, C.~Meingast, A.~H.~Nevidomskyy, K.~Okazaki, T.~Sasagawa, Y.~Sato, T.~Shimojima, S.~Shin, M.~Sigrist, T.~Tamegai, T.~Terashima, Y.~Tsutsumi, Y.~J.~Uemura, Y.~Uwatoko, T.~Watanabe, T.~Watanuki, T.~Watashige, M.~D.~Watson, T.~Wolf, and J.~Wosnitza.
We particularly thank S.~Kasahara and Y.~Mizukami for the long-term collaboration.  
We are grateful to D.~Agterberg, A.~V.~Chubkov, P.~Dai, R.~M.~Fernandes, P.~J.~Hirschfeld, R.~Ikeda, J.~S.~Kim, S.~A.~Kivelson, H.~Kontani, E.-G.~Moon, S.~Onari, R.~Prozorov, J.~Schmalian, J.~Wang, Y.~Yamakawa, and Y.~Yanase for valuable discussions.
This work was supported by Grants-in-Aid for Scientific Research (KAKENHI) (Nos.~JP19H00649, JP18H05227, JP15H02106, JP16H04024, JP25610096, and JP24244057) and Innovative Areas ``Quantum Liquid Crystals'' (No.~JP19H05824), and ``Topological Material Science" (No.~JP15H05852) from Japan Society for the Promotion of Science (JPSJ), and JST CREST (Nos.~JPMJCR19T5 and JPMJCR16F2).

\end{document}